\documentclass[a4paper]{paper}
\usepackage{jheppub,esint,shuffle,psfrag}
 \usepackage[utf8]{inputenc}
\usepackage{float}
\usepackage{varioref}
\usepackage{amsmath,amsfonts,amsthm,mathrsfs}
\usepackage{enumerate}
\usepackage{fancyvrb}
\usepackage{verbatim}
\usepackage{wrapfig}
\usepackage{appendix}
\usepackage{amstext}
\usepackage{amssymb}
\usepackage{graphicx}
\usepackage{color}
\usepackage{varioref}
\usepackage{multirow,graphics}
\usepackage{epstopdf}
\usepackage{bbm}
\usepackage{slashed}
\usepackage{relsize}
\usepackage{mathtools}

\numberwithin{equation}{section}

\def\<{\langle}
\def\>{\rangle}

\makeatletter
\@ifundefined{usebibtex}{} {}
\makeatother

\def\Tr{\text{Tr}~}

\usepackage{esint}
\usepackage{breqn}
\def \Tr {\mathop{\rm Tr}\nolimits}

\def\numberbysection{\@addtoreset{equation}{section}
                     \def\theequation{\thesection.\arabic{equation}}}

\title{Hexagonalization of Fishnet integrals. Part II. Overlaps and multi-point correlators}
\author{Enrico Olivucci$^a$}
\affiliation[a]{Perimeter Institute for Theoretical Physics, Waterloo, Ontario N2L2Y5, Canada.}
\begin{document}
\abstract{This work presents the building-blocks of an integrability-based representation for multi-point Fishnet Feynman integrals with any number of loops. Such representation relies on the quantum separation of variables (SoV) of a non-compact spin-chain with symmetry $SO(1,5)$ explained in the first paper of this series. The building-blocks of the SoV representation are overlaps of the wave-functions of the spin-chain excitations inserted along the edges of a triangular tile of Fishnet lattice. The zoology of overlaps is analyzed along with various worked out instances in order to achieve compact formulae for the generic triangular tile. The procedure of assembling the tiles into a Fishnet integral is presented exhaustively. The present analysis describes multi-point correlators with disk topology in the bi-scalar limit of planar $\gamma$-deformed $\mathcal{N}=4$ SYM theory, and it verifies some conjectural formulae for hexagonalisation of Fishnets CFTs present in the literature. The findings of this work are suitable for generalization to a wider class of Feynman diagrams.}  
\maketitle
\newpage
\section{Introduction}
\begin{figure}[t]
\includegraphics[scale=1]{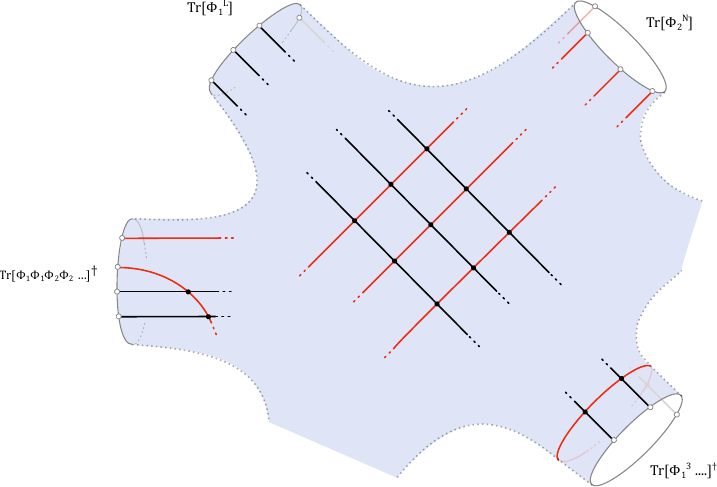}
\caption{General structure of a Feynman integral entering the weak coupling expansion of a planar Fishnet $n$ point correlator. Here we chose $n>4$ and $\langle\text{Tr}\left[\phi_1^L\right] \text{Tr}\left[\phi_2^N\right] \text{Tr}\left[\phi_1^2\phi_2^2\cdots\right]^{\dagger} \text{Tr}\left[\phi_1^3\cdots\right]^{\dagger} \cdots\rangle$.
 The graph lies on a Riemann surface homotopic to a sphere with $n$-punctures. At each puncture there is a $SU(N)$ trace of fields $\phi_1$ and $\phi_2$, represented by a grey circle. Black/red solid lines are the bare scalar propagators $c/(x-y)^2$ of fields $\phi_1$, $\phi_2$. In the bulk the Feynman diagram is a portion of square-lattice fishnet, while around the punctures there are boundary effects due to the quantum correction of the 2-point function of the single-trace operator at the puncture.}
\label{general_bulk_punc}
\end{figure}

The Feynman integrals that dominate the planar limit of multi-trace correlators in the Fishnet CFTs \cite{Gurdogan:2015csr} consist of a portion of Fishnet lattice -- a planar square-lattice made of quartic single-trace vertices $\Tr\left[\phi_1\phi_2\phi_1^{\dagger}\phi_2^{\dagger}\right]$ and bare propagators of the $SU(N)$ adjoint complex matrix fields $\phi_1$, $\phi_2$ that enter the Fishnet Lagrangian. Such portions of Fishnet lattice can be drawn on a sphere with $m$ punctures, as in figure \ref{general_bulk_punc}, where $m$ is the number of traces and whose boundary conditions are determined by the fields inside each trace. 

Interestingly, as a consequence of the non-unitarity of the quartic interaction, a single-trace planar $n$-point correlator\footnote{A single-trace operator features a product of fields at different spacetimes points inside the same $SU(N)$ trace, and can sometimes be referred to as non-local one-point correlator. Objects of this type are well-defined in the Fishnets CFTs, which do not have a gauge symmetry.} gets corrections only at a given order in the weak-coupling expansion in the form of one Fishnet integral that can be drawn on a disk hosting the $n$ spacetime points where fields are inserted on its boundary. This type of correlators is the object of the paper; nevertheless the methods derived in the following can be applied as well in order to describe the Fishnet integrals wrapped on a multi-punctured sphere (e.g. double-trace correlators, lying on a cylinder).

This work deals with the first-principle derivation of the Separation of Variables (SoV) representation for the multi-point Fishnets -- essentially making use of tools of non-compact spin-chain integrability first developed in the simpler instance of the four-point ``Basso-Dixon" correlators \cite{Basso:2015zoa, Derkachov2020}. We proceed constructively and divide the derivation into a few main steps. First, we shall cut the $n$-point disk Fishnets into triangular tiles that are computed using integrability, then compute an integrability-based representation of the tiles and  finally recover the full Feynman integral by ``gluing" its tiles. In practice, ``cutting" the integral means to insert along a column of the Fishnet lattice a resolution of the identity over the excitations of the non-compact $SO(1,5)$ spin chain that realizes Fishnet diagrams as an integrable system \cite{Gurdogan:2015csr,Gromov:2017cja,Derkachov2020}. These excitations are the so-called Fishnet \emph{mirror magnons} extensively studied in \cite{Olivucci:2021cfy}\footnote{The name mirror is related to the terminology of string theory inherited by the Fishnet CFTs from the holographic $\mathcal N=4$ SYM. The mirror magnons in Fishnet theory have been first analyzed in \cite{Basso:2018cvy} from the deformation of the analogue formalism in $\mathcal N=4$ SYM.}.
The direct computation of triangular tiles involves two steps: the replacement of a tile with its eigenvalue and the overlap between the wave-functions of mirror excitations inserted along its three edges. While the first step corresponds to the diagonalization of Fishnet integrals, a worked-out problem \cite{Derkachov2020, Derkachov:2020zvv,Derkachov:2021rrf, Olivucci:2021cfy}, the second step is quite less trivial because it involves the computation of overlaps of complicated functions of many spacetime points. Nevertheless, we will show how to carry out such computations in a simple way that rely on the continuum version of star-triangle identities \cite{Chicherin:2012yn,Derkachov:2021rrf}, and take a particularly handy graphical form. The representation of each triangular patch in the basis of mirror excitations reveals a beautiful factorization of the higher-point, high-loop order Feynman diagrams into the \emph{hexagon} form-factors as conjectured in \cite{Basso:2018cvy} from the weak-coupling limit of the corresponding quantities in $\mathcal{N}=4$ SYM \cite{Basso:2015zoa}, in turn conjectured on the basis of AdS/CFT holography. 

The paper is organized as follows. In section \ref{sec:eigenf} we introduce the Fishnet diagrams as conserved charges of an integrable, non-compact $SO(1,5)$ spin-chain and review the graphical formalism of \emph{star-triangle} computations. Finally we recall details about the eigenfunctions of the chain and their properties which are needed for later. In section \ref{sec:cut} we describe the procedure of cutting a Fishnet integral via suitable triangulations into Fishnet tiles and show how the overlaps of spin-chain wave-functions comes about in this procedure. Section \ref{sec:overs} deals with the classification of overlaps and their general formulae presented along with the computation of a few instructive instances. Finally the sections \ref{sec:tiles} and \ref{sec:SoV} are meant to collect the results for overlaps, analyzing their algebraic properties in connection with the spin-chain integrability, and to provide instructions about how to assemble the building-blocks into the representation of a Fishnet integral. We conclude the paper with a discussion of outlooks and possible applications in section \ref{sec:outlook}.
\section{Eigenfunctions of the Fishnet}
\label{sec:eigenf}
\begin{figure}[t]
\begin{center}
\includegraphics[scale=1]{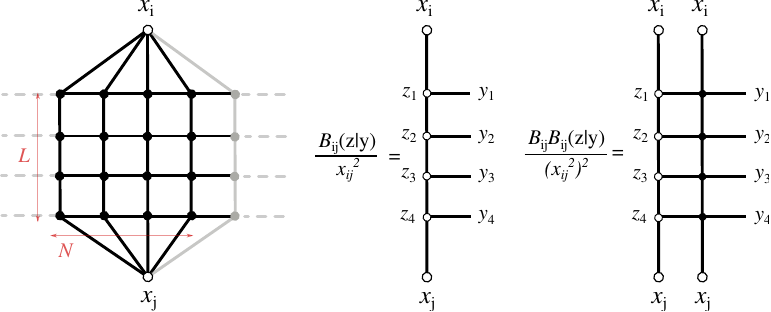}
\end{center}
\caption{The Feynman integral representation of a Fishnet integral takes the form of a square-lattice of scalar propagators (solid segments) and quartic vertices (black dots). $\textbf{(a)}$ A Fishnet square-lattice of height $N=4$ and unbounded width (dashed lines mark the continuation of the lattice). With reference to the text we enlighten a portion of size $L=4$. \textbf{(b)} The kernel of the ``graph-building" integral operator $\hat{B}_{ij}^{(4)}$ defined in \eqref{graph_building} in Feynman diagram notation. \textbf{(c)} The kernel of $\hat{B}_{ij}^{(4)}\hat{B}_{ij}^{(4)}$ forms two rows of lattice.}
\label{fish_rectangle}
\end{figure}
The starting assumption of this paper is that any rectangular portion of a Fishnet integral of the type exemplified in figure \ref{fish_rectangle}, with the upper boundary points and the lower boundary points identified with two points $x_i$, $x_j$, has a compact Mellin-like representation based on the integrability of the Fishnet square-lattice \cite{Zamolodchikov1980a} where the number of lattice columns $L$ enters as a simple exponent in the integrand without changing the number $N$ of Mellin variables and lattice rows. Let us review its main features in the following paragraphs.

A rectangular portion of Fishnet of dimensions $N\times L$ is equal to the $L$-th power of an integral ``graph-building operator" which we one defines via its action on a function of $N$ points in four-dimensional euclidean space $z_1,\dots, z_N$ as follows
\begin{equation}
\label{graph_building}
\hat{B}_{ij}^{(N)} \, f(z_1,\dots,z_N) =
{x_{ij}^2}
\int\left(\prod_{k=1}^N \frac{d^4 y_k}{(z_{k}-z_{k+1})^2(z_{k}-y_{k})^2}\right)  \frac{f(y_1,\dots,y_N)}{(x_i-z_1)^{2}(z_L-x_j)^{2}}\,.
\end{equation}
The operator $\hat{B}_{ij}^{(N)}$ turns out to be the transfer matrix of the integrable non-compact $SO(1,5)$ spin chain with $N$ sites in the unitary representation defined by unit scaling dimension and zero spins $(\Delta,s,\dot s)=(1,0,0)$, which belongs to the complementary series.  This observation was made first in the papers \cite{Derkachov2020, Derkachov:2020zvv} where it was used to compute the four-point Fishnet diagram (also known as Basso-Dixon integral) for any number $N\times L$ of loops. Later, it was formalised in the papers \cite{Derkachov:2021rrf, Olivucci:2021cfy} where it was extended to a more general family of integrable Feynman diagrams describing, for instance, some sectors of  the generalised Fishnets dubbed $\chi$-CFT$_4$ \cite{Caetano:2016ydc, Kazakov:2018gcy}. In the first work of this series \cite{Olivucci:2021cfy}, the eigenfunctions of the $SO(1,5)$ spin-chain corresponding to the Fishnets and its generalisations were found for a generic choice of principal series representation $(\Delta_k,s_k,\dot{s}_k)$ in the $k$-th site of the spin-chain, so to include inhomogeneous and spinning models.  
\subsection*{Graphic Formalism}
The integrability of the spin-chain models of interest follows from the existence of a continuum version of Onsager's star-triangle identity \cite{Onsager} which allows to construct a (factorised) solution of the Yang-Baxter equation and thus of the commuting transfer matrices of the model \cite{Chicherin:2012yn}. We shall review here this fundamental identity and contextually introduce a graphical notation that renders star-triangle computations simple to perform.
Let us introduce some fundamental objects. 
First, the bare propagator in the four-dimensional euclidean space (stripped of constants) with scaling dimension $\Delta$ and left/right spins $\ell$ and $\dot \ell$ reads
\begin{center}
\includegraphics[scale=1]{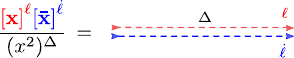}
\end{center}
where $\mathbf{x} \equiv \hat x_{\mu} \boldsymbol{\sigma}^{\mu},\, \mathbf{\bar x} = \hat x_{\mu} \boldsymbol{\bar{\sigma}}^{\mu}$ and $\boldsymbol{\sigma}$ are defined via the standard Pauli matrices $\sigma^{k}$ as
\begin{equation}
\boldsymbol{\sigma}^0= \mathbbm{1}_{2},\, \boldsymbol{\sigma}^k= i \sigma^k,\,\,\, \overline{\boldsymbol{\sigma}}^{\mu} =(\boldsymbol{\sigma}^{\mu})^{\dagger}\,, \end{equation}
and the compact notation $[\mathbf{x}]^{\ell}$ stands for the $\ell$-fold symmetric tensor product
\begin{equation}
\left([\mathbf{x}]^{\ell}\right)_{\dot{\mathbf{a}}}^{\mathbf{b}} = \left([\mathbf{x}]^{\ell}\right)_{\dot a_1,\dot a_2 \dots \dot a_{\ell}}^{b_1,b_2\dots b_{\ell}} = \mathbf{x}_{\dot a_1}^{b_1} \otimes \cdots \otimes \mathbf{x}_{\dot a_{\ell}}^{b_{\ell}} \,.
\end{equation}
The matrix $[\mathbf{x}]^{\ell}$ -- whose exponent is often omitted and clear from the context -- carries the indices $\dot{\mathbf a}$, $\mathbf b$ of $\ell$-symmetric right/left spinors, and so it encodes a symmetric traceless tensor of rank $\ell$. Similarly, the hermitian conjugate matrix $[\overline{\mathbf{x}}]^{\ell}$ carries the indices $\mathbf b$, $\dot{\mathbf a}$ of $\ell$-symmetric right/left spinors. The second building-block is denoted as $\mathbf{R}_{mn}(u)$, and it is a solution of the Yang-Baxter equation on the tensor product of $m,n$-fold symmetric spinors, obtained from the fundamental $su(2)$ $R$-matrix via the fusion procedure \cite{Kulish:1981gi} (see appendix \ref{app:Rmat} for details). When it is multiplied by some matrices $[\mathbf{x}]$ it takes the following graphical notation
\begin{center}
\includegraphics[scale=1.1]{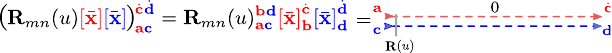}\,.
\end{center}
When the $R$-matrix appears contracted with the numerator of a spinning propagator, it reads
\begin{center}
\includegraphics[scale=1.1]{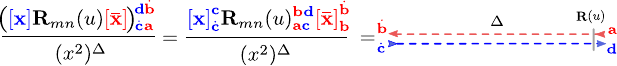}\,.
\end{center}
Having established these notations, the star-triangle identity relates a \emph{star} -- an integrated cubic vertex $x_0$ where three spinning propagators are convoluted -- with a \emph{triangle} -- the product of three spinning propagators extending between the vertices of the star -- as follows 
\begin{center}
\includegraphics[scale=1.4]{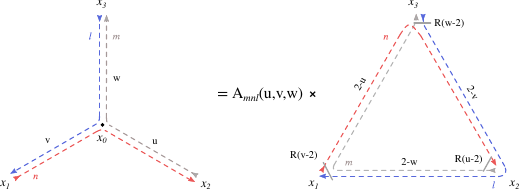}
\end{center}
and it holds under the scale-invariance condition of the integrated cubic vertex $u+v+w=4$. The coefficient $A_{mnl}(u,v,w)$ is a ratio of $\Gamma$-functions depending on the dimensions and spins of the propagators 
\begin{align}
\begin{aligned}
\label{A_functions}
&A_{mnl}(u,v,w) =\pi^{2} A_{mn}(u) A_{nl}(v)  A_{lm}(w)\,,\\ &A_{mn}(u) = \frac{(i)^{m-n}\Gamma\left(2-u+\frac{m-n}{2}\right)}{\left(u-1+\frac{m+n}{2}\right) \Gamma\left(u-1+\frac{m-n}{2}\right)}\,.
\end{aligned}
\end{align}
For completeness we shall report its ``amputated version" that is very often applied in the computations and is obtained sending the point $x_3$ to $\infty$ via an inversion map \footnote{The transformation is $x^{\mu} \mapsto (x')^{\mu}=  ({x^{\mu}-x_3^{\mu}})/{(x-x_3)^2}$. Notably, this transformation applied to the standard propagator takes a very simple form \begin{equation}
[\mathbf{x_{ab}}]^n[\overline{\mathbf{x_{ab}}}]^{\dot n}/(x_{ab})^{2\Delta}\,\, \mapsto \,\, [\mathbf{x_{3a}}\overline{\mathbf{x_{ab}}} \mathbf{x_{b3}}]^{n}  [\overline{\mathbf{x_{3a}}}\mathbf{x_{ab}} \overline{\mathbf{x_{b3}}}]^{\dot n} (x_{a3}^2 x_{b3}^2)^{\Delta}/(x_{ab})^{2\Delta}\,.
\end{equation}} around $x_3$ which also transforms the leftover points $x_1 \to x_1'$ and $x_2 \to x_2'$
\begin{center}
\includegraphics[scale=1.45]{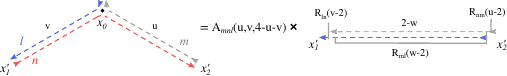}
\end{center}
The analytic form of the star-triangle identity is quite cumbersome and it is reported in the appendix for clarity (formula \eqref{amp_star}) in spite of the fact that the notation introduced so far allows to do all computations graphically. A remarkable consequence of this identity is the continuum analogue of the star-star relation (see e.g. \cite{Baxter:1982zz}), which is often applied in the computations of this paper
\begin{center}
\includegraphics[scale=1.4]{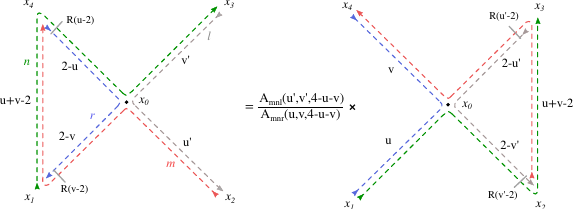}
\end{center}
The derivation of the star-triangle identity and its corollaries in the form of convolutions of propagators in $4d$ has been achieved in the papers \cite{Derkachov2020, Derkachov:2020zvv, Derkachov:2021rrf}, whereas the spinless version of star-triangle is a well known identity \cite{DEramo:1971hnd}.
\subsection*{Fishnet as a spin-chain}
\begin{figure}[b]
\begin{center}
\includegraphics[scale=1.05]{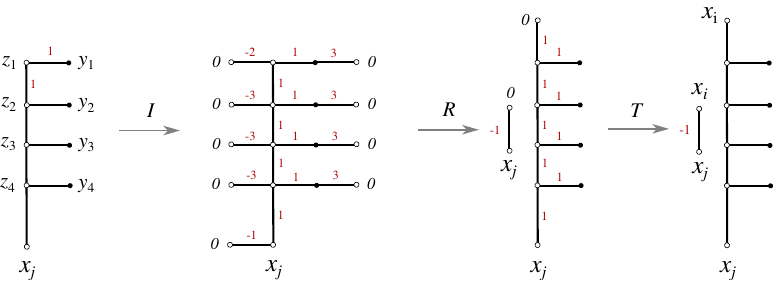}
\end{center}
\caption{The three steps of transformation $T\cdot R \cdot I$ that relates the integral operator $\hat B_j^{(N)}(\mathbf z|\mathbf y) d^{4}\mathbf y$ defined by \eqref{limit_B} with the graph-building operator for a Fishnet integral. Solid segments are propagators, circles are external points and black blobs are integration points $y_k$, each coming with a measure factor $d^4 y_k$. The scaling power of many propagators is specified by red labels.}
\label{fish_ISO}
\end{figure}

Let's review the spin-chain integrability of the bi-scalar Fishnet integrals. As already mentioned, a portion of Fishnet of size $N\times L$ and external points $x_i$ and $x_j$ enlightened in in figure \ref{fish_rectangle} for $N=4,L=4$ is equal to the following product of integral operators defined in \eqref{graph_building}
\begin{equation}
\frac{1}{(x_{ij}^{2})^L}\underbrace{\hat{B}_{ij}^{(N)}\cdot \hat{B}_{ij}^{(N)}\cdots \hat{B}_{ij}^{(N)}}_{L-times} \prod_{k=1}^N \delta^{(4)}(z_k-y_k)\,.
\end{equation}
The ``graph-building" operator $\hat{B}_{ij}^{(N)}$ is a conserved charge of the conformal $SO(1,5)$ spin chain with boundary points fixed to $x_i,x_j$ as defined in \cite{Derkachov2020,Olivucci:2021cfy}, as it belongs to the family of commuting transfer matrices of the model $\mathbf{Q}_{x_j}^{(N)}(u)$. The transfer matrix with fixed boundary can be read from sections 3-4 in \cite{Olivucci:2021cfy}, namely:
\begin{equation}
\label{Q_mat}
\mathbf{Q}_{x_j}^{(N)}(u) = \int d^4 x_a \mathcal{R}_{a1}(u) \mathcal{R}_{a2}(u) \cdots \mathcal{R}_{aN}(u) \delta^{(4)}(x_a-x_j)\,.
\end{equation} 
Here $\mathcal{R}_{ak}(u)$ is the infinite-dimensional solution of the Yang-Baxter equation acting on the $SO(1,5)$ representations $\mathbb{V}_a \otimes \mathbb{V}_k = (1,0,0) \otimes (1,0,0)$ with unitary representation $(\Delta,s,\dot s)=(1,0,0)$ both in physical spaces and in the auxiliary space\footnote{This $\mathcal{R}$-operator and the star-triangle relation, at least for zero spins $s=\dot s=0$, can be recovered in the rational limit of the R-matrix and star-triangle identities for the chiral Potts models with infinite continuous states \cite{Au_Yang_1999}. We thank J.H.H. Perk for this observation.} \cite{Chicherin:2012yn}.
In particular the amputated operator $\hat B_{j} = (z_1-x_i)^2 \hat B_{ij}$ on the left of Fig.\ref{fish_ISO} is proportional to such transfer matrix evaluated at a special point
\begin{equation}
\label{limit_B}
\hat{B}_{j}^{(N)} =\frac{x_{ij}^2}{\pi^{2}}  \lim_{u \to 1} \left(u-1\right) \mathbf{Q}_{x_j}^{(N)}(u)\,,
\end{equation}
and the actual graph-builder \eqref{graph_building} is recovered after a chain of transformations that involve inversion around zero, that is $I: z^{\mu} \mapsto z^{\mu}/z^2$, a conjugation $R$ by a factor $1/\prod_{k=1}^N (z_k^2)^{3}$ and a translation $T$ of all points by $x_i^{\mu}$
\begin{equation}
\label{the_ISO}
\hat{B}_{j}^{(N)} \xrightarrow{R \cdot I}  \frac{1}{\prod_{k=1}^N (z_k^2)^{3}} I[\hat{B}_{j}^{(N)}] \prod_{k=1}^N (z_k^2)^{3}   \xrightarrow{T}  \hat{B}_{ij}^{(N)} \,.
\end{equation}
The three steps of the latter transformation are represented for clarity in Feynman diagram notation in Fig.\ref{fish_ISO}.
It was showed to various degrees of generalities in \cite{Derkachov2020, Derkachov:2020zvv, Derkachov:2021rrf,Olivucci:2021cfy} that the spin-chain model defined by the transfer matrices \eqref{Q_mat} is formally integrable
\begin{equation}
[\mathbf{Q}_{x_j}^{(N)}(u),\mathbf{Q}_{x_j}^{(N)}(v)]=0\,,\,\,\,\, [\mathbf{Q}_{x_j}^{(N)}(u),\mathbf{Q}_{x_j}^{(N)}(v)^{\dagger}]=0\,,
\end{equation}
and the spectral problem of the chain can be solved by separation of variables (SoV). The eigenfunctions of the chain take the factorised form of the convolution of certain integral operators $\mathbf \Lambda_k(Y)$ that we dub \emph{layers}, each of them carrying the quantum numbers of one excitation of the chain $Y=(\nu,n)\in \mathbb{R}\times \mathbb{N}$ such that each layer solves the separation of variable condition
\begin{equation}
\mathbf{Q}_{x_j}^{(N)}(u) \mathbf{\Lambda}_{N}(Y) = q(u,Y)  \mathbf{\Lambda}_{N}(Y) \mathbf{Q}_{x_j}^{(N-1)}(u)\,.
\end{equation}
\begin{figure}[b]
\begin{center}
\includegraphics[scale=0.9]{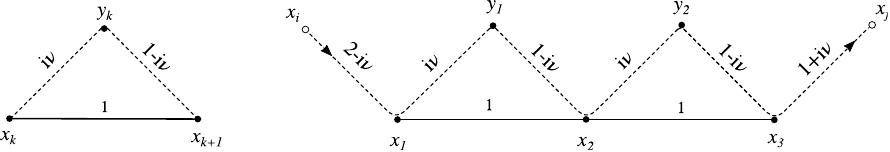}
\end{center}
\caption{\textbf{Left:} Kernel of the operator $\mathbb{R}^{(n)}_{ij}(\Delta_1/2-i\nu)$; as usual the dashed lines stand for $SU(2)$ matrices that in a product are alternatively defined as $[{\mathbf{x}}]=\boldsymbol{\sigma}_{\mu} \hat x^{\mu}$ and $[\overline{\mathbf{x}}]=\overline{\boldsymbol{\sigma}}_{\mu} \hat x^{\mu}$, to power $n$ of the symmetric representation. \textbf{Right:} Product of $\mathbb{R}$-operators defining the kernel of $\mathbf{\Lambda}_3(Y)$ according to \eqref{Layer_inhom}: the $SU(2)$ indices flow from the point $x_i$ to the point $x_j$ according to the arrows on dashed lines.}
\label{r_scalar_fig}
\end{figure}
\noindent
It follows that once the layers are defined the ket eigenfunctions read
\begin{equation}
\label{eig_biscalar}
|\Psi_{j}(\mathbf{Y})\rangle  = \Psi_{j}(\mathbf{Y}|x_1,\dots, x_N)= \mathbf  \Lambda_L(Y_N)\cdots  \mathbf \Lambda_2(Y_2) \mathbf \Lambda_1(Y_1) \prod_{k=1}^{N} {((-1)^{n_k}E(Y_k))^{k-1}}\,,
\end{equation}
where $j$ refers to the boundary points of the chain and the normalisation factors in the r.h.s. are given by $E(Y) = {\pi^2}/\left(1-i\nu+\frac{n}{2} \right)\left(i\nu+\frac{n}{2}\right)$. The spectrum of the transfer matrices is factorised over the excitation numbers $\mathbf{Y} = Y_{1},\dots, Y_N$, where here we denote $Y_k =(n_k,\nu_k)$
\begin{equation}
\label{spectrum_Q}
\prod_{k=1}^N q(u,Y_k) =  \frac{\pi^{2N+2}}{(2-u)(1-u)}  \prod_{k=1}^N \frac{\Gamma\left(u+\frac{n_k}{2}-i\nu_k \right) \Gamma\left(1-u+\frac{n_k}{2}+i\nu_k \right)}{\Gamma\left(2-u+\frac{n_k}{2} +i\nu_k \right) \Gamma\left(1+u +\frac{n_k}{2}-i\nu_k \right)}\,,
\end{equation}
and in the limit \eqref{limit_B} it becomes the Fishnet's eigenvalue
\begin{equation}
\label{spectrum_B}
\prod_{k=1}^N E(Y_k) \equiv E(\mathbf{Y})\,.
\end{equation}
The layer operators have been found in \cite{Derkachov2020} and analyzed in detail in \cite{Derkachov:2020zvv,Olivucci:2021cfy}; in formulae they read 
\begin{align}
\begin{aligned}
\label{Layer_inhom}
&\mathbf{\Lambda}_1(Y)\equiv  \mathbf \Lambda_1(n,\nu) = \frac{[\mathbf{(x-x_j)}]^{n}}{(x-x_j)^{2\left(1+i\nu \right)}}\,,\\&
\mathbf \Lambda_k(Y)\equiv \mathbf \Lambda_k(n,\nu) = \mathbb{R}^{(n)}_{12}\left(\frac{1}{2}-i\nu\right)\cdots \mathbb{R}^{(n)}_{k-1k}\left(\frac{1}{2}-i\nu\right)\frac{[\mathbf{x_k-x_j}]^{n}}{(x_k-x_j)^{2\left(1+i\nu \right)}}\,,\end{aligned}
\end{align}
where the symbols $\mathbb{R}^{(n)}_{ij}(u)$ are a compact notation for the integral operators
\begin{align}
\begin{aligned}
&[\mathbb{R}^{(n)}_{ij}(u)] \Phi(x_i,x_j) = \int d^4 y \,{R}^{(n)}_u (x_i,x_j|y)  \Phi(y,x_j) \\
&{R}^{(n)}_u (x_i,x_j|y) = \frac{[\mathbf{(x_1-y)(\overline{y-x_2})}]^n}{(x_i-x_j)^{2}(x_i-y)^{2\left(-u +\frac{1}{2} \right)}(y-x_j)^{2\left(u+\frac{1}{2}\right)}} \,,
\end{aligned}
\end{align}
represented in figure \ref{r_scalar_fig} (left).
The eigenfunctions of $\hat B^{(N)}_{ij}$ are recovered from $\Psi_{j}$ essentially by the transformation \eqref{the_ISO}. It is convenient to perform on the resulting eigenfunctions $ \Psi_{ij}$ a few transformations that set them in a nicer form, namely a rotation from the right by $\mathbf{\overline{x}}_{ij}$ and a rescaling $1/(x_{ij}^2)^{1+i\nu}$ on each layer,
\begin{equation}
 \Psi_{ij}(\mathbf{Y}|\mathbf{x}) \mapsto \Psi_{ij}(\mathbf{Y}|\mathbf{x}) \cdot  \frac{[\mathbf{\overline{x}}_{ij}]^{n_1}}{(x_{ij}^2)^{1+i\nu_1}}\otimes \cdots \otimes \frac{[\mathbf{\overline{x}}_{ij}]^{n_N}}{(x_{ij}^2)^{1+i\nu_N}}
 \end{equation}
followed by a charge-conjugation and transposition of each of the layers,
$$\mathbf{\Lambda}_k(n,\nu) \mapsto \left(\boldsymbol{\varepsilon}^n \mathbf{\Lambda}_k(n,\nu) \boldsymbol{\varepsilon}^{n}\right)^{t_n}.$$ The final form of the eigenfunction is
\begin{equation}
\label{eig_biscalar_ij}
|\Psi_{ij}(\mathbf{Y})\rangle  = \Psi_{ij}(\mathbf{Y}|x_1,\dots, x_N)= \mathbf  \Lambda_L(Y_N)\cdots  \mathbf \Lambda_2(Y_2) \mathbf \Lambda_1(Y_1)  \prod_{k=1}^{N} {((-1)^{n_k}E(Y_k))^{k-1}} \,,
\end{equation}
with $\boldsymbol{\Lambda}$ the transformed layers exemplified in the right picture of Fig.\ref{r_scalar_fig}
\begin{align}
\begin{aligned}
\label{Layer_ISO}
&\mathbf{\Lambda}_1(Y)\equiv  \mathbf \Lambda_1(n,\nu) = \frac{[\mathbf{\overline{(x_i-x)}(x-x_j)}]^{n}}{(x_i-x)^{2\left(2-i\nu \right)}(x-x_j)^{2\left(1+i\nu \right)}}\,,\\&
\mathbf \Lambda_k(Y)\equiv \mathbf \Lambda_k(n,\nu) =\frac{[(\overline{\mathbf{x_i-x_1}})]^{n}}{(x_i-x_1)^{2\left(2-i\nu \right)}} \mathbb{R}^{(n)}_{12}\left(\frac{1}{2}-i\nu\right)\cdots \mathbb{R}^{(n)}_{k-1k}\left(\frac{1}{2}-i\nu\right)\frac{[\mathbf{x_k-x_j}]^{n}}{(x_k-x_j)^{2\left(1+i\nu \right)}}\,.\end{aligned}
\end{align}
Notice that the spectrum of the graph-builder \eqref{spectrum_B} is left untouched by the transformation $T\cdot R \cdot I$ and also by the following cosmetic adjustments of the eigenfunctions, wrt to which the operator $\hat{B}^{(N)}_{ij}$ is invariant. These latter modifications are done just in order to work with such a basis for which the form of star-triangle transformation given here is enough for the later computations.

\begin{figure}[t]
\begin{center}
\includegraphics[scale=0.9]{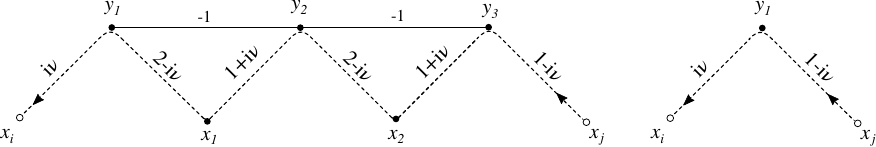}
\end{center}
\caption{Kernels of the bra layers $\overline{\mathbf \Lambda}_3(Y)$ and $\overline{\mathbf \Lambda}_1(Y)$ in Feynman diagram notation.}
\label{bra_layer}
\end{figure}

In conclusion one should spell the form of the dual of the eigenfunctions \eqref{eig_biscalar_ij} with respect to the scalar product of the complementary series, a.k.a. \emph{bra} eigenfunctions,
\begin{equation}
\label{eig_biscalar_BRA}
\langle \Psi_{ij}(\mathbf{Y})|  = \overline{\Psi_{ij}}(\mathbf{Y}|x_1,\dots, x_N)= \prod_{k=1}^{N} \frac{E(Y_k)^{1-k}}{(-1)^{k \times n_k}} \, \overline{\mathbf \Lambda}_1(Y_N)  \overline{\mathbf \Lambda}_2(Y_{N-1}) \cdots  \overline{\mathbf  \Lambda}_L(Y_1)  \,,
\end{equation}
for \emph{conjugated} layers defined graphically in Fig.\ref{bra_layer}.
\subsection*{Properties of the eigenfunctions}
The exchange of two excitations $Y_i\leftrightarrow Y_j$ in the eigenfunctions takes simple form due to the following property of the layers
\begin{equation}
\label{symm_scalar}
 \mathbf \Lambda_{k+1}(Y)\cdot  \mathbf \Lambda_{k}(Y') = (-1)^{n+n'}\frac{E_{k}(Y')}{E_k(Y)}\times  \mathbf{R}(Y'|Y) \mathbf \Lambda_{k+1}(Y')\,  \mathbf \Lambda_{k}(Y)\mathbf{R}(Y|Y')\,,
\end{equation}
and the same identity upon interchange $E(Y) \leftrightarrow E(Y')$ holds for the conjugated layers. The symmetry of eigenfunctions wrt to the exchange of quantum numbers $Y,Y'$ follows
\begin{align}
\begin{aligned}
\label{exchange_waves}
&\Psi_{ij}(\dots Y,Y' \dots|\mathbf{x}) = \mathbf{R}(Y'|Y) {\Psi}_{ij}(\dots Y',Y \dots|\mathbf{x}) \mathbf{R}(Y|Y')\,,\\
&\overline{\Psi}_{ij}(\dots Y,Y' \dots|\mathbf{x}) = \mathbf{R}(Y'|Y) \overline{\Psi}_{ij}(\dots Y',Y \dots|\mathbf{x}) \mathbf{R}(Y|Y')\,.
\end{aligned}
\end{align}
The eigenfunctions in Feynman diagram notation take the shape of a pyramid of layer operators of decreasing lengths; after the exchange of two excitations $Y_2,Y_3$, the pyramid on the left gets transformed into the pyramid on the right:
\begin{center}
\includegraphics[scale=0.65]{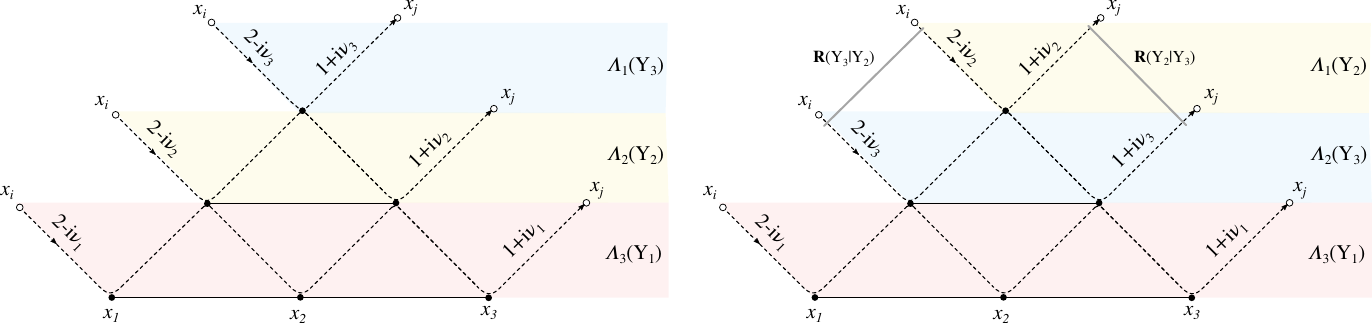}
\end{center}
\noindent
The scalar product of two eigenfunctions that carry the excitations $\mathbf{Y}=(Y_1,\dots Y_N)$ and $\mathbf{Y}'=(Y'_1,\dots Y'_N)$ was first computed in \cite{Derkachov2020}, where it was showed that such eigenfunctions are orthogonal whenever $\mathbf{Y}\neq \pi \mathbf {Y}'$ for any permutation $\pi$ of $N$ elements.  
The latter computation allows to extract the measure $\rho(\mathbf{Y})$ over the spectrum of excitations (i.e. over the separated variables),
\begin{equation}
\label{rho_def}
\rho(Y_1,\dots,Y_N) = \frac{1}{N!} \prod_{j=1}^N \frac{(n_j+1)}{2\pi^{2N+1}} \Big / \prod_{k
\neq j}^N {H_{n_k,n_h}(\nu_k,\nu_h)}\,,
\end{equation}
where the non-factorisable term reads
\begin{equation}
\label{H_factor}
H_{m,n}(\mu,\nu)=\frac{(-1)^n \Gamma\left(1+\frac{m}{2}- i\mu \right)\Gamma\left(1+\frac{n}{2}+ i\nu \right)\Gamma\left(\frac{m-n}{2}+ i\mu-i\nu \right)}{\Gamma\left(1+\frac{m}{2}+ i\mu \right)\Gamma\left(1+\frac{n}{2}- i\nu \right)\Gamma\left(1+\frac{m-n}{2}- i\mu+i\nu \right)\left(1+\frac{m+n}{2}- i\mu+i\nu \right)}\,.
\end{equation}
The measure enters the conjectural completeness relation of the Fishnet's eigenfunctions
\begin{align}
\begin{aligned}
\label{SoV_completeness}
\Phi(x_1,\dots,x_N) = (x_{ij}^2)^{2N}\!\!\!\sum_{n_1,\dots,n_N=0}^{\infty} \int d \nu_1 \cdots d \nu_N\, \rho(\mathbf{Y}) \Psi_{ij}(\mathbf{Y}|\mathbf{x})_{\mathbf{a}}^{\mathbf{b}} \int \! d^4 {z}_1\cdots d^4 z_N \bar{\Psi}_{ij}(\mathbf{Y}|\mathbf{z})_{\mathbf{b}}^{\mathbf{a}} \Phi(\mathbf{z})\,,
\end{aligned}
\end{align}
where the matrix indices over symmetric spinors $\mathbf{a}=\{\mathbf{a}_1,\dots \mathbf{a}_N\}$ in $\Psi(\mathbf{x}|\mathbf{Y})_{\mathbf{a}}^{\mathbf{b}}\bar{\Psi}(\mathbf{Y}|\mathbf{x'})_{\mathbf{b}}^{\mathbf{a}}$ are contracted forming a trace.
This completeness relation is invariant under any permutation of layers' excitations due to the property \eqref{symm_scalar} and to the unitarity $\mathbf{R}(Y|Y')\mathbf{R}(Y'|Y)=\mathbbm{1}$ of the $R$-matrix. This fact provides an important consistency check in any computation of quantities involving the insertion of a resolution of the identity based on \eqref{SoV_completeness}, as it constrains the result to be invariant wrt permutations of excitations $Y_i$.  The insertion of the resolution of the identity over the spectrum of separated variables is an ubiquitous operation in this paper starting from the next section.

\section{Cutting the Fishnet}
\label{sec:cut}
The spin-chain technology reviewed in the previous section can be used to transform any multi-loop Fishnet Feynman diagram from the space of coordinate to the space of the mirror excitations $\mathbf{Y}$, which can be referred to Separation of Variables (SoV) representation. In order to do so, a Fishnet shall first be factorized into patches of square-lattice via a suitable triangulation procedure described in this section. 

\begin{figure}[b]
\begin{center}
\includegraphics[scale=0.85]{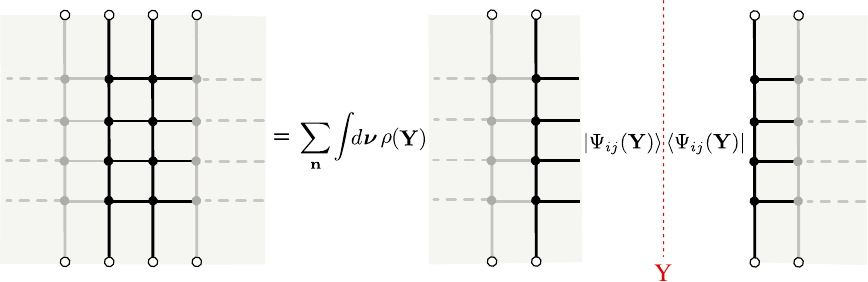}
\caption{Cutting the Fishnet: the insertion of completeness relation \eqref{complete_delta} between two graph-building operators $\hat{B}^{(M)}_{kl} \hat{B}^{(M)}_{ij}$ factorises their product separating two portions of Fishnet diagram. In compact notation we denote such cuts by a red dashed line associated to a set of mirror excitations $\mathbf{Y}$ re-summed by the resolution of identity.}
\label{cutting_general}
\end{center}
\end{figure}

\begin{figure}[t]
\begin{center}
\includegraphics[scale=0.85]{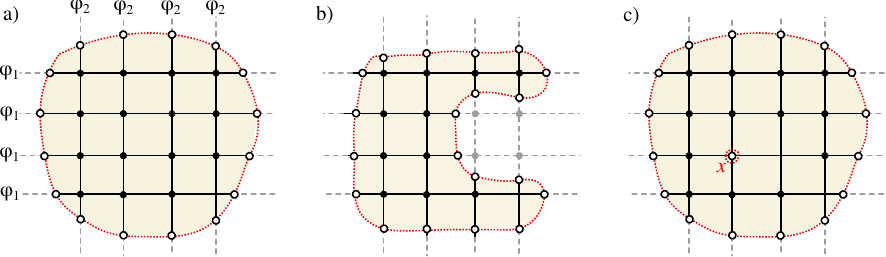}
\caption{Starting from an infinite square lattice on the plane made of scalar propagators and quartic vertices $\phi_1\phi_2\phi_1^{\dagger}\phi_2^{\dagger}$, a concrete Fishnet integral is obtained drawing a circle (red dashed line) such that it crosses the legs into external points (white circles). The external points are in general different, giving a multi-point Feynman diagram with disk topology. It can be useful to distinguish the case of a convex shape (a) from the case with concavities (b) or even the difference of two disks (c) where some points in the ``bulk" of the Fishnet square-lattice, e.g. $x$, are fixed as external points of the diagram.}
\label{cutplane}
\end{center}
\end{figure}

The topology of Fishnet integrals is that of a square-lattice of scalar propagators of fields $\phi_1$ and $\phi_2$ and quartic vertices $\phi_1\phi_2\phi_1^{\dagger}\phi_2^{\dagger}$ as in the figures \ref{general_bulk_punc},\ref{cutting_general},\ref{cutplane}. 
Any given bundle of horizontal propagators (fields $\phi_1$) crosses a number of vertical propagators (fields $\phi_2$). The general rule for cutting the Fishnet is that of identifying the external points of the diagram -- white circles in figures \ref{cutting_general}, \ref{cutplane} -- and then to triangulate the surface on which the diagram lies by drawing a cut parallel to the propagator $\phi_2$ that connects two external points $x_i,x_j$. Such a cut crosses the bundle of propagators $\phi_1$ flowing between $x_i$ and $x_j$. Any cut is characterised by a width $M$ that is the number of propagators $\phi_1$ in such bundle.
It should be clear that this prescription for cutting is not unique, as for instance another good triangulation is obtained exchanging the role of fields $\phi_2$ and $\phi_1$.
Each cut (red dashed lines in Fig.\ref{cutting_general}) corresponds to the insertion of a resolution of the identity
\begin{align}
\begin{aligned}
\label{complete_delta}
(x_{ij}^2)^{2M}\sum_{\mathbf{n}}\int \! \! d\boldsymbol{\nu} \,\rho(\mathbf{Y})& \Psi_{ij}(\mathbf{Y}|y_1,\dots,y_M)_{\mathbf{a}}^{\mathbf{b}}\,
\overline{\Psi}_{ij}(\mathbf{Y}|z_1,\dots,z_M)_{\mathbf{b}}^{\mathbf{a}}=\\&=\delta^{(4)}(y_1-z_1)  \delta^{(4)}(y_2-z_2) \cdots \delta^{(4)}(y_M-z_M) \,,
\end{aligned}
\end{align}
based on the completeness relation \eqref{SoV_completeness}, and therefore it carries a set of excitations' quantum numbers $\mathbf{Y}=(Y_1,\dots, Y_M)$. In the spin-chain language, the width of the cut $M$ is the length of the graph building operator insisting on the cut ($M=4$ in the figure \ref{cutting_general}).
\begin{figure}[b]
\begin{center}
\includegraphics[scale=0.85]{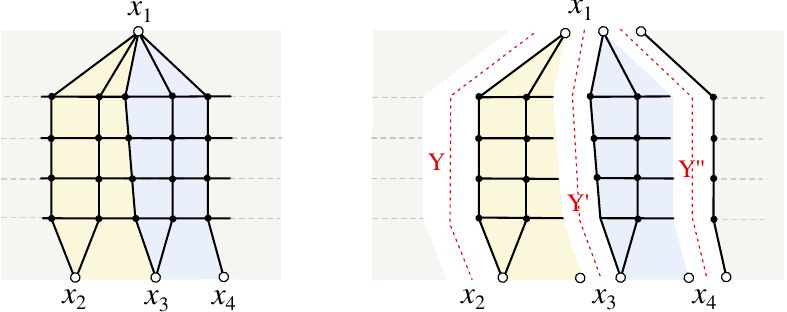}
\caption{A local tiling of a Fishnet with two triangles (yellow, blue) whose the bases $x_{23}$, $x_{34}$ lie on the same boundary of the square lattice. Each red, dashed line denotes the insertion of a resolution of identity with quantum numbers $\mathbf{Y}$,$\mathbf{Y}'$ ,$\mathbf{Y}''$ respectively.} 
\label{trisame}
\end{center}
\end{figure}

\begin{figure}[t]
\begin{center}
\includegraphics[scale=0.85]{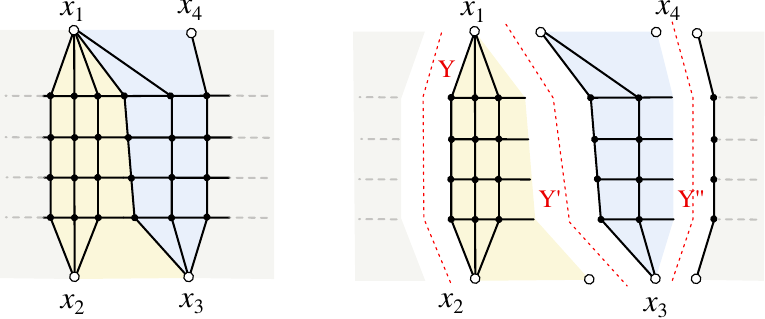}
\caption{A local tiling of a Fishnet with two triangles (yellow, blue) whose the bases $x_{23}$, $x_{14}$ lie on opposite boundaries of the square lattice. Each red, dashed line denotes the insertion of a resolution of identity with quantum numbers $\mathbf{Y}$,$\mathbf{Y}'$ ,$\mathbf{Y}''$ respectively.}
\label{triflip}
\end{center}
\end{figure}

Depending on the topology of the Fishnet integral, i.e. on the boundary condition of the square-lattice of propagators, one can distinguish tiles of the triangulation with different features. In the case of a convex portion of Fishnet lattice, such as that of figure \ref{cutplane} (a), the tiles are of those types showed in figure \ref{trisame} and \ref{triflip}. In the example of figure \ref{trisame} the cuts extend between points $x_{1}x_{2}$, $x_{1}x_{3}$ and $x_{1}x_{4}$ drawing two triangles (yellow and blue). Each triangle has two edges coinciding with a cut and the third one, $x_{23}$ and $x_{34}$ respectively, coinciding with the boundary of the Fishnet. The operator expression for the Fishnet contained inside the two tiles is $\hat B^{(4)}_{12} \hat B^{(4)}_{12}$ and $\hat B^{(4)}_{13} \hat B_{13}^{(4)}$. The SoV representation of the operators is achieved after insertion of \eqref{complete_delta} along the cuts:
\begin{align}
\begin{aligned}
\label{trisame_SoV}
\langle \Psi_{12}(\mathbf Y) |\hat B^{(4)}_{12} \hat B^{(4)}_{12}| \Psi_{13}(\mathbf Y') \rangle & \langle \Psi_{13}(\mathbf Y')|  \hat B^{(4)}_{13} \hat B_{13}^{(4)} | \Psi_{14}(\mathbf Y'') \rangle = \\ &=E(\mathbf Y )^2  E(\mathbf{Y}')^2 \times \langle \Psi_{12}(\mathbf Y) | \Psi_{13}(\mathbf Y') \rangle \langle \Psi_{13}(\mathbf Y')  | \Psi_{14}(\mathbf Y'') \rangle \,.
\end{aligned}
\end{align}
The loop integrations in coordinate space are replaced by a certain power of the Fishnet's eigenvalues $E(\mathbf Y)$ and a product of overlaps of two eigenfunctions (bra/ket) relative to graph-builder with different boundaries, e.g. $\langle \Psi_{12}(\mathbf Y) | \Psi_{13}(\mathbf Y') \rangle$ for cuts $x_{1}x_{2}$ and $x_{1}x_{3}$.

In the example of figure \ref{triflip} the cuts extend between points $x_{1}x_{2}$, $x_{1}x_{3}$ and $x_{3}x_{4}$ and the two triangles have two edges on cuts and one edge ($x_{23}$ and $x_{14}$ respectively) on the boundary. These two edges stand on opposite boundaries of the Fishnet, whereas they were adjacent in figure \ref{trisame}, and the blue triangle is ``flipped". The Fishnet contained into the triangles is given by the operators $\hat B^{(4)}_{12} \hat B^{(4)}_{12}\hat B^{(4)}_{12}$ and $\hat B^{(4)}_{13} \hat B_{13}^{(4)}$. The transformation to SoV representation reads:
\begin{align}
\begin{aligned}
\label{triflip_SoV}
\langle \Psi_{12}(\mathbf Y) |\hat B^{(4)}_{12} \hat B^{(4)}_{12} \hat B^{(4)}_{12}| \Psi_{13}(\mathbf Y') \rangle & \langle \Psi_{13}(\mathbf Y')|  \hat B^{(4)}_{13} \hat B_{13}^{(4)} | \Psi_{34}(\mathbf Y'') \rangle = \\ &=E(\mathbf Y )^3 E(\mathbf{Y}')^2 \times \langle \Psi_{12}(\mathbf Y) | \Psi_{13}(\mathbf Y') \rangle \langle \Psi_{13}(\mathbf Y')  | \Psi_{34}(\mathbf Y'') \rangle \,.
\end{aligned}
\end{align}
The difference between the expressions in the rhs of \eqref{trisame_SoV} and \eqref{triflip_SoV} essentially sits in the overlap between excitations $\mathbf{Y}'$ and $\mathbf{Y}''$, respectively
\begin{equation}
\langle \Psi_{13}(\mathbf Y')  | \Psi_{14}(\mathbf Y'') \rangle \,\,\,\text{and} \,\,\,\, \langle \Psi_{13}(\mathbf Y')  | \Psi_{34}(\mathbf Y'') \rangle\,,
\end{equation}
that differs because the eigenfunctions have a defined orientation, i.e. are not symmetric w.r.t. the flip of the external points $\langle \Psi_{ab}(\mathbf Y) | \neq  \langle \Psi_{ba}(\mathbf Y) |$.
In spite of the fact that the tiles analysed so far are enough to describe the rich class of convex Fishnets on the disk, it is interesting to push the analysis to concavities in the square-lattice as in figure \ref{cutplane} (b) or even of ``defects", non integrated points in the middle of the disk as in figure \ref{cutplane} (c).

In the case of a concave Fishnet it happens that somewhere a bundle of $M$ propagators $\phi_1$ gets split into two bundles of $M_1$, $M_2$ propagators such that $M_1+M_2 \leq  M$ as in figure \ref{cutplane} (b). Whenever this happens, the triangulation of the Fishnet integral involves a new type of tiles featuring a cut on each of its three edges.
\begin{figure}[t]
\begin{center}
\includegraphics[scale=0.85]{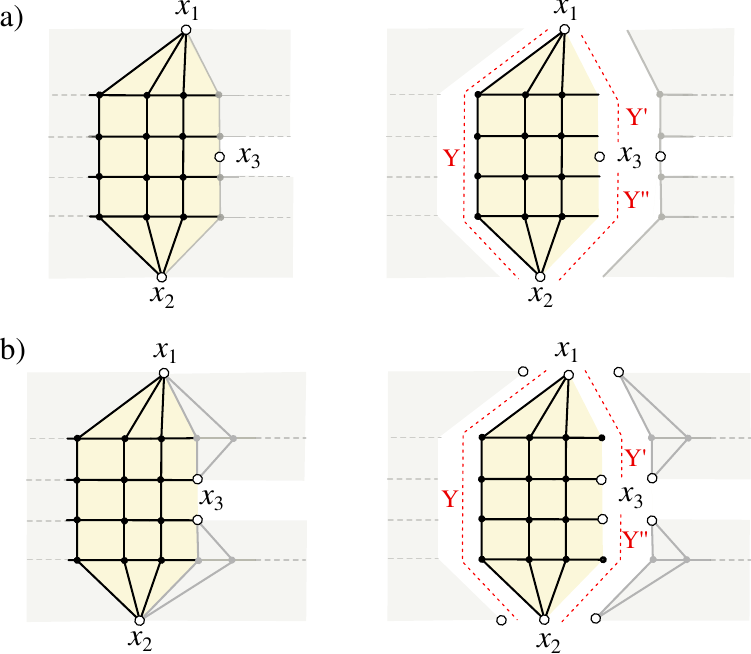}
\caption{Local tilings of a Fishnet with one triangle (yellow) whose three edges coincide with cuts of the triangulation. Case $(a)$: $M=4$ horizontal propagators between points $x_1,x_2$ are split into two bundles of size $M_1=M_2=2$ flowing between $x_1,x_3$ and $x_3,x_2$ respectively. Case $(b)$: $M=4$ horizontal propagators between points $x_1,x_2$ are split into two bundles of size $M_1=M_2=1$ flowing between $x_1,x_3$ and $x_3,x_2$ respectively. The leftover $M-M_1-M_2=2$ propagators terminate at point $x_3$.}
\label{triconvex}
\end{center}
\end{figure}
Take the example of figure \ref{triconvex} (a). The operator expression of the yellow triangle is $\hat B_{12}^{(4)}\hat B_{12}^{(4)}$ and inserting the resolution of the identity along cuts $x_1x_2$, $x_1x_3$ and $x_2x_3$ one gets the SoV representation of the tile as a function of three sets of excitations $\mathbf{Y}$, $\mathbf{Y}'$ and $\mathbf{Y}''$:
\begin{align}
\begin{aligned}
\label{convex_example_no_delta}
 \langle \Psi_{12}(\mathbf Y) |\hat B^{(4)}_{12} \hat B^{(4)}_{12}| \Psi_{13}(\mathbf Y') \otimes  \Psi_{32}(\mathbf Y'')\rangle =E(\mathbf{Y})^2 \times   \langle \Psi_{12}(\mathbf Y) |\Psi_{13}(\mathbf Y') \otimes \Psi_{32}(\mathbf Y'') \rangle \,.
\end{aligned}
\end{align}
The novelty of \eqref{convex_example_no_delta} is that the overlap appearing in its rhs involves one bra wave-function of length $M=4$, $\langle \Psi_{12}(\mathbf Y)|$ with two ket wave-functions of length $M_1,M_2=2$, namely $|\Psi_{13}(\mathbf Y') \rangle$ and $| \Psi_{32}(\mathbf Y'') \rangle$.
Another nontrivial example is that of figure \ref{triconvex} (b). The operator expression for the yellow triangle is $\hat B_{12}^{(4)}\hat B_{12}^{(4)}\hat B_{12}^{(4)}$ and the SoV representation of the triangle, as a function of quantum numbers $\mathbf{Y}$, $\mathbf{Y}'$ and $\mathbf{Y}''$, reads
\begin{align}
\begin{aligned}
\label{convex_example}
 \langle \Psi_{12}(\mathbf Y) |\hat B^{(4)}_{12} \hat B^{(4)}_{12} \hat B^{(4)}_{12}&| \Psi_{13}(\mathbf Y') \otimes \delta_{x_3}  \otimes \delta_{x_3} \otimes \Psi_{32}(\mathbf Y'')\rangle = \\ &=E(\mathbf{Y})^3 \times   \langle \Psi_{12}(\mathbf Y) |\Psi_{13}(\mathbf Y') \otimes \delta_{x_3}  \otimes \delta_{x_3} \otimes \Psi_{32}(\mathbf Y'') \rangle \,.
\end{aligned}
\end{align}
The novelty of \eqref{convex_example} is that the overlap $\langle \Psi_{12}(\mathbf Y) |\Psi_{13}(\mathbf Y') \otimes \delta_{x_3}  \otimes \delta_{x_3} \otimes \Psi_{32}(\mathbf Y'') \rangle$ pairs the bra function of length $M=4$ with ket functions of lengths $M_1,M_2=1$ and also with $2=M-M_1-M_2$ delta-functions in the point $x_3$. This type of overlap that involves three sets of excitations and also some reduction to a point is the most general instance one can encounter.

\begin{figure}[t]
\begin{center}
\includegraphics[scale=0.8]{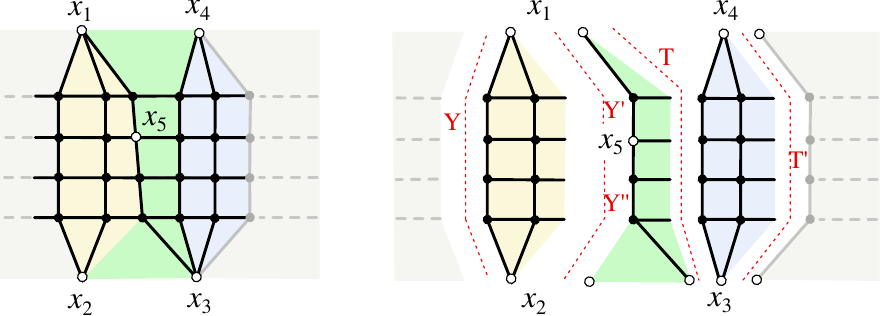}
\caption{Local tilings of a Fishnet with a ``defect", i.e. a non-integrated point in the square-lattice.}
\label{tridef}
\end{center}
\end{figure}

In order to conclude the analysis it is interesting to look at what happens with Fishnet that lies on a disk with one or more internal points of the square-lattice that are not integrated, as the one in figure \ref{cutplane} (c). The triangulation in this case is slightly more involved and a primer of its tilings is given in figure \ref{tridef}. The operator expression for the yellow, green and blue tiles is respectively $\hat{B}^{(4)}_{12}\hat{B}^{(4)}_{12}$, $\hat{B}^{(4)}_{13}$ and $\hat{B}^{(4)}_{43}\hat{B}^{(4)}_{43}$.
Notice that the yellow and green triangles carry some excitation on all the three edges, respectively $(\mathbf{Y},\mathbf{Y}',\mathbf{Y}'')$ and $(\mathbf{Y}',\mathbf{Y}'',\mathbf{T})$. After cutting, the coloured portion of Fishnet gets factorised into three tiles in SoV representation
\begin{align}
\begin{aligned}
\langle \Psi_{12}(\mathbf{Y})|\hat{B}^{(4)}_{12}\hat{B}^{(4)}_{12} |{\Psi}_{15}(Y') \otimes \delta_{x_5}  \otimes \Psi_{52}(\mathbf{Y}'') \rangle  & \times  \langle {\Psi}_{15}(Y') \otimes \delta_{x_5}  \otimes \Psi_{52}(\mathbf{Y}'') |\hat{B}^{(4)}_{13}|{\Psi}_{13}(\mathbf{T})\rangle \times \\& \times \langle{\Psi}_{13}(\mathbf{T})|\hat{B}^{(4)}_{43}\hat{B}^{(4)}_{43}| \Psi_{43}(\mathbf{T'}) \rangle\,,
\label{punc_SoV_I}
\end{aligned}
\end{align}
which reduces to a product of overlaps and eigenvalues 
\begin{align}
\begin{aligned}
\label{punctured_SoV}
\eqref{punc_SoV_I}&=E(\mathbf{Y})^2 E(\mathbf{T}) E(\mathbf{T}')^2 \times \langle \Psi_{12}(\mathbf{Y})  |{\Psi}_{15}(Y') \otimes \delta_{x_5}  \otimes \Psi_{52}(\mathbf{Y}'') \rangle  \times \\&\;\;\;\;\;\times \langle {\Psi}_{15}(Y') \otimes \delta_{x_5}  \otimes \Psi_{52}(\mathbf{Y}'') |{\Psi}_{13}(\mathbf{T})\rangle \times \langle{\Psi}_{13}(\mathbf{T})| \Psi_{43}(\mathbf{T'}) \rangle\,.
\end{aligned}
\end{align}
The overlaps appearing in \eqref{punctured_SoV} belong to the classes already encountered in the previous examples.  

At this point an advantage of the SoV representation is already showing up itself: the $M\times L$ loop integrations in four-dimensional coordinate space for a triangular tile containing a lattice of dimensions $M, L$ are replaced by a summation over only a number $\sim M$ of \emph{mirror} excitations $\mathbf{Y}$, for any value of $L$. The non-trivial part of this representation, left to work out, are the overlaps that appear in formulae \eqref{trisame_SoV}-\eqref{punctured_SoV}. Since the latter feature a certain variety of cases the next sections deal with their classification and with the derivation of a general, compact expression for the overlaps. 

\section{Overlaps}
\label{sec:overs}
The overlaps of Fishnet wave-functions relative to different cuts -- i.e. eigenfunctions of graph-building operators defined with different boundary points $x_i x_j$ in \eqref{Layer_inhom} -- are the non trivial building-blocks in order to write the SoV representation of a generic planar Fishnet integral. In fact, such overlaps can be computed using the star-triangle transformations of section \ref{sec:eigenf}. In the following the overlaps are classified and a few examples are computed in support of their general formulae which are presented in the end. The latter are valid for any size of the Fishnet tilings and their reliability for any size is based on the iterative nature of their calculation. A sample of detailed computations is reported in the appendices \ref{app:firstover} to \ref{app:lastover}. 
\subsection*{Classification of Overlaps}
The overlaps are classified by increasing complexity into four cases, essentially distinguished by the different boundaries of the triangular tiles. As discussed in the previous section, given a Fishnet integral there are a few different type of overlaps that can arise after its triangulation:
\begin{itemize}
\item[I)] Overlap of a bra/ket wave-function with the same number $N$ of magnons $\mathbf{Y}_n$ and $\mathbf{Y}_m$ inserted along two cuts with a point in common, for instance $x_1x_2$ and $x_1x_3$. 
\begin{align}
\begin{aligned}
&\langle \Psi_{1 2} (\mathbf{Y}_n) |\Psi_{1 3}(\mathbf{Y}_m)\rangle\,.
\end{aligned}
\end{align}
This case occurs whenever the triangulation of the Fishnet has the type of tile depicted on the left, and the corresponding overlap of eigenfunctions is the figure on the right (here, $x_1 \equiv \infty$):
\begin{center}
\includegraphics[scale=0.7]{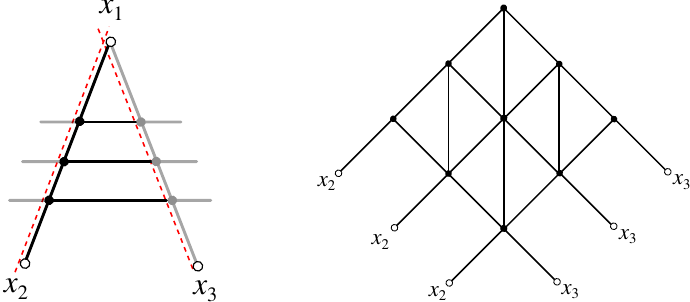}
\end{center}

\item[II)] Overlap of a bra/ket wave-function with different widths, i.e. different number of magnons $N\neq M$ inserted along two adjacent cuts, say $x_1x_2$ and $x_1x_3$. If the bra wave-function has $N-M$ sites more than the ket, these $N-M$ sites are identified with the extremal point $x_3$ of the second cut:
\begin{align}
\begin{aligned}
&\langle \Psi_{1 2} (\mathbf{Y}_n) |\Psi_{1 3}(\mathbf{Y}_m) \otimes \underbrace{\delta^{(4)}_{x_3} \otimes \cdots \otimes \delta^{(4)}_{x_3}}_{n}\rangle\,.
\end{aligned}
\end{align}
In pictures, for the cases $N=3$ and $M=2,1$, we have the tiles (left) and the corresponding overlaps (right, depicted for $x_1 \equiv \infty$):
\begin{center}
\includegraphics[scale=0.7]{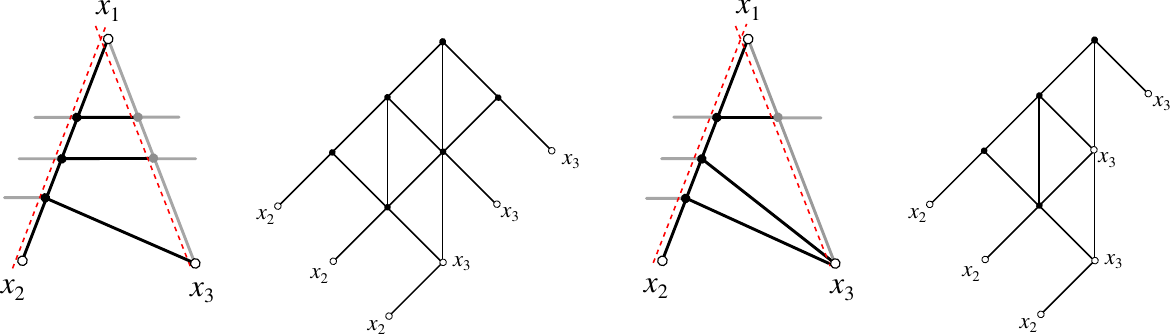}
\end{center}
In the limiting case $M=0$ this type of overlap is what occurs on the right boundary of Basso-Dixon Fishnet integrals (see section (8.3) in \cite{Derkachov:2020zvv}), that is the ``reduction" of the eigenfunction to one point $x_3$, namely
\begin{align}
\begin{aligned}
&\langle \Psi_{1 2} (\mathbf{Y}_n) | \underbrace{\delta^{(4)}_{x_3} \otimes \cdots \otimes \delta^{(4)}_{x_3}}_{N}\rangle\,,
\end{aligned}
\end{align}
\begin{center}
\includegraphics[scale=0.7]{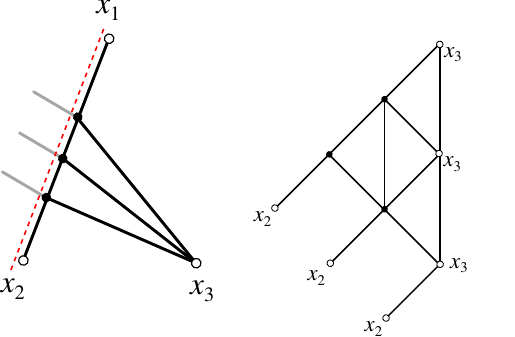}
\end{center}
The counterpart of the latter overlaps is when the width of the ket is larger than that of the bra, $N\neq M$. The $M-N$ propagators $\phi_1$ in excess have their left extremal points reduced to the point $x_2$, after the amputation of the $\phi_2$ propagators in the definition of the graph-builder \eqref{graph_building} that connects the last $N-M$ sites. In figures, the tile for $M=N+1$, and the tile with $M=N+2$ regarded as the amputation of a graph-building operator of length $M=3$:
\begin{center}
\includegraphics[scale=0.7]{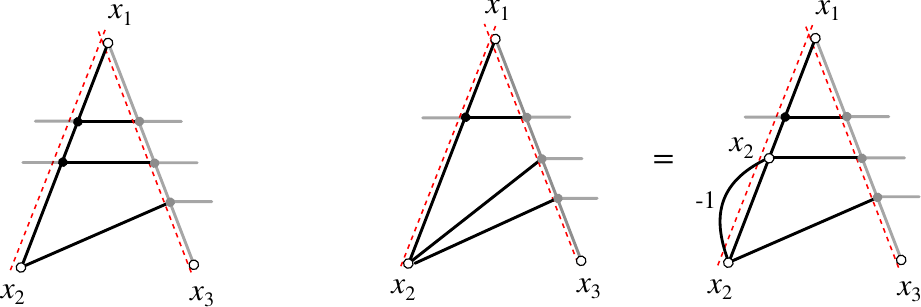}
\end{center}
In formulae, the overlap $M>N$ is
\begin{align}
\label{II_bk}
&\langle \Psi_{1 2} (\mathbf{Y}_n) \otimes  \underbrace{\delta^{(4)}_{x_2} \otimes \cdots \otimes \delta^{(4)}_{x_2}}_{M-N}  |\prod_{j=N}^{M-1} (z_{j,j+1}^2) |  \Psi_{1 3}(\mathbf{Y}_m)\rangle\,.
\end{align}
\item[III)]  Overlap of three wave-functions, one bra function relative to the cut $x_{1}x_{2}$ with $N$ magnons $\mathbf{Y}_n$ and two ket wave-functions with $M,R$ magnons $\mathbf{Y}_m$ and $\mathbf{Y}_r$ relative to cuts $x_1 x_3$ and $x_{2}x_{3}$ such that $M+R=N$. This case happens when a bundle of $N$ propagators traversing the cut $x_1x_2$ of the triangulation gets split into two bundles across $x_{1}x_3$ and $x_2 x_3$,
\begin{align}
\begin{aligned}
&\langle \Psi_{1 2} (\mathbf{Y}_n) |\Psi_{1 3}(\mathbf{Y}_m) \otimes  \Psi_{2 3}(\mathbf{Y}_r)\rangle\,.
\end{aligned}
\end{align}
For example, when $N=3$, $M=2$ and $R=1$, a possible tile of the triangulation and the overlap have the form (with the usual identification $x_1 \equiv \infty$):
\begin{center}
\includegraphics[scale=0.7]{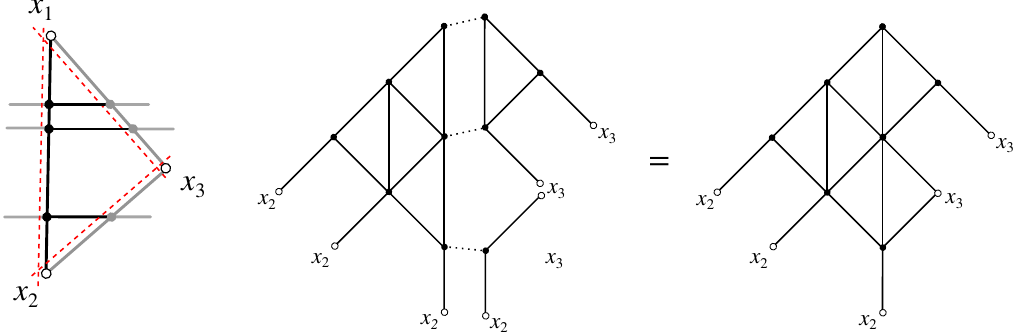}
\end{center}
For completeness, we include in case III the situation with exchanged bra/ket, namely the overlap of one ket function relative to the cut $x_{1}x_{3}$ with $M$ magnons $\mathbf{Y}_m$ with two bra wave-functions with $N,R$ magnons $\mathbf{Y}_n$ and $\mathbf{Y}_r$ relative to cuts $x_1 x_2$ and $x_{2}x_{3}$ such that $N+R=M$.
\begin{align}
\label{III_bk}
&\langle \Psi_{1 2} (\mathbf{Y}_n) \otimes \Psi_{2 3}(\mathbf{Y}_r) |  \Psi_{1 3}(\mathbf{Y}_m)\rangle\,,
\end{align}
and its picture is essentially a reflection of the last one.
\item[IV)]  Overlap of three wave-functions where one bra function relative to the cut $x_{12}$ has $N$ magnons and two ket wave-functions carry $M,R$ magnons relative to cuts $x_{1}x_{3}$ and $x_{3}x_{2}$ respectively, such that $N>M+R$. In this case, the leftover $n=N-M-R$ propagators $\phi_1$ propagators terminate in $x_3$, i.e. some ``reduction" happens
\begin{align}
\begin{aligned}
&\langle \Psi_{1 2} (\mathbf{Y}_n) |\Psi_{1 3}(\mathbf{Y}_m) \otimes  \underbrace{\delta^{(4)}_{x_3} \otimes \cdots \otimes \delta^{(4)}_{x_3}} \otimes \Psi_{2 3}(\mathbf{Y}_r)\rangle\,.
\end{aligned}
\end{align}
The triangular tile is of the type depicted for $N=3,M=R=1$ and $N=4,M=R=1$, together with the overlap of wave-functions to compute:
\begin{center}
\includegraphics[scale=0.7]{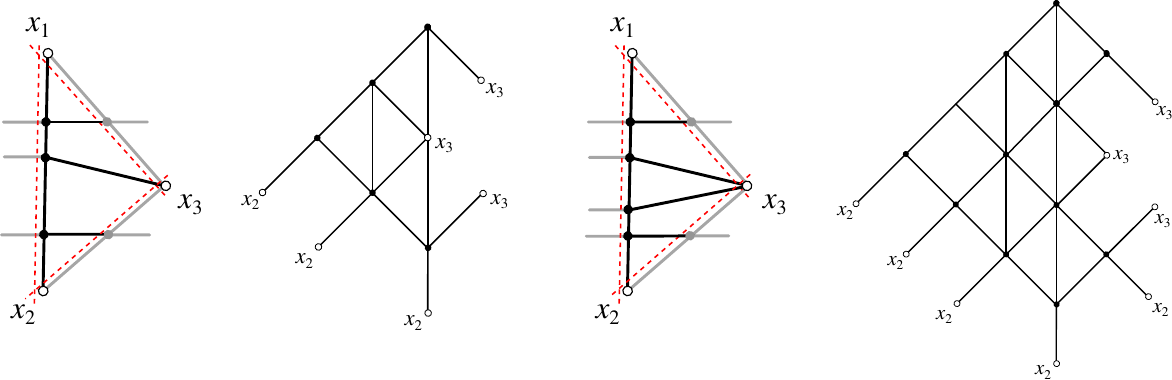}
\end{center}
As for the overlap of case II, exchanging bra/ket leads to a slightly different overlap formula. Indeed, for two edges of widths $N,R$ carrying two bra, and one edge of  width $M$ with a ket, the reductions of $n=M-N-R$ propagators $\phi_2$ to the point $x_2$ is reached after an amputation of graph-building operator in the tile, leading to the following formula:
\begin{align}
\label{IV_bk}
&\langle \Psi_{1 2} (\mathbf{Y}_n) \otimes  \underbrace{\delta^{(4)}_{x_2} \otimes \cdots \otimes \delta^{(4)}_{x_2}}_{M-N-R} \otimes \Psi_{2 3}(\mathbf{Y}_r)  |\prod_{j=N}^{M-R-1} (z_{j,j+1}^2)  |  \Psi_{1 3}(\mathbf{Y}_m)\rangle\,.
\end{align}
\end{itemize}
The overlaps described in case (IV) cover the most general occurrence, comprehending (III) whenever there are no reductions ($n=0$) as well as the cases (II) whenever $M=0$, $R=0$ or $N=0$. Finally the case (I) is recovered from (II) for $n=0$. The distinction of overlaps into cases is useful to work with simpler expressions whenever possible, and to break down the general derivation into easier steps.

In the following a few primers of computations of overlaps are presented -- case by case -- via the graphical notation of section \ref{sec:eigenf}. 
Dealing with a tile of vertices $x_1,x_2,x_3$, it is convenient to send the point $x_1$ to $\infty$ in order to deal with slightly lighter expressions and drawings. This allows to draw slightly lighter computations by replacing some eigenfunctions $\Psi_{ij}$ defined in \eqref{eig_biscalar_ij} with $\Psi_j$ given in \eqref{eig_biscalar} by the chain of transformations $T\cdot R \cdot I$ explained throughly in section \ref{sec:eigenf}. 
For the sake of generality we will sometimes recover the expression of the overlap at $x_1\neq \infty$. \footnote{A further simplification of the result would require to perform an extra conformal transformation that aligns the three points $x_1,x_2,x_3$ along the same direction as $(\infty,0,1)$. In this case the result of the overlap is stripped of any spacetime dependence. The ladder needs to be anyways restored when gluing patches of the triangulation, so we prefer to keep $x_2,x_3$ generic.} 
\subsection*{Examples of Overlaps}
\subsubsection*{Case I}
The overlap of case I for wave-functions of length $N$ reads
\begin{equation}
\int d^4 {y_1} \cdots d^4 {y_N} \,\overline{\Psi}_{2}(\mathbf{Y}_n; y_1,\dots,y_N) {\Psi}_{3}(\mathbf{Y}_m; y_1,\dots,y_N)\,.
\end{equation}
The computation of this overlap is straightforwardly achieved by one application of the star-triangle identity (in its ``amputated version"), with magnons $Y_n =(\nu,n)$
 and $Y_m=(\mu,m)$:
 \begin{center}
\includegraphics[scale=0.9]{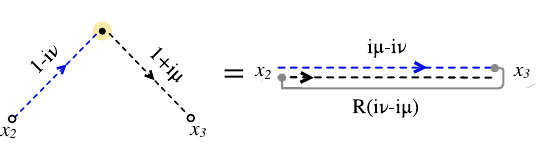}
\end{center}
In formulae, the star-triangle computation reads:
\begin{align}
\begin{aligned}
\langle\Psi_{2}(Y_n)|\Psi_{3}(Y_m)\rangle &= \!\! \int d^4 x \frac{[\mathbf{(x_2-x)}]^n [\overline{\mathbf{(x-x_3)}}]^m}{(x-x_2)^{2(1-i \nu)}(x-x_3)^{2(1+i \mu)}}= \frac{[\mathbf{x_{23}}]^n \mathbf{R}_{n,m}(i\nu-i\mu) [\overline{\mathbf{x_{23}}}]^m} {(x_2-x_3)^{2(i\mu- i \nu)}} \times \\  \!\!  \!\!  \!\! & \!\!  \!\!  \!\!  \!\!  \!\! \times\! \frac{\pi^2 (-1)^n \Gamma\left(1+\frac{m}{2}-i\mu\right) \Gamma\left(1+\frac{n}{2}+i\nu\right) \Gamma\left(\frac{m-n}{2}+i\mu-i\nu\right)}{\Gamma\left(1+\frac{m}{2}+i\mu\right) \Gamma\left(1+\frac{n}{2}-i\nu\right) \Gamma\left(1+\frac{m-n}{2}-i\mu+i\nu\right) \! \left(1+\frac{m+n}{2}-i\mu+i\nu\right)}\,.
\end{aligned}
\end{align}
Here and in the following we use the function $H(Y_m|Y_n) \equiv H_{m,n}(\mu,\nu)$ defined in \eqref{H_factor}, and the notation $\mathbf{R}(Y_m|Y_n)=\mathbf{R}_{m,n}(i\mu-i\nu)$ so that the latter overlap takes the compact form
\begin{align}
\begin{aligned}
\langle\Psi_{1 2}(Y_n)|\Psi_{1 3}(Y_m)\rangle = \pi^2 \frac{[\mathbf{x_{23}}]^n \mathbf{R}(Y_n|Y_m) [\overline{\mathbf{x_{23}}}]^m}{(x_2-x_3)^{2(i\mu- i \nu)}} \times H(Y_m|Y_n)\,.
\end{aligned}
\end{align}
The result for generic $x_1$ is restored performing the transformation corresponding to $|\Psi_{3}\rangle \mapsto |\Psi_{1 3}\rangle$, $\langle \Psi_{2}| \mapsto  \langle \Psi_{12}|$ and it reads
\begin{align}
\begin{aligned}
\langle\Psi_{1 2}(Y_n)|\Psi_{1 3}(Y_m)\rangle&= \pi^2  \frac{[\mathbf{\overline{x_{23}}x_{31}}]^n \mathbf{R}(Y_n|Y_m) [\mathbf{\overline{x_{12}}x_{23}}]^m} {(x_{12}^2)^{1-i\mu} (x_{13}^2)^{1+i\nu} (x_{23}^2)^{(i\mu-i \nu)}}  \times H(Y_m|Y_n) \,.
\end{aligned}
\end{align}

When $N\geq 2$ the computation of the overlap is more complicated, involving more star-triangle transformations. In figures, for $x_1\equiv \infty$, the result for overlap with $N=2$ is given by
\begin{center}
\includegraphics[scale=0.9]{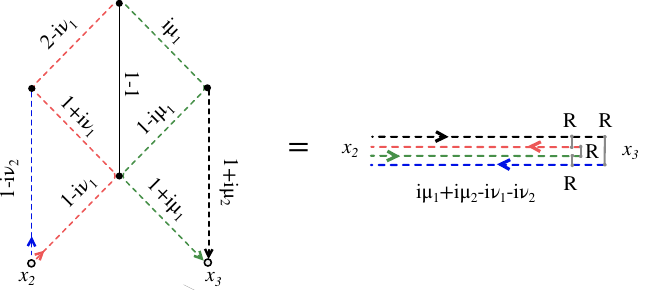}
\label{N2_over}
\end{center}
The rhs in the last picture has the following analytic expression
\begin{equation}
\frac{[\mathbf{x_{23}}]^{n_1}\! [\mathbf{x_{23}}]^{n_2}\! \mathbf{R}(Y_{n_1}|Y_{m_2})\mathbf{R}(Y_{n_1}|Y_{m_1})\mathbf{R}(Y_{n_2}|Y_{m_2})\mathbf{R}(Y_{n_2}|Y_{m_1}) \! [\overline{\mathbf{x_{23}}}]^{m_1} \! [\overline{\mathbf{x_{23}}}]^{m_2}}{(x_2-x_3)^{2(i\mu_1+i\mu_2-i\nu_1-i\nu_2)}}\,\! ,
\end{equation}
and it is multiplied by a factor -- due to the A-functions \eqref{A_functions} produced at each star-triangle transformation -- that together with the wave-functions normalisation in \eqref{eig_biscalar} equals to
\begin{equation}
\pi^8 H_{m_1,n_1} (\mu_1,\nu_1)H_{m_1,n_2} (\mu_1,\nu_2)H_{m_2,n_1} (\mu_2,\nu_1)H_{m_2,n_2} (\mu_2,\nu_2)\,.
\end{equation}
It compact form, the overlap of case I for generic width $N$ is proportional to the product of $N$ bra and $N$ ket wave-functions of length one, each one carrying an excitation $Y_{n_h}$ or $Y_{m_k}$. The spin indices of these functions are contracted with a product of R-matrices $\mathbf R(Y_{n_k}|Y_{m_h})$ and by a scalar factor $H(Y_{m_k}|Y_{n_h})$ for each pair of excitations $h,k=1,\dots, N$:
\begin{align}
\begin{aligned}
\label{T_overlap_2}
&\pi^{2N^2} {H}(\mathbf{Y}_m|\mathbf{Y}_n)\times  \left(\prod_{k=1}^{N}\frac{ [\mathbf{x_{23}}]^{n_k}}{(x_{23}^2)^{i\nu_k}} \right) \mathbf{R}(\mathbf{Y}_n|\mathbf{Y}_m)\left(\prod_{h=1}^{N}\frac{ [\mathbf{\overline{x_{23}}}]^{m_h}}{(x_{23}^2)^{i\mu_h}}\right)\,,
\end{aligned}
\end{align}
where the bold notation $\mathbf{Y}$ stands for a products of R-matrices
\begin{equation}
\label{compact_R}
\mathbf{R}(\mathbf{Y}_n|\mathbf{Y}_m)=\prod_{k=1}^N \prod_{h=1}^N \mathbf{R}(Y_{n_{k}}|Y_{m_{N-h}})\,,
\end{equation}
and for a product of scalar terms ${H}(\mathbf{Y}_n|\mathbf{Y}_m)=\prod_{k=1}^N \prod_{h=1}^N H(Y_{n_k}|Y_{m_h})$, both of them factorised over the pairs of bra/ket excitations.
The derivation of equation \eqref{T_overlap_2} is given explicitly in appendix \ref{app:firstover} for $N=2$; the procedure relies on a few iterative applications of star-triangle identities and can be repeated systematically for $N>2$.
\subsubsection*{Case II} The case I generalises to case II is simple.
 Let us consider the case $N=M+1$, namely
\begin{align}
\begin{aligned}
\langle\Psi_{2}(\mathbf{Y_{n}})|\Psi_{3}(\mathbf{Y_m}) \otimes \delta_{x_3}^{(4)}\rangle &= \!\! \int\! d^{4M} y_k\, \,\overline{\Psi}_{2}(\mathbf{Y_n}|y_1,\cdots ,y_{N-1},x_3)  {\Psi}_{3}(\mathbf{Y_m}|y_1, \dots, y_{N-1})\,.
\end{aligned}
\end{align}
For $N=2$, in pictures the overlap follows from two subsequent star-triangle transformations relative to the enlightened integration points
\begin{center}
\includegraphics[scale=1]{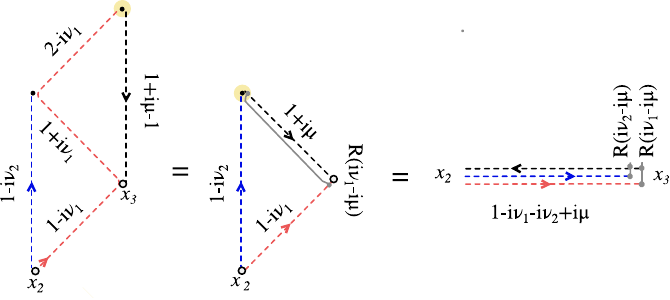}
\end{center}
In formulae, accounting for the normalisation of the wave-functions \eqref{eig_biscalar}, the overlap results in the expression
\begin{align}
\begin{aligned}
& \langle \Psi_{2} ({Y}_{n_1},Y_{n_2}) |\Psi_{3}({Y}_m) \otimes  \delta^{(4)}_{x_3} \rangle =  \\ &= \pi^{2N M} \frac{H(Y_m|Y_{n_2}) H(Y_m|Y_{n_2})}{E(Y_m)}\times
 \frac{   [\overline{\mathbf{x_{23}}}]^{n_1}  [\overline{\mathbf{x_{23}}}]^{n_2} \mathbf{R}_{n_1,m}(i\nu_1-i\mu) \mathbf{R}_{n_2,m}(i\nu_2-i\mu) [\mathbf{x_{23}}]^{m}   }{(x_{23}^2)^{1+i \mu-i\nu_1-i\nu_2}}  \,.
\end{aligned}
\end{align}
The details of the computation for $N=2,3$ are reported in the appendix \ref{NMp1}. For general $N$ the computation can be repeated systematically and leads to the following formula 
\begin{align*}
\begin{aligned}
\langle \Psi_{2} (\mathbf{Y}_{n}) |\Psi_{3}(\mathbf{Y}_m) \otimes  \delta^{(4)}_{x_3} \rangle = \pi^{2N M} \frac{{H}(\mathbf{Y}_m|\mathbf{Y}_n)}{{E}(\mathbf{Y}_m)^{N-M}}\! \times\!
 \frac{   [\overline{\mathbf{x_{23}}}]^{\mathbf n}  \mathbf{R}(\mathbf{Y}_n| \mathbf{Y}_m)[\mathbf{x_{23}}]^{\mathbf m} }{(x_{23}^2)^{1+i \boldsymbol{\mu}-i\boldsymbol{\nu}}}.
\end{aligned}
\end{align*}
Proceeding by increasing generality to the cases $N=M+2$, one discovers that the result is modified by an additional power of ${E}(\mathbf{Y}_m)^{-1} ={E}({Y}_{m_1})^{-1}  \cdots {E}({Y}_{m_M})^{-1}$ and the same happens for higher values of $N-M$, that is
\begin{align}
\begin{aligned}
\langle\Psi_{2}(\mathbf{Y}_{n})|\Psi_{3}(\mathbf{Y}_m) \otimes \delta_{x_3}^{(4)}\rangle &= \!\! \int\! d^{4M} y_k\, \overline{\Psi}_{2}(\mathbf Y_{n};y_1,\cdots ,y_{M},x_3,\cdots,x_3)  {\Psi}_{3}(\mathbf Y_{m};y_1, \dots, y_{M})\,,
\end{aligned}
\end{align}
that are described by the general formula
\begin{align}
\begin{aligned}
\label{CaseII}
\langle \Psi_{2} (\mathbf{Y}_{n}) |\Psi_{3}(\mathbf{Y}_m) \otimes  \underbrace{\delta^{(4)}_{x_3} \otimes  \cdots \otimes  \delta^{(4)}_{x_3}}_{N-M} \rangle &=\\ =  \pi^{2N M} \frac{{H}(\mathbf{Y}_m|\mathbf{Y}_n)}{{E}(\mathbf{Y}_m)^{N-M}}\! &\times\!
 \frac{[\overline{\mathbf{x_{23}}}]^{\mathbf n}  \mathbf{R}(\mathbf{Y}_n| \mathbf{Y}_m)[\mathbf{x_{23}}]^{\mathbf m} }{(x_{23}^2)^{N-M+i \boldsymbol{\mu}-i\boldsymbol{\nu}}}\,.
\end{aligned}
\end{align}
In the general situation $x_1\neq 0$ the last formula improves to 
\begin{align}
\begin{aligned}
\label{CaseII_gen}
&\langle \Psi_{12} (\mathbf{Y}_{n}) |\Psi_{13}(\mathbf{Y}_m) \otimes  \underbrace{\delta^{(4)}_{x_3} \otimes  \cdots \otimes  \delta^{(4)}_{x_3}}_{N-M} \rangle =\\ &=  \pi^{2N M}\frac{{H}(\mathbf{Y}_m|\mathbf{Y}_n)}{{E}(\mathbf{Y}_m)^{N-M}}\!\times\!
\frac{[\mathbf{x_{23}\overline{x_{31}}}]^{\mathbf n} \mathbf{R}(Y_n|Y_m) [\mathbf{x_{12}\overline{x_{23}}}]^{\mathbf m}}{(x_{12}^2)^{M-i\boldsymbol{\mu}} (x_{13}^2)^{2M-N+i\boldsymbol{\nu}}   (x_{23}^2)^{N-M+i \boldsymbol{\mu}-i\boldsymbol{\nu}}}\,.
\end{aligned}
\end{align}
A particular case of \eqref{CaseII} is given by $M=0$, for which the overlap reduces the bra eigenfunction of length $N$ into a product of $N$ bra eigenfunctions of length $1$, each one carrying one excitation

\begin{align}
\begin{aligned}
\langle \Psi_{1 2} (\mathbf{Y}_{n}) |\underbrace{\delta^{(4)}_{x_3} \otimes  \cdots \otimes  \delta^{(4)}_{x_3}}_{N} \rangle = \frac{[\mathbf{\overline{x_{23}}}]^{\mathbf n}}{(x_{23}^2)^{N-i\boldsymbol{\nu}}}\,,
\end{aligned}
\end{align}
which is a familiar result in the context of Basso-Dixon integrals \cite{Derkachov2020}.
\subsubsection*{Case III}
\begin{figure}[t]
\begin{center}
\includegraphics[scale=0.7]{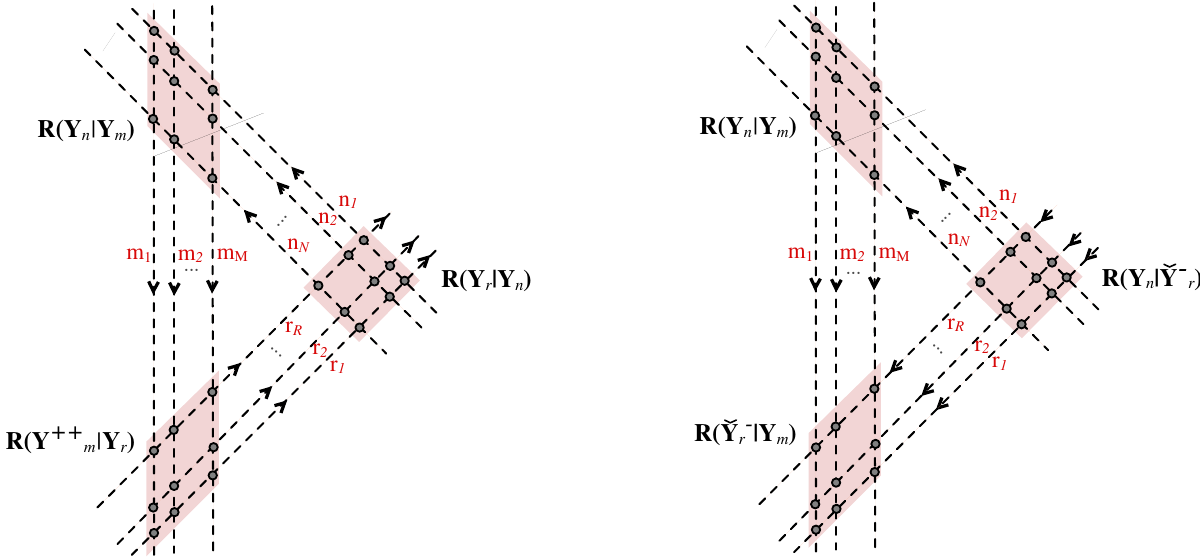}
\caption{\textbf{Left:} The structure of $\mathbf{R}$ matrices emerging from the computation of overlaps \eqref{CaseIII} is that of a partition function of a system of lines with three slopes. Each slope is associated to a wave-function and each line with an excitation, whose spin number is denoted in red. The interaction vertex when $a_i$ (in-coming arrow) and $b_j$ (out-coming arrow) is the matrix $\mathbf{R}({Y}_{a_i}|{Y}_{b_j})$. The orientation of lines fixes the order of matrices in the product. When a line ${m_i}$ crosses a line ${r_j}$ there is a shift $Y_{m}^{++}$ in the rapidity. \textbf{Right:} The same picture after the charge-conjugation and transposition of the spaces of $Y_r$'s. In this case the cyclic orientation of arrows is broken in favour of an orientation right-to-left of the arrows that permits to write formula \eqref{matrix_part} without spinor indices.}
\label{partition_F}
\end{center}
\end{figure}

The overlaps of case III are quite more general than previous cases, since here all the three edges of a tile are traversed by at least by one propagator $\phi_1$ and therefore carry at least one mirror excitation. Starting from the simplest case $N=2$ and $M=R=1$, that is
\begin{align}
\begin{aligned}
\langle \Psi_{2} ({Y}_{n_1},Y_{n_2}) | \Psi_{3} ({Y}_{m}) \otimes   \Psi_{32} ({Y}_{r}) \rangle =\int d^4{y_1} d^4{y_2}\, \overline{\Psi}_{2} ({Y}_{n_1},Y_{n_2}| y_1,y_2) \Psi_{3} ({Y}_{m}|y_1) \Psi_{32} ({Y}_{r}|y_2)\,,
\end{aligned}
\end{align}
It is relatively simple to compute the overlap in the graphical notation via a chain of five star-triangle transformations, where we enlighten to which integrated point or to which triangle this identity is applied:
 \begin{center}
\includegraphics[scale=0.9]{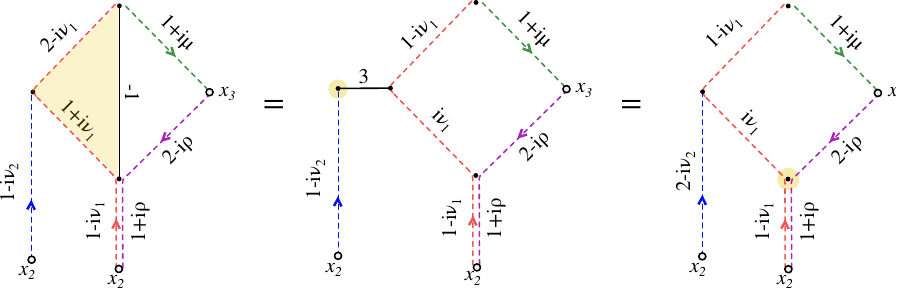}
\end{center}
\begin{center}
\includegraphics[scale=0.9]{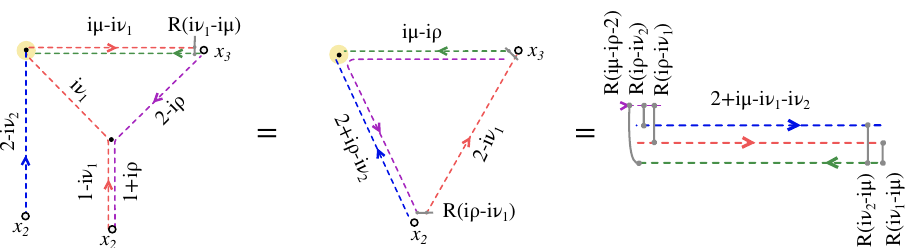}
\end{center}
The rhs of the last picture shall be expressed in formulae. First, notice that in order to write the matrix part without explicit spinor indices it is needed to use transposition $t_r$ of the spinor indices carried by the magnons $\mathbf{Y}_r$, namely
\begin{equation}
\label{matrix_part}
\left(\mathbf{R}_{rn_1}(i\rho-i\nu_1)^{t_r} \mathbf{R}_{rn_2}(i\rho-i\nu_2)^{t_r}  \mathbf{R}_{n_1m}(i\nu_1-i\mu) \mathbf{R}_{n_2m}(i\nu_2-i\mu) \mathbf{R}_{rm}(i\mu-i\rho-2)^{t_r}\right)^{t_r}.
\end{equation}
The $R$-matrices satisfy a notorious crossing identity after transposition and charge conjugation,
\begin{equation}
\label{transp_id}
\mathbf{R}_{mn}(u)^{t_m} = c_{mn}(u) \times  \boldsymbol{\varepsilon}^{ m} \mathbf{R}_{mn}(-u-1) \boldsymbol{\varepsilon}^{ m}  \,,
\end{equation}
where $\boldsymbol{\varepsilon}^m = (\sigma_2)^{\otimes m}$ is the charge-conjugation matrix (details in appendix \ref{app:Rmat}). 
It follows that \eqref{matrix_part} can be rewritten using the crossing property and introducing the notation $Y_r^{\pm} = (r,\rho\pm i)$ as
\begin{equation}
\label{matrix_part_II}
\boldsymbol{\varepsilon}^{r} \left( \mathbf{R}(Y_{n_1}|Y_r^-)\mathbf{R}(Y_{n_2}|Y_r^-) \mathbf{R}(Y_{n_1}|Y_m) \mathbf{R}(Y_{n_2}|Y_m) \mathbf{R}(Y_{r}^-|Y_m) \right)^{t_r} \boldsymbol{\varepsilon}^{r} \,.
\end{equation}
Now, there is a factor that multiplies the matrix part as a consequence of the crossing property \eqref{transp_id}, of the star-triangle transformations and of the wave-functions normalisation. Overall, this factor reads
\begin{align}
\begin{aligned}
&\frac{H(Y_m|Y_{n_1}) H(Y_m|Y_{n_2}) H(Y_{n_1}|Y_{r})  H(Y_{n_2}|Y_{r}) }{H(Y_m| Y_r)} \frac{E(Y_r)}{E(Y_{n_1})E(Y_{n_2})}\frac{c_{m r}(i\mu-i\rho-1) ( \pi^2)^{N M + N R -M R}}{c_{n_1 r}(i\nu_1-i\rho-1)c_{n_2 r}(i\nu_2-i\rho-1)}\,.
\end{aligned}
\end{align}
Using the compact notation \textbf{Y}, the full expression of the overlap reads
\begin{align}
\begin{aligned}
&\frac{H(Y_m|\mathbf{Y}_n) H(\mathbf{Y}_n|Y_r)}{H(Y_m|Y_r)} \frac{E(Y_r)}{E(\mathbf Y_{n})} \frac{c({Y}_m|Y_r^-)}{{c}(\mathbf{Y}_n|Y_r^-)}\,\frac{\boldsymbol{\varepsilon}^{ r} [\mathbf{x_{23}}]^{m} \left(\mathbf{R}(\mathbf{Y}_n|Y_r^{-}) \mathbf{R}(\mathbf{Y}_n|Y_m) \mathbf{R}({Y}_r^{-}|Y_m)\right)^{t_r} [\overline{\mathbf{x_{23}}}]^{\mathbf n}  \boldsymbol{\varepsilon}^{ r}}{\pi^{2(M R -N M -N R)} (x_{23}^2)^{2+i\mu-i\nu_1-i\nu_2}}\,.
\end{aligned}
\end{align}
It is simple to repeat the latter computation for different values of $N,M,R$ (see for a few detailed cases the appendix \ref{app:lastover}), as the technique relies on iterations of star-triangle transformations, allowing to derive a general formula for the overlaps of case III \begin{align}
\begin{aligned}
\label{CaseIII}
\langle \Psi_{2} (\mathbf{Y}_{n}) | \Psi_{3} (\mathbf{Y}_{m}) \otimes   \Psi_{32} (\mathbf{Y}_{r}) \rangle &= E(\mathbf Y_m)^M E(\mathbf Y_{r})^R \frac{\tilde H(\mathbf Y_m|\mathbf{Y}_n) \tilde H(\mathbf{Y}_n|\mathbf Y_r)}{\tilde H(\mathbf Y_m|\mathbf Y_r)} \frac{c(\mathbf{Y}_r|\mathbf Y_n)}{c(\mathbf{Y}_r|\mathbf Y_m)} \times \\ &\times \frac{\boldsymbol{\varepsilon}^{ r} [\mathbf{x_{23}}]^{\mathbf m} (\mathbf{ R}(\mathbf{Y}_n|\mathbf{\check{Y}}_r^{-}) \mathbf{R}(\mathbf{Y}_n|\mathbf Y_m) \mathbf{ R}(\check{\mathbf{Y}}_r^{-}|\mathbf Y_m))^{t_r}[\overline{\mathbf{x_{23}}}]^{\mathbf n} \boldsymbol{\varepsilon}^{ r}}{(x_{23}^2)^{N-M+R+i \boldsymbol{\mu}-i\boldsymbol{\nu}}}\,.
\end{aligned}
\end{align}
Here the notation $\tilde H(Y|Y') = \pi^2 H(Y|Y')/E(Y) $ is meant to get a more transparent structure of energy factors $\mathbf{E}(Y)$, given only by the excitation carried by the ket eigenfunctions to a power given by the width of the corresponding cut. The symbol $\mathbf{\check {Y}}_r$ stands for a set of excitation in reverted order $\mathbf{\check{Y}} = (Y_{r_R}, \dots, Y_{r_2},Y_{r_1})$ as a consequence of the transpositions $t_r$, so that together with the definition \eqref{compact_R} one has:
\begin{equation}
 \mathbf{R}(\check{\mathbf{Y}}_r^{-}|\mathbf Y_m) = \prod_{k=1}^R \prod_{h=1}^M  \mathbf{R}(Y_{r_{R-k}}^{-}| Y_{m_{M-h}})\, , \qquad
\mathbf{R}(\mathbf{Y}_n|\check{\mathbf{Y}}_r^{-}) = \prod_{k=1}^N \prod_{h=1}^R  \mathbf{R}(Y_{n_k}| Y_{r_{h}}^{-})\,.
\end{equation}
A neat picture of the finite-dimensional partition functions of the type $\mathbf{R}(\mathbf{Y}|\mathbf{Y}')$ appearing \eqref{CaseIII} is given in Figure \ref{partition_F}.  
Thanks to the \textbf{bold} notation which stands for product over the magnons, the structure of the overlap is identical for all the width $M,R,N$ of the cuts. This factorization property is a consequence of the integrability of the spin-chain model, a striking advantage of working within the SoV representation of the Fishnet.
\subsubsection*{Case IV}
This latter case is the general one, as it comprehends cases I, II and III as particular corners. The type of computations implied in the derivation of such overlaps are the same as those illustrated so far.
The overlap of a bra wave-function with $N$ magnons $\mathbf{Y}_n$ along the cut $x_{1}x_{2}$ that overlaps with two ket wave-functions of respectively $M$ magnons  $\mathbf{Y}_m$ along $x_{1}x_{3}$ and $R$ magnons  $\mathbf{Y}_r$ along $x_{2}x_{3}$ such that $N>M+R$, and the leftover $N-M-R$ propagators of $\phi_1$ terminate in the point $x_3$, that is
\begin{equation}
\label{overlap4_d}
\langle \Psi_{2} (\mathbf{Y}_n) |\Psi_{3}(\mathbf{Y}_m) \otimes  \underbrace{\delta^{(4)}_{x_3} \otimes \cdots \otimes \delta^{(4)}_{x_3}}_{N-M-R} \otimes \Psi_{32}(\mathbf{Y}_r)\rangle\,.
\end{equation}
The result provides a general formula for overlaps of type bra/ket$^2$ which reads
\begin{align}
\begin{aligned}
\label{full_over}
\eqref{overlap4_d}=&\frac{\tilde H(\mathbf Y_m|\mathbf{Y}_n) \tilde H(\mathbf{Y}_n|\mathbf Y_r)}{\tilde H(\mathbf Y_m|\mathbf Y_r)} \frac{c(\mathbf{Y}_r|\mathbf Y_n)}{c(\mathbf{Y}_r|\mathbf Y_m)}E(\mathbf Y_r)^R E(\mathbf{Y}_m)^{M}  \times \\ &\times  \frac{\boldsymbol{\varepsilon}^{ r} [\mathbf{x_{23}}]^{\mathbf m} (\mathbf{ R}(\mathbf{Y}_n|\mathbf{\check{Y}}_r^{-}) \mathbf{R}(\mathbf{Y}_n|\mathbf Y_m) \mathbf{ R}(\check{\mathbf{Y}}_r^{-}|\mathbf Y_m))^{t_r}[\overline{\mathbf{x_{23}}}]^{\mathbf n} \boldsymbol{\varepsilon}^{ r}}{(x_{23}^2)^{(N-M+R+i \boldsymbol{\mu}-i\boldsymbol{\nu})}}\,.
\end{aligned}
\end{align}
\subsection*{Bra/ket exchange}
The formulae \eqref{CaseII}, \eqref{CaseIII} and \eqref{full_over} cover the case of the overlaps that occur in a triangular tile of type bra/ket$^2$, i.e. with bra wave-function of width $N$ which is bigger or equal to that of two ket wave-functions $N\geq M+R$. In order to describe completely the zoology of overlaps, one should work out the cases II-IV whenever two bra wave-functions of width $N,R$ are overlapped with one ket wave-function of width $M\geq N+R$ classified in \eqref{II_bk},\eqref{III_bk} and \eqref{IV_bk}.
The overlap in the most comprehensive case (IV) takes the compact form -- where the $\delta=1$ if $M+R<N$ and zero otherwise:
\begin{align}
\begin{aligned}
\label{full_over_bk}
\eqref{IV_bk}=&\frac{\tilde H(\mathbf Y_m|\mathbf{Y}_n) \tilde H(\mathbf{Y}_r|\mathbf Y_m)}{\tilde H(\mathbf Y_r|\mathbf Y_n)} \frac{c(\mathbf{Y}_m|\mathbf Y_r)}{c(\mathbf{Y}_n|\mathbf Y_r)} {E(\mathbf{Y}_m)^{M}} (E(\mathbf Y_r) E(\mathbf{Y}_n)/E(\mathbf{Y}_m))^{\delta}  \times \\ &\times \frac{\boldsymbol{\varepsilon}^{ r} [\mathbf{x_{23}}]^{\mathbf m} (\mathbf{R}(\mathbf{Y}_n|\check{\mathbf Y}_r^{+}) \mathbf{R}(\mathbf{Y}_n|\mathbf Y_m) \mathbf{R}(\check{\mathbf{Y}}_r^{+}|\mathbf Y_m))^{t_r}[\overline{\mathbf{x_{23}}}]^{\mathbf n} \boldsymbol{\varepsilon}^{ r}}{(x_{23}^2)^{M-N-R+i \boldsymbol{\mu}-i\boldsymbol{\nu}}}\,.
\end{aligned}
\end{align}
The reduction of \eqref{full_over_bk} for $N=R=0$ is the situation that occurs at the left boundary of Basso-Dixon integrals, computed in details in \cite{Derkachov2020}. In this case $\delta=1$, the dynamical and matrix factors disappear due to absence of interactions:\begin{equation}
\langle  \underbrace{\delta^{(4)}_{x_2} \otimes \cdots \otimes \delta^{(4)}_{x_2}}_{M}  |\prod_{j=1}^{M-1} (z_{j,j+1}^2) |  \Psi_{1 3}(\mathbf{Y}_m)\rangle = E(\mathbf{Y}_m)^{M-1}\frac{ [{\mathbf{x_{23}}}]^{\mathbf m}}{(x_{23}^2)^{i\boldsymbol{\mu}}}\,.
\end{equation}
\section{Tiles: General Formulae}
\label{sec:tiles}
\begin{figure}
\begin{center}
\includegraphics[scale=0.9]{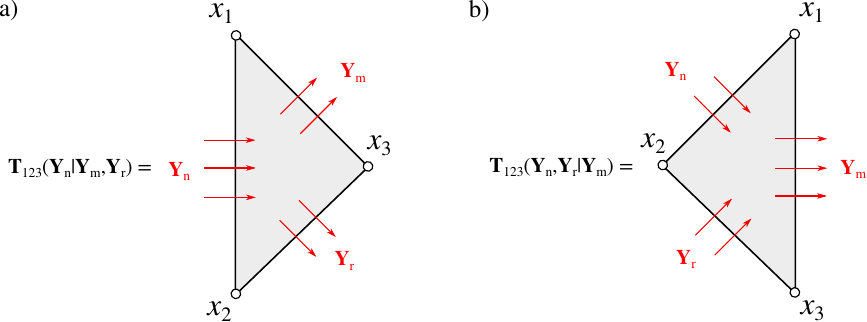}
\caption{(a) A tile of type bra/ket$^2$, hosting magnons $\mathbf{Y}_n$ on the bra edge $x_1 x_2$ and magnons $\mathbf{Y}_m$, $\mathbf{Y}_r$ on the ket edges $x_1 x_3$, $x_2 x_3$ respectively. (b) A tile of type ket/bra$^2$, hosting magnons $\mathbf{Y}_m$ on the ket edge $x_1 x_3$ and magnons $\mathbf{Y}_n$, $\mathbf{Y}_r$ on the bra edges $x_1 x_2$ and $x_2 x_3$.}
\label{two_tiles}
\end{center}
\end{figure}

The discussion of overlaps of the previous sections can be wrapped-up in a few \emph{boxed} formulae of general validity. A triangular tile of the Fishnet is characterised by the three widths $\{N,M,R\}$ of its three edges, say $x_{1}x_{2},\,x_{1}x_{3}$ and $x_{2}x_{3}$. 
The edge of a tile can host the bra or the ket wave-function coming from the insertion of a resolution of the identity \eqref{cutting_general} along the edge itself. There exist two types of tiles: the first with two edges with a ket and one edge with the bra (bra/ket$^2$) and the second with two bras and one ket (bra$^2$/ket). They are represented respectively in figure \ref{two_tiles} (a) and (b). The width $N$ of the edge equals the number of magnons inserted along the cut, and we refer to their quantum numbers with the compact notation
\begin{equation}
\mathbf{Y}_n = (Y_{n_1},Y_{n_2},\dots, Y_{n_N})\,,\,
\end{equation}
recalling that each label $Y_n=(\nu,n)$ defines the rapidity $\nu$ and the symmetric traceless representation $\left(\frac{n}{2},\frac{n}{2}\right)$ of the excitation. Adopting the notation $\mathbf{T}(\mathbf{Y}_{bra}|\mathbf{Y}_{ket})$, the two types of tiles are denoted
\begin{equation}
\label{the_tiles}
\mathbf{T}_{123}(\mathbf{Y}_n|\mathbf{Y}_m,\mathbf{Y}_r)\,,\,\,\text{and}\, \mathbf{T}_{123}(\mathbf{Y}_n,\mathbf{Y}_r|\mathbf{Y}_m)\,.
\end{equation}
The formulae \eqref{full_over} and \eqref{full_over_bk} describe the two types of tile with two general expressions to be analysed. Starting from formula \eqref{full_over}, it can be separated into an interaction part -- depending on pairs of magnons inserted along different cuts -- and a part that is factorized over contributions of single magnons.
We start the analysis from the first one. The part that contains interactions between magnons of different edges features a matrix part and a scalar ``dynamical" part. 
The dynamical part contains all such factors that depend on the quantum numbers of pairs of magnons. Such factors appear in the overlap and couple pairs of magnons inserted along different cuts, namely
   \begin{equation}
   \label{Dyn_part_I}
\boxed{\mathcal{D}(\mathbf{Y}_n|\mathbf{Y}_m,\mathbf{Y}_r)=\frac{\tilde H(\mathbf Y_m|\mathbf{Y}_n) \tilde H(\mathbf{Y}_n|\mathbf Y_r)}{\tilde H(\mathbf Y_m|\mathbf Y_r)}}\,,
  \end{equation}
  where the explicit form of $H(\mathbf{Y}|\mathbf{Y}')$ is given after formula \eqref{compact_R}.
The matrix part acts on the symmetric spinor indices of the excitations; we break it into a product of crossing factors and a product of $R$-matrices:
\begin{equation}
\label{Matrix_part_I}
\boxed{\mathcal{M}(\mathbf{Y}_n|\mathbf{Y}_m,\mathbf{Y}_r)=\frac{ c(\mathbf{Y}_r|\mathbf Y_n)}{{c}(\mathbf{Y}_r|\mathbf Y_m)}\times \mathbf{R}(\mathbf{Y}_n|\check{\mathbf Y}_r^{-}) \mathbf{R}(\mathbf{Y}_n|\mathbf Y_m) \mathbf{R}(\check{\mathbf{Y}}_r^{-}|\mathbf Y_m)}\,.
\end{equation}
The order of $R$-matrices in the matrix part can be interchanged using the Yang-Baxter equation
\begin{equation}
\label{YBE_Mat_part}
\mathbf{R}(\mathbf{Y}_n|\check{\mathbf{Y}}_r^{-}) \mathbf{R}(\mathbf{Y}_n|\mathbf Y_m) \mathbf{R}(\check{\mathbf{Y}}_r^{-}|\mathbf Y_m)=\mathbf{R}(\check{\mathbf{Y}}_r^{-}|\mathbf{Y}_m) \mathbf{R}(\mathbf{Y}_n|\mathbf Y_m) \mathbf{R}(\mathbf{Y}_n|\check{\mathbf Y}_{r}^{-})\,.
\end{equation}
The symmetry under permutation of excitations expressed by  \eqref{exchange_waves} allows to easily re-shuffle the order of $Y's$ inside \eqref{Matrix_part_I}. For instance, moving $Y_{m_h}$ by $k$ positions to the left in the string $\mathbf{Y}_m= (Y_{m_1},\cdots , Y_{m_M})$ the matrix part involving the reshuffled set of excitations $\tilde{\mathbf{Y}}_m$ reads
\begin{equation}
\label{Matrix_part_Exch_I}
\mathcal{M}(\mathbf{Y}_n|\mathbf{Y}_m,\mathbf{Y}_r)=\left(\prod_{j=h-k}^{h-1} \mathbf{R}(Y_{m_h}|Y_{m_j})\right) \mathcal{M}(\mathbf{Y}_n|\tilde{\mathbf{Y}}_m,\mathbf{Y}_r) \left(\prod_{j=h-k}^{h-1} \mathbf{R}(Y_{m_j}|Y_{m_h})\right) \,,
\end{equation}
or again, moving $Y_{m_h}$ by $k$ positions to the right, the matrix part transforms as:
\begin{equation}
\label{Matrix_part_Exch_II}
\mathcal{M}(\mathbf{Y}_n|\mathbf{Y}_m,\mathbf{Y}_r)=\left(\prod_{j=h+1}^{h+k} \mathbf{R}(Y_{m_j}|Y_{m_h})\right) \mathcal{M}(\mathbf{Y}_n|\tilde{\mathbf{Y}}_m,\mathbf{Y}_r) \left(\prod_{j=h+1}^{h+k} \mathbf{R}(Y_{m_h}|Y_{m_j})\right) \,.
\end{equation}
Another remarkable property of \eqref{Matrix_part_I} is that whenever one magnon in the bra and one in the ket have coinciding quantum numbers, e.g. $Y_n= Y_m$, the pair \emph{decouples} from the matrix part. At such decoupling point the $R$-matrix reduces to the exchange of their spinor indices, $\mathbf{R}(Y_n|Y_n)=\mathbf{R}_{n,m=n}(0)= \mathbb{P}_{n,m}$ (see also appendix \ref{app:Rmat}) and the net effect is the decoupling of excitations $Y_n, Y_m$ from the overlap in \eqref{Matrix_part_I} and the identification of the two. For one magnon on each edge this means
\begin{align}
\begin{aligned}
\label{simpli_Mat}
&\mathcal{M}(Y_n|Y_n,Y_r)= \mathbb{P}_{n,m}\,,
\end{aligned}
\end{align}
and the two excitations are identified under trace of spinor indices, eventually appearing in the resolution of the identity, following the property $\text{Tr}_{m,n} A_m\otimes B_n \mathbb{P}_{m,n}=\text{Tr}_{m} A_m B_m = \text{Tr}_{n} A_n\otimes B_n$ that holds whenever $m=n$. Similarly, the bra/ket decoupling happens also at $Y_{r}=Y_{n}$ giving
\begin{align}
\begin{aligned}
\label{simpli_Mat_II}
&\mathcal{M}(Y_n|Y_m,Y_n)=  \varepsilon^{r} \varepsilon^{n}  \mathbb{P}_{n,r} \,.
\end{aligned}
\end{align}
For larger widths of the tile, the exchange of the two excitations $Y_{n_h},Y_{m_k}$ affects the interaction of these two with other magnons; the matrix part can still be remarkably simplified by use of \eqref{simpli_Mat} and Yang-Baxter equations. Let ${\mathbf{Y}}'_m, {\mathbf{Y}}'_n$ stand for the sets of excitations without $Y_{m_M}$ and $Y_{n_N}$; upon decoupling and identification of the two spaces the matrix part reads
\begin{align}
\begin{aligned}
\label{DECA_nice}
\mathcal{M}(\mathbf Y_n| \mathbf Y_m,\mathbf Y_r)_{Y_{n_N}=Y_{m_M}}&= \mathbf{R}(Y_{m_M}|\mathbf{Y}_{m}')  \mathcal{M}({\mathbf Y}'_n|{\mathbf Y}'_m,\mathbf Y_r)  \mathbf{R}(\mathbf{Y}'_{n}|Y_{m_M})
\end{aligned}
\end{align}
The formula \eqref{DECA_nice} generalizes to $Y_{n_h}=Y_{m_k}$ easily due to the exchange symmetry \eqref{exchange_waves} of an eigenfunctions excitations. Concretely, it can be derived by first moving $Y_{n_h},Y_{m_k}$ to the position of $Y_{n_N},Y_{m_M}$ via \eqref{Matrix_part_Exch_I}-\eqref{Matrix_part_Exch_II} and then to make use of \eqref{DECA_nice}. The resulting formula reads
\begin{align}
\begin{aligned}
\label{DECA_nice_general}
\mathcal{M}(\mathbf Y_n| \mathbf Y_m,\mathbf Y_r)_{Y_{n_h}=Y_{m_k}}&= \mathbf{R}(\{Y_{n_{h+1}},\dots ,Y_{n_{N}}\}|Y_{m_k}) \mathbf{R}(Y_{m_k}|\{Y_{m_1},\dots ,Y_{m_{k-1}}\}) \times \\&\times \mathcal{M}({\mathbf Y}'_n|{\mathbf Y}'_m,\mathbf Y_r) \times \\&\times  \mathbf{R}(\{Y_{n_{1}},\dots ,Y_{n_{h-1}}\}|Y_{m_k}) \mathbf{R}(Y_{m_k}|\{Y_{m_{k+1}},\dots ,Y_{m_{M}}\})\,.
\end{aligned}
\end{align}
\begin{figure}[t]
\begin{center}
\includegraphics[scale=0.63]{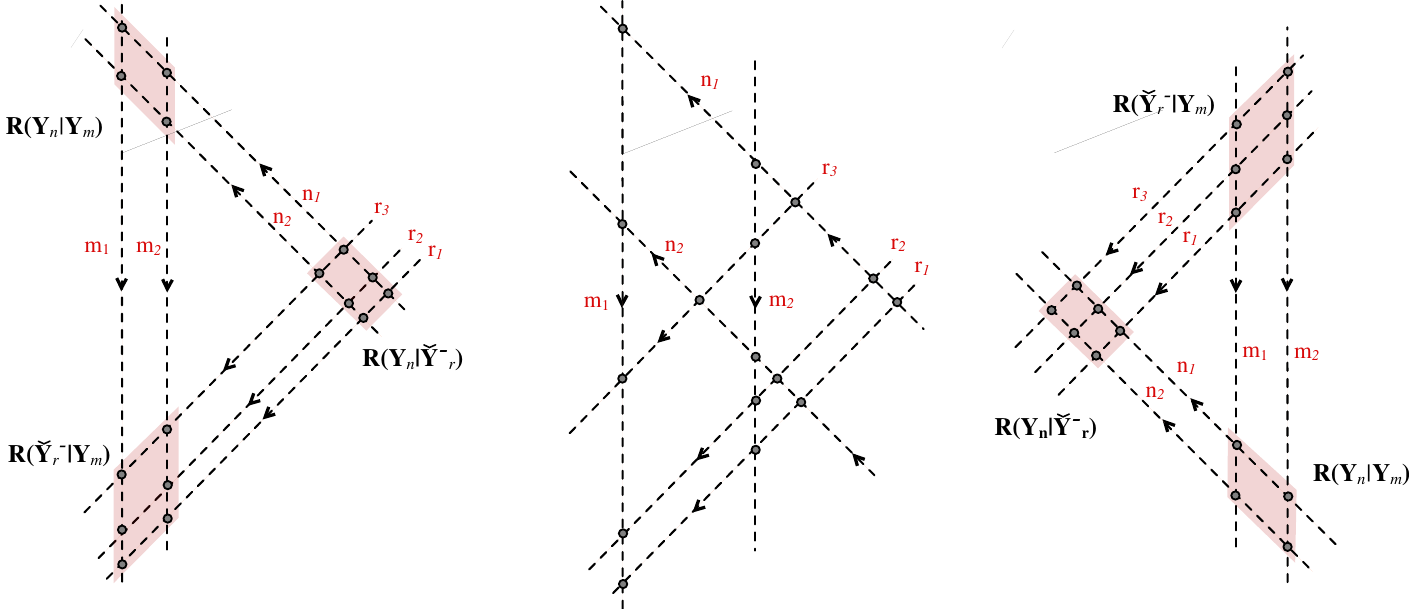}
\caption{Starting from the partition function in the l.h.s. of \eqref{YBE_Mat_part} with $N=M=2$ and $R=3$ one can apply the Yang-Baxter equation to perform rigid moves of the lines. The central picture is obtained from the left by the replacement $\mathbf{R}({Y}_{n_2}|{Y}_{r_3}^{-})\mathbf{R}({Y}_{n_2}|{Y}_{m_2})\mathbf{R}({Y}_{r_3}^{-}|{Y}_{m_2})=\mathbf{R}({Y}_{r_3}^{-}|{Y}_{m_2})\mathbf{R}({Y}_{m_2}|{Y}_{n_2})\mathbf{R}({Y}_{n_2}|{Y}_{r_3}^{-})$. Iterating the YBE the lines can be moved to the configuration in the right picture $\mathbf{R}(\mathbf{Y}_r^{-}|\mathbf{Y}_m)\mathbf{R}(\mathbf{Y}_n|\mathbf{Y}_m)\mathbf{R}(\mathbf{Y}_n|\mathbf{Y}_r^{-})$ realizing the r.h.s. of \eqref{YBE_Mat_part}.}
\label{partition_F_YBE}
\end{center}
\end{figure}
\begin{figure}[t]
\begin{center}
\includegraphics[scale=0.63]{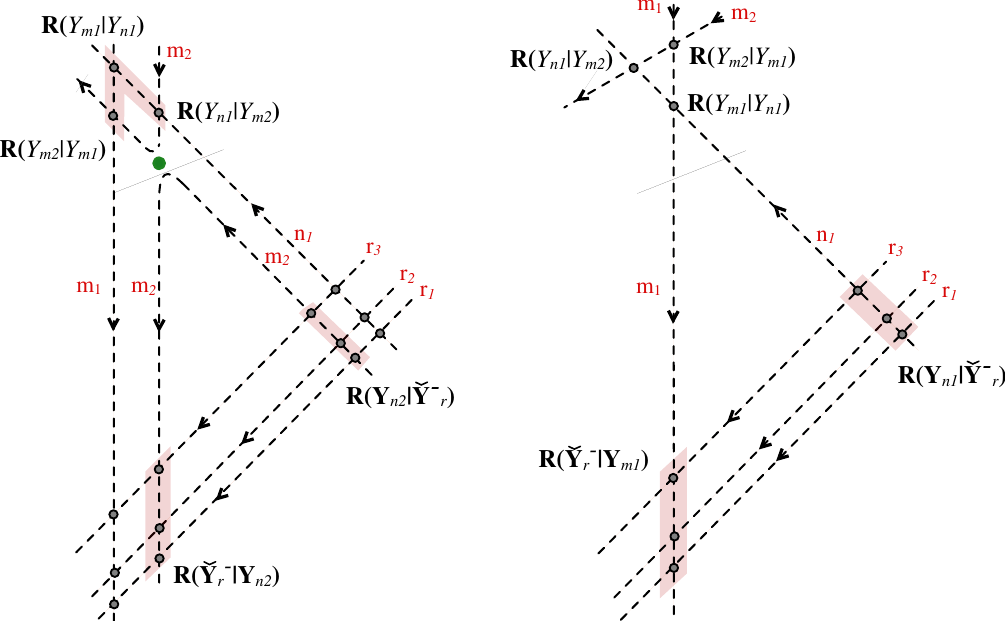}
\caption{Starting from the partition function in the l.h.s. of \eqref{YBE_Mat_part} with $N=M=2$ and $R=3$ at the decoupling point $Y_{n_2}=Y_{m_2}$ the corresponding $R$-matrix boils down to the permutation operator $\mathbb{P}_{n_2,m_2}$ (green blob). Next, due to $R$-matrix unitarity $\mathbf{R}(Y_{n_2}|\mathbf{Y}_{r}^{-}) \mathbf{R}(\mathbf{Y}_{r}^{-}|Y_{n_2}) = \mathbbm{1}$ the excitation $Y_{n_2}$ decouple from the partition function.}
\label{partition_F_DECA}
\end{center}
\end{figure}
At this point it is left only to analyze the part of a tile which is fully factorised over the set of magnons. It consists of a coordinate-dependent part and of the energy term. 
The first one comes from the overlap \eqref{full_over} and its numerator multiplies with the matrix-part in the spaces of symmetric spinors:
\begin{align}
\begin{aligned}
\label{theMatpart}
\frac{ [\mathbf{x_{3}}]^{\mathbf r}[\mathbf{{x_{2}}}]^{\mathbf n} [\mathbf{x_{32}}]^{\mathbf m} \boldsymbol{\varepsilon}^{ r} \mathcal{M}(\mathbf{Y}_n|\mathbf{Y}_m,\mathbf{Y}_r)^{t_r}  \boldsymbol{\varepsilon}^{r}  [\overline{\mathbf{x_{32}}}]^{\mathbf n} [\mathbf{\overline{{x_{3}}}}]^{\mathbf m}  [\mathbf{\overline{x_{2}}}]^{\mathbf r} }{(x_{23}^2)^{N-M+R+i \boldsymbol{\mu}-i\boldsymbol{\nu}}}\,.
\end{aligned}
\end{align}
Last expression has been dressed wrt to the raw result in \eqref{full_over} with a multiplication from the left by the unitary spin matrices $[\mathbf{x_{3}}]^{\mathbf r}[\mathbf{{x_{2}}}]^{\mathbf n}$ and from the right by $[\mathbf{\overline{{x_{3}}}}]^{\mathbf m} [\mathbf{\overline{x_{2}}}]^{\mathbf r}$. This operation is a spatial rotation of the single tile which eventually does not affect the Fishnet upon gluing back together the tiles. Indeed, the resolution of identity \eqref{complete_delta} is invariant wrt spacetime isometries such as a simultaneous rotation of bra and ket functions. Notice that \eqref{theMatpart} is valid for $x_1= \infty$; for the sake of generality the dependence on $x_1$ shall be restored. This is done by an inversion $x^{\mu}\to x^{\mu}/x^2$ followed by a translation $x \to x+x_1$, which delivers the following nice expression\footnote{Here the overlap has been further multiplied by some factors of the type $(x^2)^{i\nu}$, rescaling in a invariant manner the bra and ket wave-functions entering the cuts.} modulo transposition and conjugation in the spinor indices of $\mathbf{Y}_r$ magnons
\begin{equation}
\boxed{e^{i \boldsymbol{\mu} \log \frac{x^2_{12}}{x^2_{23}}} e^{i  \boldsymbol{\rho} \log \frac{x^2_{13}}{x^2_{12}}} e^{i \boldsymbol{\nu} \log \frac{x^2_{23}}{x^2_{13}}}}
\times \boxed{[\mathbf{x_{23}\overline{x_{31}}}]^{\mathbf n} [\mathbf{x_{23}\overline{x_{31}}}]^{\mathbf r} \mathcal{M}(\mathbf{Y}_n|\mathbf{Y}_m,\mathbf{Y}_r) [\mathbf{x_{12}\overline{x_{23}}}]^{\mathbf m} [\mathbf{x_{12}\overline{x_{23}}}]^{\mathbf r}   }\,.
\end{equation}
The last contribution to a tile that we shall account for is the energy term, i.e. powers of $E(Y)$. This contribution depends crucially on how many magnons flow across the cuts and how many columns of square lattice, say $L$, are contained in the tile:
\begin{equation}
\boxed{ E(\mathbf Y_n)^L  E(\mathbf Y_m)^M E(\mathbf Y_r)^R}\,.
\end{equation}
Altogether the formula for the generic tile of type bra/ket$^2$ reads
\begin{align}
\begin{aligned}
\label{Tile_I}
\mathbf{T}_{123}(\mathbf{Y}_n|\mathbf{Y}_m,\mathbf{Y}_r)& = 
e^{i \boldsymbol{\mu} \log \frac{x^2_{12}}{x^2_{23}}} e^{i  \boldsymbol{\rho} \log \frac{x^2_{13}}{x^2_{12}}} e^{i \boldsymbol{\nu} \log \frac{x^2_{23}}{x^2_{13}}}
 \times  \\ &\times  \mathcal{D}(\mathbf{Y}_n|\mathbf{Y}_m,\mathbf{Y}_r) E(\mathbf Y_n)^L  E(\mathbf Y_m)^M E(\mathbf Y_r)^R \times \\ & \times \left(\boldsymbol{\varepsilon}^{r} [\mathbf{x_{23}\overline{x_{31}}}]^{\mathbf n} [\mathbf{x_{23}\overline{x_{31}}}]^{\mathbf r} \mathcal{M}(\mathbf{Y}_n|\mathbf{Y}_m,\mathbf{Y}_r) [\mathbf{x_{12}\overline{x_{23}}}]^{\mathbf m} [\mathbf{x_{12}\overline{x_{23}}}]^{\mathbf r} \boldsymbol{\varepsilon}^{ r}\right)^{t_r}\,,
\end{aligned}
\end{align}
where the sign of transposition and the charge-conjugation matrices $\boldsymbol{\varepsilon}$ have been restored.

Similarly to \eqref{Tile_I} we shall write the contribution of the generic tile bra$^2$/ket $\mathbf{T}_{123}(\mathbf{Y}_n,\mathbf{Y}_r|\mathbf{Y}_m)$, which is essentially contained into \eqref{full_over_bk}. The dynamical factor is
  \begin{equation}
   \label{Dyn_part_II}
\boxed{\mathcal{D}(\mathbf{Y}_n,\mathbf{Y}_r|\mathbf{Y}_m)=\frac{\tilde H(\mathbf Y_m|\mathbf{Y}_n) \tilde H(\mathbf{Y}_r|\mathbf Y_m)}{\tilde H(\mathbf Y_r|\mathbf Y_n)}}\,,
  \end{equation}
  the matrix part is
\begin{align}
\begin{aligned}
\label{Mat_part_II}
\boxed{\mathcal{M}(\mathbf{Y}_n,\mathbf{Y}_r|\mathbf{Y}_m)= \frac{\mathbf c(\mathbf{Y}_m|\mathbf Y_r)}{\mathbf{c}(\mathbf{Y}_n|\mathbf Y_r)} \times\mathbf{R}(\mathbf{Y}_n|\mathbf Y_r^{+})  \mathbf{R}(\mathbf{Y}_n|\mathbf Y_m)  \mathbf{R}(\mathbf{Y}_r^{+}|\mathbf Y_m)} \,.
\end{aligned}
\end{align}
The coordinate dependent part after restoring of generic point $x_1$ takes the form
\begin{equation}
\boxed{e^{i \boldsymbol{\mu} \log \frac{x^2_{12}}{x^2_{23}}} e^{i  \boldsymbol{\rho} \log \frac{x^2_{13}}{x^2_{12}}} e^{i \boldsymbol{\nu} \log \frac{x^2_{23}}{x^2_{13}}}}\times \boxed{[\mathbf{\overline{x_{23}}x_{31}}]^{\mathbf n} [\mathbf{\overline{x_{23}}x_{31}}]^{\mathbf r}  \mathcal{M}(\mathbf{Y}_n,\mathbf{Y}_r|\mathbf{Y}_m)[\mathbf{x_{12}\overline{x_{23}}}]^{\mathbf m}[\mathbf{x_{12}\overline{x_{23}}}]^{\mathbf r}}\,.
\end{equation}
The contribution of the energy term in this case includes a factor related to the number of amputations performed to the Fishnet lattice, featuring an exponent $\delta=1$ in presence of reductions, and zero otherwise:
\begin{equation}
\boxed{E(\mathbf Y_m)^{M+L} (E(\mathbf Y_r) E(\mathbf{Y}_n)/E(\mathbf{Y}_m))^{\delta}}  \,.
\end{equation}
The complete formula for the generic tile of type bra$^2$/ket reads
\begin{align}
\begin{aligned}
\label{Tile_I|}
\mathbf{T}_{123}(\mathbf{Y}_n,\mathbf{Y}_r|\mathbf{Y}_m)& ={e^{i \boldsymbol{\mu} \log \frac{x^2_{12}}{x^2_{23}}} e^{i  \boldsymbol{\rho} \log \frac{x^2_{13}}{x^2_{12}}} e^{i \boldsymbol{\nu} \log \frac{x^2_{23}}{x^2_{13}}}}
\times  \\ &\times  \mathcal{D}(\mathbf{Y}_n,\mathbf{Y}_r|\mathbf{Y}_m)  E(\mathbf Y_m)^{M+L} (E(\mathbf Y_r) E(\mathbf{Y}_n)/E(\mathbf{Y}_m))^{\delta} \times \\ & \times \left(\boldsymbol{\varepsilon}^{r} [\mathbf{\overline{x_{23}}x_{31}}]^{\mathbf n} [\mathbf{\overline{x_{23}}x_{31}}]^{\mathbf r}  \mathcal{M}(\mathbf{Y}_n,\mathbf{Y}_r|\mathbf{Y}_m)[\mathbf{x_{12}\overline{x_{23}}}]^{\mathbf m}[\mathbf{x_{12}\overline{x_{23}}}]^{\mathbf r} \boldsymbol{\varepsilon}^{ r} \right)^{t_r}\,.
\end{aligned}
\end{align}
In fact, the derivation of overlaps can be repeated exploiting the symmetries of star-triangle identities, so to cast the result in different yet equivalent forms:
\begin{align}
\begin{aligned}
\label{moving_spins}
 &\left( [\mathbf{x_{23}\overline{x_{31}}}]^{\mathbf n} [\boldsymbol{\varepsilon}\mathbf{x_{23}\overline{x_{31}}}]^{\mathbf r} \mathcal{M}(\mathbf{Y}_n,\mathbf{Y}_r|\mathbf{Y}_m) [\mathbf{x_{12}\overline{x_{23}}}]^{\mathbf m} [\mathbf{x_{12}\overline{x_{23}}}\boldsymbol{\varepsilon}]^{\mathbf r} \right)^{t_r} = \\ = &\left( [\boldsymbol{\varepsilon}\mathbf{x_{12}\overline{x_{23}}}]^{\mathbf n} [\mathbf{x_{12}\overline{x_{23}}}]^{\mathbf m} \mathcal{M}(\mathbf{Y}_m|\mathbf{Y}_r,\mathbf{Y}_n) [\mathbf{x_{31}\overline{x_{12}}}\boldsymbol{\varepsilon}]^{\mathbf n} [\mathbf{x_{31}\overline{x_{12}}}]^{\mathbf r}\right)^{t_n} =
  \\ = & \left( [\boldsymbol{\varepsilon} \mathbf{x_{31}\overline{x_{12}}}]^{\mathbf m} [\mathbf{x_{31}\overline{x_{12}}}]^{\mathbf r} \mathcal{M}(\mathbf{Y}_r^+,\mathbf{Y}_m^-|\mathbf{Y}_n^-) [\mathbf{x_{13}\overline{x_{31}}}]^{\mathbf n} [\mathbf{x_{13}\overline{x_{31}}}\boldsymbol{\varepsilon}]^{\mathbf m} \right)^{t_m} \,.
\end{aligned}
\end{align}
The property \eqref{moving_spins} will prove useful when assembling back the tiles of a Fishnet, since it relates three forms of the same object where the coordinate dependence over a certain set of magnons - e.g. $\mathbf{Y}_r$ -- multiplies the matrix part from both sides, only from the left and only from the right. In this regard another useful property of the matrix part follows from crossing \eqref{transp_id}, and reads
\begin{equation}
\boldsymbol{\varepsilon}^{m}  \mathcal{M}(\mathbf{Y}_n,\mathbf{Y}_r|\mathbf{Y}_m)^{t_m} \boldsymbol{\varepsilon}^{m}=\mathcal{M}(\mathbf{Y}_m|\mathbf{Y}_r,\mathbf{Y}_n)\,.
\end{equation}
The proof of \eqref{moving_spins} is given in appendix \ref{app:str_insights}.

\section{Disk integrals: SoV representation}
\label{sec:SoV}
The procedure of cutting the Fishnet integrals with disk topology into a collection of triangular tiles allows to write an elegant representation of this integral, in principle for any size and any loop order, over the basis of mirror excitations of the Fishnet. In the section \ref{sec:eigenf} we reviewed how to perform a spectral transformation that diagonalises any rectangular square-lattice Fishnet diagram with fixed boundary conditions on the up/down edges. Then, in sections \ref{sec:cut}-\ref{sec:overs} we computed what happens when the wave-functions of the spectral representation -- a.k.a. the \emph{mirror excitations} of the Fishnet -- defined w.r.t. different boundary conditions are overlapped. In order to achieve the SoV representation of Fishnet integrals it is left to ``glue" together the tiles.
\subsection*{Gluing two tiles}
\begin{figure}[t]
\begin{center}
\includegraphics[scale=0.7]{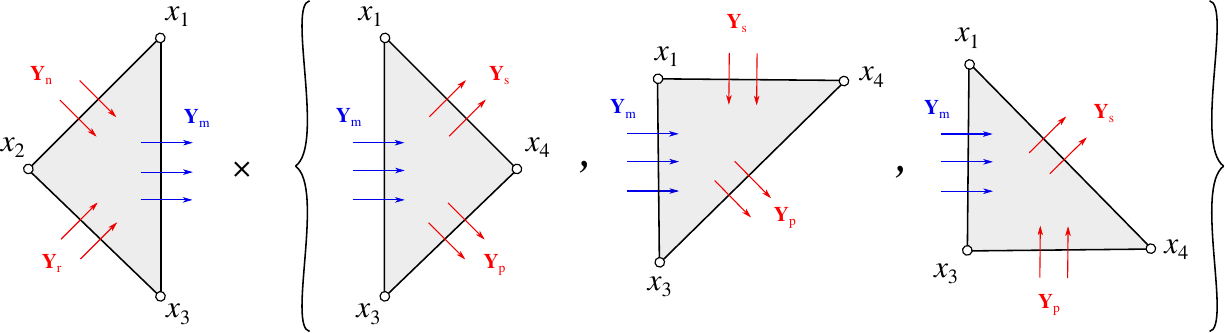}
\caption{A tile $\mathbf{T}_{123}(\mathbf{Y}_n,\mathbf{Y}_r| \color{blue}\mathbf{Y}_{m}\color{black})$ can be glued along the cut $x_{2}x_{3}$ with another tile in three ways. The excitations $\color{blue}\mathbf{Y}_{m}$ \color{black} can enter either a tile with one bra $\mathbf{T}_{134}( \color{blue}\mathbf{Y}_{m}\color{black}|\mathbf{Y}_s , \mathbf{Y}_p)$ or a tile with two bras, thus in two possible ways $\mathbf{T}_{413}(\mathbf{Y}_s, \color{blue}\mathbf{Y}_{m}\color{black}|\mathbf{Y}_p)$ and $\mathbf{T}_{134}( \color{blue}\mathbf{Y}_{m}\color{black},\mathbf{Y}_p|\mathbf{Y}_s)$.}
\label{gluing_tiles}
\end{center}
\end{figure}

We can now turn to what happens in the product of two tiles that have one cut in common, referring to the figure \ref{gluing_tiles}. In this notation the left tile in figure \ref{gluing_tiles} is of the type $\mathbf{T}_{123}(\mathbf{Y}_n,\mathbf{Y}_r|\mathbf{Y}_m)$ and it is glued along the cut $x_{1}x_{3}$ with a right tile which is chosen among one of three possibilities $\mathbf{T}_{134}(\mathbf{Y}_{m}|\mathbf{Y}_s , \mathbf{Y}_p)\,,\,\mathbf{T}_{413}(\mathbf{Y}_s, \mathbf{Y}_{m}|\mathbf{Y}_p)$ and $\mathbf{T}_{134}(\mathbf{Y}_{m},\mathbf{Y}_p|\mathbf{Y}_s)$.
The operation of gluing correspond to the summation over the spectrum of quantum numbers $\mathbf{Y}_m$ of the magnons inserted along the cuts according to the completeness relation \eqref{complete_delta},
\begin{align}
\begin{aligned}
\label{compl_glue}
&\sum_{\mathbf{m}}\int \! \! d\boldsymbol{\mu} \,\rho(\mathbf{Y}_m) \text{Tr}_{\mathbf m}\left[\mathbf{T}_{123}(\mathbf{Y}_n,\mathbf{Y}_r|\mathbf{Y}_m) \mathbf{T}_{\dots13\dots}(\dots \mathbf{Y}_{m} \dots)\right] \,,
\end{aligned}
\end{align}
where the dots on the right tile leave the freedom for the three choices in figure \ref{gluing_tiles} and the trace $\text{Tr}_{\mathbf m}=\text{Tr}_{ m_1}\cdots \text{Tr}_{ m_M}$ runs over the symmetric spinor indices of the excitations $Y_{m_1},\dots, Y_{m_N}$. In the following we consider the gluing with $\mathbf{T}_{134}(\mathbf{Y}_{m}|\mathbf{Y}_s , \mathbf{Y}_p)$ and obtain the expression for a quadrilateral patch in compact form. The other two possible choices for the right tile lead to a similar result and will be analysed successively.
First, the dynamical parts of the product of the tiles reads
\begin{equation}
\frac{\tilde H(\mathbf Y_m|\mathbf{Y}_n) \tilde H(\mathbf{Y}_r|\mathbf Y_m)}{\tilde H(\mathbf Y_r|\mathbf Y_n)} \frac{\tilde H(\mathbf Y_s|\mathbf{Y}_m) \tilde H(\mathbf{Y}_m|\mathbf Y_p)}{\tilde H(\mathbf Y_s|\mathbf Y_p)}\,.
\end{equation}
The coordinate-dependent part is given by a scalar factor,
\begin{align}
\begin{aligned}
\label{scalar_fact_quad}
&e^{i \boldsymbol{\rho} \log \frac{x_{13}^2}{x_{12}^2}} e^{ i \boldsymbol{\nu} \log \frac{x_{23}^2}{x_{13}^2}}  \times  e^{ i \boldsymbol{\mu}  \log \frac{x_{12}^2 x_{34}^2}{x_{14}^2 x_{23}^2}}  \times e^{ i \boldsymbol{\sigma} \log \frac{x_{13}^2}{x_{34}^2}} e^{ i \boldsymbol{\pi} \log \frac{x_{14}^2}{x_{13}^2}}\,,
\end{aligned}
\end{align}
and by the spinning $SU(2)$ matrix term, namely
\begin{align}
\begin{aligned}
\label{glue_1}
 &\left([\boldsymbol{\varepsilon}\mathbf{x_{23}\overline{x_{31}}}]^{\mathbf r} [\mathbf{x_{23}\overline{x_{31}}}]^{\mathbf n}[\boldsymbol{\varepsilon} \mathbf{x_{34}\overline{x_{41}}}]^{\mathbf p}  \right. \times\\ &\qquad \qquad\times
\text{Tr}_{\mathbf{m}}\left[ \mathcal{M}(\mathbf{Y}_m|\mathbf{Y}_s,\mathbf{Y}_p)   \mathcal{M}(\mathbf{Y}_n,\mathbf{Y}_r|\mathbf{Y}_m)  [\mathbf{x_{12}\overline{x_{23}}x_{34} \overline{x_{41}}}]^{\mathbf m}  \right] \times \\&\qquad  \qquad \qquad \qquad \qquad\qquad\qquad\qquad\qquad \left. \times  [\mathbf{\overline{x_{12}} x_{23}}\boldsymbol{\varepsilon}]^{\mathbf r}  [\mathbf{\overline{x_{13}} x_{34}}]^{\mathbf s} [\mathbf{\overline{x_{13}} x_{34}}\boldsymbol{\varepsilon}]^{\mathbf p} \right)^{t_r\, t_p}\,.
\end{aligned}
\end{align}
In last formula we distinguish a first row with the spinning part of the ``incoming" magnons $\mathbf{Y}_n,\mathbf{Y}_r$, the second line with the interactions and the exchanged magnon $\mathbf{Y}_m$ and a last line with the spinning part of the ``outgoing" magnons $\mathbf{Y}_s,\mathbf{Y}_p$. The integrand of \eqref{compl_glue} is completed by the energy factors 
\begin{equation}
E(\mathbf{Y}_m)^{L_1+L_2+M} \left(\frac{E(\mathbf Y_n) E(\mathbf Y_r)}{E(\mathbf Y_m)}\right)^{\delta}  E(\mathbf Y_s)^S E(\mathbf Y_p)^P \,,
\end{equation}
here $L_1,L_2$ are the number of ``columns" of Fishnet square lattice -- i.e. the number of propagators $\phi_2$ -- contained in the two tiles.

Let's turn to the case $\mathbf{T}_{413}(\mathbf{Y}_s, \mathbf{Y}_{m}|\mathbf{Y}_p)$;  the product of the dynamical parts read
\begin{equation}
\frac{\tilde H(\mathbf Y_m|\mathbf{Y}_n) \tilde H(\mathbf{Y}_r|\mathbf Y_m)}{\tilde H(\mathbf Y_r|\mathbf Y_n)} \frac{\tilde H(\mathbf Y_p|\mathbf{Y}_s) \tilde H(\mathbf{Y}_m|\mathbf Y_p)}{\tilde H(\mathbf Y_m|\mathbf Y_s)} \,.
\end{equation}
The scalar part of the coordinate dependence is equal to \eqref{scalar_fact_quad}, whereas the spinning term differs. Indeed, the spinning term that comes with the matrix part of $\mathbf{T}_{413}$, as it appears in \eqref{Tile_I|}, contains a transposition of the spin indices of excitations $\mathbf{Y}_m$, which is not optimal for gluing. It is convenient to use the property \eqref{moving_spins} on the second tile:
\begin{align}
\begin{aligned}
 &\left( [\mathbf{x_{13}\overline{x_{34}}}]^{\mathbf s} [\boldsymbol{\varepsilon}\mathbf{x_{13}\overline{x_{34}}}]^{\mathbf m} \mathcal{M}(\mathbf{Y}_s,\mathbf{Y}_m|\mathbf{Y}_p) [\mathbf{x_{41}\overline{x_{13}}}]^{\mathbf p} [\mathbf{x_{41}\overline{x_{13}}}\boldsymbol{\varepsilon}]^{\mathbf m} \right)^{t_m} =
  \\ = & \left( [\boldsymbol{\varepsilon} \mathbf{x_{34}\overline{x_{41}}}]^{\mathbf p} [\mathbf{x_{34}\overline{x_{41}}}]^{\mathbf m} \mathcal{M}(\mathbf{Y}_m^+,\mathbf{Y}_p^-|\mathbf{Y}_s^-) [\mathbf{x_{41}\overline{x_{13}}}]^{\mathbf s} [\mathbf{x_{41}\overline{x_{13}}}\boldsymbol{\varepsilon}]^{\mathbf p} \right)^{t_p} \,,
\end{aligned}
\end{align}
in order to group the spinning term of $\mathbf{Y}_m$ into one single $SU(2)$ matrix
\begin{align}
\begin{aligned}
\label{glue_2}
 &\left([\boldsymbol{\varepsilon}\mathbf{x_{23}\overline{x_{31}}}]^{\mathbf r} [\mathbf{x_{23}\overline{x_{31}}}]^{\mathbf n}[\boldsymbol{\varepsilon} \mathbf{x_{34}\overline{x_{41}}}]^{\mathbf p}  \right. \times\\ &\qquad \qquad\times
\text{Tr}_{\mathbf{m}}\left[ \mathcal{M}(\mathbf{Y}_n,\mathbf{Y}_r|\mathbf{Y}_m)  [\mathbf{x_{12}\overline{x_{23}}}\mathbf{x_{34}\overline{x_{41}}}]^{\mathbf m} \mathcal{M}(\mathbf{Y}_m|\mathbf{Y}_s^{--},\mathbf{Y}_p^{-})  \right] \times \\&\qquad  \qquad \qquad \qquad \qquad\qquad\qquad\qquad\qquad \left. \times  [\mathbf{\overline{x_{12}} x_{23}}\boldsymbol{\varepsilon}]^{\mathbf r}  [\mathbf{x_{41}\overline{x_{13}}}]^{\mathbf s} [\mathbf{x_{41}\overline{x_{13}}}\boldsymbol{\varepsilon}]^{\mathbf p}  \right)^{t_r\,,t_p}\,.
\end{aligned}
\end{align}
Finally, the energy factors depend on the presence of reductions $\delta_1,\delta_2 =0,1$ and on the number $L_1,L_2$ of columns of Fishnet lattice in the two tiles 
\begin{equation}
E(\mathbf{Y}_m)^{L_1+M} E(\mathbf{Y}_p)^{L_2+P} \left(\frac{E(\mathbf Y_n) E(\mathbf Y_r)}{E(\mathbf Y_m)}\right)^{\delta_1}  \left(\frac{E(\mathbf Y_m) E(\mathbf Y_s)}{E(\mathbf Y_p)}\right)^{\delta_2} \,. 
\end{equation}

The last possibility in figure \ref{gluing_tiles} is the gluing with the right tile $\mathbf{T}_{134}(\mathbf{Y}_m, \mathbf{Y}_{p}|\mathbf{Y}_s)$; the product of the dynamical parts read
\begin{equation}
\frac{\tilde H(\mathbf Y_m|\mathbf{Y}_n) \tilde H(\mathbf{Y}_r|\mathbf Y_m)}{\tilde H(\mathbf Y_r|\mathbf Y_n)} \frac{\tilde H(\mathbf Y_p|\mathbf{Y}_s) \tilde H(\mathbf{Y}_m|\mathbf Y_p)}{\tilde H(\mathbf Y_m|\mathbf Y_s)} \,.
\end{equation}
 The coordinate-dependent parts is given by the scalar factor \eqref{scalar_fact_quad} and by the spinning term, together with the matrix part of the tiles, read
\begin{align}
\begin{aligned}
\label{glue_3}
 &\left([\boldsymbol{\varepsilon}\mathbf{x_{23}\overline{x_{31}}}]^{\mathbf r} [\mathbf{x_{23}\overline{x_{31}}}]^{\mathbf n}[\boldsymbol{\varepsilon} \mathbf{x_{34}\overline{x_{41}}}]^{\mathbf p}  \right. \times\\ &  \qquad \qquad \times
 \mathcal{M}(\mathbf{Y}_n,\mathbf{Y}_r|\mathbf{Y}_m)  [\mathbf{x_{12}\overline{x_{23}}x_{34} \overline{x_{41}}}]^{\mathbf m}  \text{Tr}_{\mathbf{m}}\left[ \mathcal{M}(\mathbf{Y}_m,\mathbf{Y}_p|\mathbf{Y}_s)   \right] \times \\& \qquad \qquad\qquad \qquad\qquad\qquad\qquad\qquad\qquad \left. \times  [\mathbf{\overline{x_{12}} x_{23}}\boldsymbol{\varepsilon}]^{\mathbf r}  [\mathbf{\overline{x_{13}} x_{34}}]^{\mathbf s} [\mathbf{\overline{x_{13}} x_{34}}\boldsymbol{\varepsilon}]^{\mathbf p} \right)^{t_r\, t_p}\,.
\end{aligned}
\end{align}
Finally, the energy factors read:
\begin{equation}
E(\mathbf{Y}_m)^{L_1+M} E(\mathbf{Y}_s)^{L_2+S} \left(\frac{E(\mathbf Y_n) E(\mathbf Y_r)}{E(\mathbf Y_m)}\right)^{\delta_1}  \left(\frac{E(\mathbf Y_m) E(\mathbf Y_p)}{E(\mathbf Y_s)}\right)^{\delta_2} \,. 
\end{equation}
The analysis of this section can be repeated for the case that a tile bra$^2$/ket is glued to another tile bra$^2$/ket, instead of bra/ket$^2$. In this case, the difference in the dynamical and energy factors of the result follow straightforwardly from the formulae in section \ref{sec:tiles}. The take here is that by gluing two tiles along a cut $x_ix_j$ one forms a quadrilateral that is, clockwise, $x_ix_kx_jx_l$. The conformal invariants associated to the quadrilateral are
\begin{equation}
Z \bar {Z} = \frac{x_{il}^2 x_{jk}^2}{x_{ik}^2 x_{jl}^2}\,,\, (1-Z)(1-\bar {Z}) =  \frac{x_{ij}^2 x_{kl}^2}{x_{ik}^2 x_{jl}^2}\,.
\end{equation}
The quadrilateral contributes to the SoV representation of the Fishnet diagram, for magnons $\mathbf{Y}_m$ along the cut, with the product of the measure $\rho(\mathbf{Y}_m)$, by the \emph{momentum factor}
\begin{equation}
e^{i \boldsymbol{\mu} Z \bar Z}= \prod_{k=1}^M e^{i \boldsymbol{\mu_k} Z \bar Z}\,,
\end{equation}
and by an \emph{energy factor}
\begin{equation}
E(\mathbf{Y}_m)^{L+M} = \prod_{k=1}^M E(\mathbf{Y}_{m_k})^{L+M}\,,
\end{equation}
where $L$ is the number of columns of Fishnet lattice of extremal points $x_ix_j$ in the Feynman integral under consideration. Furthermore, these contributions multiply the matrix parts and the dynamical parts which come with the two tiles and contain the interaction between magnons inserted along different cuts. In particular it is useful to notice that the coordinate-dependent term that appears in \eqref{glue_1}, \eqref{glue_2}, \eqref{glue_3} inside the trace has the form
\begin{equation}
\text{Tr}_{\mathbf{n}}\,\left[ \mathcal{M}(\dots \mathbf{Y}_{m})  [\mathbf{x_{ik}\overline{x_{kj}}x_{jl}\overline{x_{li}}}]^{\mathbf m}  \mathcal{M}(\mathbf{Y}_{m}\dots)\right]\,,
\end{equation}
that is a product of $SU(2)$ matrices in the representations defined by the excitation numbers $\mathbf{Y}_m$,
\begin{equation}
 [\mathbf{x_{ik}\overline{x_{kj}}x_{jl}\overline{x_{li}}}]^{\mathbf{m}} =  [\mathbf{x_{ik}\overline{x_{kj}}x_{jl}\overline{x_{li}}}]^{{m}_1} \otimes \cdots \otimes [\mathbf{x_{ik}\overline{x_{kj}}x_{jl}\overline{x_{li}}}]^{{m}_M}\,.
 \end{equation}
Each of these matrices can be compactly written as
 \begin{equation}
 [\mathbf{x_{ik}\overline{x_{kj}}x_{jl}\overline{x_{li}}}]^{{m_h}} =e^{i \theta \cdot \hat{J}^{(m_h)}}\,,
 \end{equation}
where $\hat{J}^{(m)}$ is the vector of generators of $3d$ rotations acting on left (or right) $su(2)$ spinors of spin $\tfrac{m}{2}$, and defined by the angles $\theta=(\theta_x,\theta_y,\theta_z)$. Upon a conformal transformation that sends the four points to the complex plane $(0,1,\infty,Z)$, this rotation matrix becomes a function of the conformal invariants only \begin{equation}
 e^{\frac{i}{2} \log Z/\bar Z  \hat{J}^{(m)}_0}\,,
 \end{equation}
 with $\hat{J}^{(m)}_0$ the generator of rotations on the $(Z,\bar Z)$ plane in the symmetric representation of spin $m/2$.

\section{Outlooks}
\label{sec:outlook}
The formulae obtained in the present work achieve an integrability-based representation of multi-point Fishnet Feynman diagrams, in terms of the separated variables of the Fishnet lattice. The latter are the rapidities and the spin of the mirror excitations of Fishnet CFT, that is the quantum numbers of the eigenfunctions of a non-compact $SO(1,5)$ spin-chain with fixed boundary conditions \cite{Derkachov2020, Derkachov:2020zvv}. The derivation of this representation for a generic multi-point Fishnet integral -- inspired by the hexagonalisation procedure for correlation functions in $\mathcal{N}=4$ SYM \cite{Basso:2015zoa} -- consists first in defining a cutting of the Fishnet into triangular tiles, that are computed in general form in this paper, then of the SoV transformation of the Fishnet contained in each tile, and finally of assembling back the tiles. The SoV representation has been applied successfully in \cite{Derkachov2020} for the simpler case of four-point Fishnets at any loop order, providing a first-principle derivation of the BMN representation by Basso and Dixon \cite{Basso:2017jwq}. The results of the present work allow to go beyond four-point, and provide the setting for an efficient way of computing higher-loop multi-point Fishnet integrals with disk topology, promptly applied by F.~Aprile and the author to the five-point Ladder integrals in \cite{Aprile:2023gnh}. The latter class of integrals is practically unknown, with few exceptions at the lowest orders and in special kinematics, in spite of the fact that they play an important role in the perturbative computations of higher-point Feynman integrals in massless QFT. 
A simplified starting point in such a direction could be the planar light-cone polygon kinematics of \cite{Bercini:2020msp, Olivucci:2022aza, Kostov:2022vup}, where we expect a significant trivialization of matrix part of SoV representation.

Throughout the paper we referred mostly to disk Fishnets, while the SoV representation is valid for any finite multi-trace and multi-point Fishnet integral. The cutting-and-gluing procedure for a correlators here is inspired by the hexagonalisation technique developed for $\mathcal{N}=4$ SYM and later adapted to the Fishnet CFTs in \cite{Basso:2018cvy}.
The scope of this work is to provide a unique first-principles derivation of hexagon form-factors from an integrable spin-chain framework, ultimately relying on star-triangle identities. By showing how hexagonalisation in Fishnet theories is obtained by direct derivations, this work completes the picture of Fishnets as a special limit of $\mathcal{N}=4$ SYM theory where the integrability emerges from the structure of Feynman diagrams and it is proved and realized rigorously. 

There are a few foreseen directions for developing our results. First, to extend the computation of overlaps and Fishnet tiles to the more general setting of a spinning and inhomogeneous Fishnet lattice as the one studied in \cite{Olivucci:2022aza}. This would achieve, for instance, a SoV representation of certain sectors in the generalized Fishnet theories \cite{Caetano:2016ydc, Kazakov:2018gcy}. Secondly, the class of integrable Loom CFTs recently introduced are expected to enjoy a similar SoV representation as their integrability is based on the star-triangle identities. The latter theories feature Feynman diagrams with Yangian symmetry \cite{Kazakov:2023nyu} (see \cite{Loebbert:2022nfu} for a review) as it was discovered first in the bi-scalar Fishnet, thus providing a set of constraints which is tempting to combine with a SoV representation in order to solve them. The theories from the Loom are defined in any spacetime dimension; at least in the case of bi-scalar theory the mirror excitations of the Fishnet are already at our disposal \cite{Derkachov:2021ufp} and in the simplest case of $d=2$ have been used for the computation of four-point Fishnet integrals via separation of variables \cite{Derkachov:2018rot}. The computation of overlaps in this case may prove a very good primer of hexagonalization for a CFT$_2$ hypothetically appearing as a strong-deformation limit of the Higgs branch of AdS$_3$/CFT$_2$ \cite{OhlssonSax:2014jtq}.

A very compelling direction of study is to understand integrability from a spin-chain perspective in the general $\gamma$-deformed $\mathcal{N}=4$ SYM, both for infinite twists and then for finite ones. Related to this, we do not know yet a description of the Feynman diagrams of the ``dynamical" Fishnet CFTs or of the Eclectic action \cite{Ahn:2020zly} as conserved charges of a non-compact integrable spin-chain. The ultimate goal in this context is a full and rigorous understanding of the integrability of $\mathcal{N}=4$ SYM theory. Recently appeared results \cite{Ferrando:2023ogg} suggest us to try to formulate an SoV representation for new spin-chain models that describe the Feynman integrals in certain operator-based large-twist limits of the theory, repeating the ``Fishnet approach" to any operators in $\mathcal{N}=4$ SYM theory.
\section*{Acknowledgements}
I express my gratitude to B.~Basso and S.~Derkachov for inspiring discussions on related topics. I thank F.~Aprile and V.~Kazakov for useful comments about the draft. Research at the Perimeter Institute is supported in part by the Government of Canada through NSERC and by the Province of Ontario through MRI.  I would like to thank Laboratoire de Physique - ENS of Paris for the kind hospitality in February 2023, where I held fruitful discussions about this work. Part of this work was carried out during the authors' stay at the NCCR SwissMAP workshop `Integrability in Condensed Matter Physics and QFT' (3rd to 12th of February 2023) which took place at the SwissMAP Research Station. The authors would like to thank the Swiss National Science Foundation, which funds SwissMAP (grant number 205607) and, in addition, supported the event via the grant IZSEZ0_215085.
This work was additionally supported by a grant from the Simons Foundation (Simons Collaboration on the Nonperturbative Bootstrap \#488661) and ICTP-SAIFR FAPESP grant 2016/01343-7 and FAPESP grant 2019/24277-8.
\appendix
\section{The fused $SU(2)$ $\mathbf{R}$-matrix}
\label{app:Rmat}
Let us recall that the matrix $R(u)$ acting on the tensor product of $\mathbb{C}^2\otimes \mathbb{C}^2$ and defined by
\begin{equation}
R(u)_{ab}^{cd} = \frac{1}{u+1} \left(u\, \delta_{a}^{{c}}\delta_{b}^{d} + \delta_{a}^{d}\delta_{b}^{c}\right)\,,\,\,\,\, a,b,c,d\in \{1,2\}\,,
\end{equation}
is a $SU(2)$-invariant solution of the Yang-Baxter equation (repeated indices are contracted)
\begin{equation}
R(u-v)_{ab}^{kl} R(u)_{kc}^{hs}R(v)_{ls}^{mr}=R(v)_{bc}^{ls} R(u)_{as}^{kr}R(u-v)_{kl}^{hm}\,,
\end{equation}
which in addition satisfies unitarity and crossing-symmetry properties
\begin{equation}
R(u)_{ab}^{rs}R(-u)_{rs}^{cd} = \delta_{a}^{c}\delta_{b}^{d}\,,\,\,\,\, R(u)_{ab}^{cd} =  \frac{u}{u+1} \varepsilon_{b}^{r} R(-u-1)_{as}^{cr} \varepsilon_{s}^{d}\,,
\end{equation}
where the charge-conjugation matrix is $\varepsilon_{a}^b =  \begin{pmatrix}
0 && -i \\ i &&0 \end{pmatrix}$ satisfy the properties
\begin{equation}
\varepsilon^2 = \mathbbm{1}\,,\,\,\,\varepsilon \boldsymbol{\sigma}^{k} \varepsilon = (\overline{\boldsymbol{\sigma}}^{k})^t\,,\,\,\,\, k=1,2,3\,.
\end{equation}
We shall define $\mathbf{R}_{nm}(u)$ the $R$-matrix acting either on the tensor product $(\frac{n}{2},0) \otimes (\frac{m}{2},0)$ or $(0,\frac{n}{2}) \otimes (0,\frac{m}{2})$ of the spaces of two irreducible symmetric representations of $SU(2)$ of dimension $n+1$ and $m+1$ respectively\footnote{With an abuse of notation we do not distinguish the notation for the $R$-matrix acting on dotted/undotted
indices, since the two matrices are in identical and the type of indices shall be clear from the context.}
\begin{equation}
\mathbf{R}_{mn}(u)_{(a_1 \dots a_n) (b_1,\dots,b_m)}^{({c}_1 \dots {c}_n) ({d}_1,\dots, {d}_m)}\,,
\end{equation}
where the bracket notation for indices stands for symmetryzation $(\dots a_i\dots a_j \dots)=(\dots a_j\dots a_i\dots)$. Such a matrix is defined by fusion procedure and satisfies unitarity and crossing-symmetry, here reported in matrix form (i.e. without indices)
\begin{equation}
\label{unitarity_crossing}
\mathbf{R}(u)_{mn} \mathbf{R}(-u)_{mn} = \mathbbm{1}_m \otimes \mathbbm{1}_n \,,\,\,\,\, \mathbf{R}_{mn}(u) = c_{mn}(u) \varepsilon^{\otimes n} \mathbf{R}_{mn}^{t_{n}}(-u-1) \varepsilon^{\otimes n}\,,
\end{equation}
where $t_n$ stands for the transposition of indices in the space $\mathbb{V}_n$ and the crossing-symmetry factor reads
\begin{equation}
c_{mn}(u) = 
 \frac{\Gamma\left(u+\frac{m-n}{2}+1 \right)\Gamma\left(u-\frac{m-n}{2}+1\right)}{\Gamma\left(u+\frac{m+n}{2}+1\right)\Gamma\left(u-\frac{m+n}{2}+1\right)}\,,
\end{equation}
and satisfies $c_{mn}(u)c_{mn}(-u-1)=1$.
A consequence of the relations \eqref{unitarity_crossing} and of the identity $\varepsilon_{a}^{b}\varepsilon_{b}^{c}=  \delta_{a}^c$ is the following property
\begin{equation}
\label{unitarity_crossing_II}
\mathbf{R}(-u-2)^{t_n}_{mn} \mathbf{R}^{t_n}(u)_{mn} = r_{mn}(u) \mathbbm{1}_m \otimes  \mathbbm{1}_n \,,\,\,\, r_{mn}(u) =  \frac{\left(u+\frac{m+n}{2}+1 \right)\left(u-\frac{m+n}{2}+1\right)}{\left(u+\frac{m-n}{2}+1\right)\left(u-\frac{m-n}{2}+1\right)}\,,
\end{equation}
and it holds that $r_{mn}(u) = c_{mn}(u) c_{mn}(-u-2)$.
Another property of the $\mathbf{R}$-matrix which play an important role in the SoV representation of Fishnet integrals is the decoupling at $u=0$, that is
\begin{equation}
\mathbf{R}_{n,n}(0) = \mathbb{P}_{n,n}\,,
\end{equation}
where $\mathbb{P}_{n,n}$ is the permutation operator acting on the two factors of the tensor product of $n$-fold symmetric spinors $\mathbb{V}_n \otimes \mathbb{V}_n$.
This property is simple to check for $n=1$, that is with explicit indices: $R(0)_{ab}^{cd} = \delta_{a}^{d}\delta_{b}^{c} =(\mathbb{P}_{1,1})_{ab}^{cd}$. When dealing with higher-spins it is rather convenient to check the action of $\mathbf{R}_{m,n}(0)$ in the Bargmann-Fock basis of symmetric spinors $|\alpha\rangle \in \mathbb{V}_n$
\begin{equation}
|\alpha\rangle = \underbrace{\begin{pmatrix} \alpha_1 \\ \alpha_2\end{pmatrix} \otimes \cdots \otimes  \begin{pmatrix} \alpha_1 \\ \alpha_2\end{pmatrix}}_{n-\text{fold}}\,,\,\,\,\,  \alpha_j  \in \mathbb{C}\,,
\end{equation}
based on the following completeness relation over symmetric spinors
\begin{equation}
\frac{1}{n! \pi^2}\int \!d^2\alpha_1\!\int \! d^2 \alpha_2 \,\,e^{-(\alpha_1^2 +\alpha_2^2)} \,|\alpha \rangle_{a_1 \dots a_n} \langle \alpha|_{b_1 \dots b_n}  = \delta_{a_1 b_1}\cdots  \delta_{a_n b_n} + \text{permutations}\,.
\end{equation}
In this basis one has \cite{Derkachov2020, Derkachov:2020zvv}
\begin{align}
\begin{aligned}
\langle \alpha_1|\otimes \langle \alpha_2 |\mathbf{R}_{n,n}(u)&|\beta_1\rangle\otimes |\beta_2 \rangle =\\&= \rho_{nn}(u)\langle \alpha_1| \beta_2 \rangle^{-u+n}\partial_s^n \partial_t^n\left( \langle \alpha_1| \beta_2\rangle+s \langle \alpha_2  |\beta_2\rangle +t \langle \alpha_1| \beta_1\rangle+s t \langle \alpha_1 | \beta_2\rangle\right)^{u+n}\,.
 \end{aligned}
 \end{align}
 It is straightforward to check that for $u=0$ the previous formula reduces to 
 \begin{equation}
\langle \alpha_1|\otimes \langle \alpha_2 |\mathbf{R}_{n,n}(0)|\beta_1\rangle \otimes|\beta_2 \rangle =  \langle \alpha_1  |\beta_2\rangle \langle \alpha_2| \beta_1\rangle = \langle \alpha_1  |\otimes \langle \alpha_2| \mathbb{P}_{n,n}|  \beta_1\rangle \otimes |\beta_2\rangle \,.
 \end{equation}
The decoupling property plays a role in expressions -- ubiquitous in the SoV representation of Fishnets -- of the type
\begin{equation}
\text{Tr}_{m}\left( [\mathbf{x}]^n \mathbf{R}_{nm}(i\nu-i\mu) [\overline{\mathbf{y}}]^m \right)= \text{Tr}_{m}\left( [\mathbf{x}]^n \mathbf{R}(Y_n|Y_m) [\overline{\mathbf{y}}]^m \right)\,,
\end{equation}
implying that when two mirror excitations have coinciding rapidities $\nu=\mu$ and spins $n=m$ the following simplification happens
\begin{equation}
\text{Tr}_{m}\left( [\mathbf{x}]^n \mathbf{R}_{nm}(0) [\overline{\mathbf{y}}]^m \right)= \text{Tr}_{m}\left( [\mathbf{x}]^n \mathbb{P}_{n,n} [\overline{\mathbf{y}}]^m \right)= [\mathbf{x} \overline{\mathbf{y}}]^n\,.
\end{equation}
\section{Star-triangle insights}
\label{app:str_insights}
The star-triangle relation presented in section \ref{sec:eigenf} takes a nice symmetric form with respect to the three spinorial structures involved and depicted by dashed lines of different colours. Nevertheless this is not the only possible way to spell the identity and another instance was that given in figure 8 of \cite{Derkachov:2021rrf} that is the same as figure 35 of \cite{Olivucci:2021cfy}. These formulae differ from the one in this paper because the three dashed lines are not oriented all clockwise (or all anticlockwise); the relation between the two forms of star-triangle is derived in the following figure \ref{from_str_to_str}. The direction of the flow of the line with spin $l$ can be inverted by an $SU(2)$ conjugation
\begin{equation}
[\mathbf{x_{20}\overline{x_{01}}}]^{\ell} \,\mapsto\, [\mathbf{\overline{x_{12}} x_{20}\overline{x_{01}}x_{12}}]^{\ell}=[\mathbf{\overline{x_{10}} x_{02}\overline{x_{12}}x_{12}}]^{\ell} =[\mathbf{\overline{x_{10}} x_{02}}]^{\ell} \,,
\end{equation} 
where we used $[\mathbf{x_{ij}\overline{x_{jk}}x_{ki}}]=[\mathbf{x_{ki}\overline{x_{kj}}x_{ji}}]$, following from the Clifford algerbra 
and the unitarity property of $SU(2)$ matrices $[\mathbf{x}][\mathbf{ x}]^{\dagger}=[\mathbf{x}][\mathbf{\overline x}] =\mathbbm{1}$.
\begin{figure}[t]
\begin{center}
\includegraphics[scale=1.2]{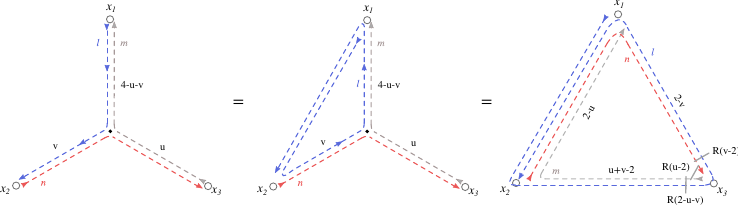}
\caption{The star-integral of section \ref{sec:eigenf} with the dashed (spinning) lines oriented clockwise is related to the star-integral of figure 8 in \cite{Derkachov:2021rrf} by $SU(2)$ conjugation $[\mathbf{x_{12}}]^{l}$. Using this second form of star-triangle transformation one obtains the corresponding triangle in the rhs.}
\label{from_str_to_str}
\end{center}
\end{figure}
What is nice about the rhs of figure \ref{from_str_to_str} is that the $R$-matrices in the product are all close to each-other in the product
\begin{equation}
\label{matrix_part_rhs_I}
\left(\mathbf{R}_{n,l}(v-2)^{t_l} \mathbf{R}_{m,n}(u-2) \mathbf{R}_{m,l}(2-u-v)^{t_l}\right)^{t_l}\,.
\end{equation}
On the other hand, starting from the lhs in figure \ref{from_str_to_str} one could perform also the conjugation by $[\mathbf{x_{13}}]^{m}$, or by $[\mathbf{x_{32}}]^{n}$, which allow respectively to change the direction of the dashed lines in grey (spin $m$) and in red (spin $n$). The rhs of the star-triangle identity looks different in the three cases, which allows to state an identity between three outcomes of the same star-triangle integral, as stated in figure \ref{the_three_rhs}.
\begin{figure}[b]
\begin{center}
\includegraphics[scale=1.2]{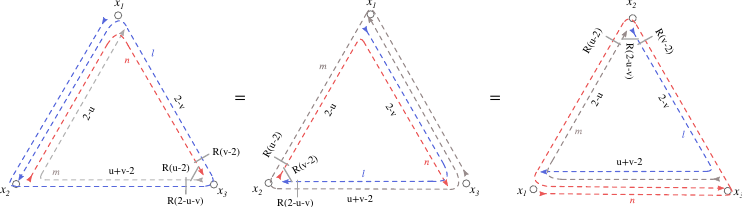}
\caption{Three equivalent results of the star-integral in the lhs of figure \ref{from_str_to_str}, achieved by three different $SU(2)$ conjugations: $[\mathbf{x_{21}}]^{l}$, $[\mathbf{x_{13}}]^{m}$ and $[\mathbf{x_{32}}]^{n}$.}
\label{the_three_rhs}
\end{center}
\end{figure}
Let us elaborate further these rhs, first completing \eqref{matrix_part_rhs_I} with the multiplications by coordinate-dependent spin matrices in the rhs:
\begin{align}
\begin{aligned}
\label{rhs1}
&[\mathbf{\overline{x_{21}} x_{13}}]^{n}  \left([\mathbf{\overline{x_{31}} x_{12}}]^{t_{l}}  \mathbf{R}_{n,l}(v-2)^{t_l} \mathbf{R}_{m,n}(u-2) \mathbf{R}_{m,l}(2-u-v)^{t_l} [\mathbf{\overline{x_{12}} x_{23}}]^{t_{l}} \right)^{t_l} [\mathbf{\overline{x_{32}} x_{21}}]^{m}=\\
= &{c_{m,l}(2-u-v)} {c_{n,l}(v-2)}\times \\ &  \qquad \times[\mathbf{\overline{x_{21}} x_{13}}]^{n} \left( [\boldsymbol{\varepsilon} \mathbf{\overline{x_{21}} x_{13}}]^{l}  \mathbf{R}_{n,l}(1-v) \mathbf{R}_{m,n}(u-2) \mathbf{R}_{m,l}(u+v-3) [\mathbf{\overline{x_{32}} x_{21}}\boldsymbol{\varepsilon}]^{l} \right)^{t_l} [\mathbf{\overline{x_{32}} x_{21}}]^{m} \,.
\end{aligned}
\end{align}
The latter expression is of the type appearing in any of the three lines of \eqref{moving_spins}. A similar expression can be written for the middle and right pictures in figure \ref{the_three_rhs}, namely
\begin{align}
\begin{aligned}
\label{rhs2}
&[\mathbf{\overline{x_{13}} x_{32}}]^{l}  \left([\mathbf{\overline{x_{23}} x_{31}}]^{t_{m}}   \mathbf{R}_{m,l}(2-u-v)^{t_m} \mathbf{R}_{n,l}(v-2) \mathbf{R}_{m,n}(u-2)^{t_m}  [\mathbf{\overline{x_{31}} x_{12}}]^{t_{m}} \right)^{t_m} [\mathbf{\overline{x_{21}} x_{13}}]^{n} =\\
= &{c_{m,l}(2-u-v)} {c_{m,n}(u-2)}\times \\ &  \qquad \times[\mathbf{\overline{x_{13}} x_{32}}]^{l}  \left([\boldsymbol{\varepsilon}\mathbf{\overline{x_{23}} x_{31}}]^{m}   \mathbf{R}_{m,l}(u+v-3) \mathbf{R}_{n,l}(v-2) \mathbf{R}_{m,n}(1-u)  [\mathbf{\overline{x_{21}} x_{13}}\boldsymbol{\varepsilon}]^{m} \right)^{t_m} [\mathbf{\overline{x_{21}} x_{13}}]^{n}  \,.
\end{aligned}
\end{align}
and, respectively:
\begin{align}
\begin{aligned}
\label{rhs3}
&[\mathbf{\overline{x_{31}} x_{13}}]^{m}  \left([\mathbf{\overline{x_{21}} x_{13}}]^{t_n}   \mathbf{R}_{m,n}(u-2)^{t_n} \mathbf{R}_{m,l}(2-u-v) \mathbf{R}_{n,l}(v-2)^{t_n}  [\mathbf{\overline{x_{23}} x_{31}}]^{t_{n}} \right)^{t_n} [\mathbf{\overline{x_{23}} x_{31}}]^{l} =\\
= &{c_{m,n}(u-2)} {c_{n,l}(v-2)}\times \\ &  \qquad \times [\mathbf{\overline{x_{31}} x_{13}}]^{m}  \left([\boldsymbol{\varepsilon}\mathbf{\overline{x_{31}} x_{12}}]^{n}   \mathbf{R}_{m,n}(1-u) \mathbf{R}_{m,l}(2-u-v) \mathbf{R}_{n,l}(1-v)  [\mathbf{\overline{x_{13}} x_{32}}\boldsymbol{\varepsilon}]^{n} \right)^{t_n} [\mathbf{\overline{x_{23}} x_{31}}]^{l}  \,.
\end{aligned}
\end{align}
The equality between the right hand side of expressions \eqref{rhs1}-\eqref{rhs3} is indeed the identity \eqref{moving_spins}, spelled in the text in the more compact language of matrix parts $\mathcal{M}$ and where $u,v$ are replaced by linear combinations of the rapidities of mirror excitations.
\section{Bra/ket overlap of $N=M$ magnons}
\label{app:firstover}
Let us consider the overlap of the wave-functions of an equal number of mirror magnons $N=M$ relative to two cuts with a point in common, say $x_1x_2$ and $x_2x_3$
\begin{equation}
\label{NN_over}
\langle\Psi_{12}(\mathbf{Y}_n)|\Psi_{13}(\mathbf{Y}_m)\rangle\,.
\end{equation}
with quantum numbers $\mathbf{Y}_n = (Y_{n_1},\dots, Y_{n_N})$ and $\mathbf{Y}_m = (Y_{m_1},\dots, Y_{m_N})$ where $Y_n = (n,\nu)$ and $Y_m = (m,\mu)$. In the following we compute explicitly a few cases of this overlap.
\subsection*{$N=M=1$}
The simplest overlap is the case $N=M=1$. Its computation is further simplified by considering the point $x_1\to \infty$ via the inversion map
\begin{equation}
\label{to_inf}
x^{\mu} \, \to \, \frac{x^{\mu}-x_1^{\mu}}{(x-x_1)^2}\,,
\end{equation}
 so that the wave-function is just a spinning propagator
\begin{equation}
|\Psi_{13}(Y_m)\rangle =  \frac{\left[\,\overline{\mathbf{x-x_3}}\,\right]^m}{(x-x_3)^{2(1+i \mu)}}\,.
\end{equation}
The overlap \eqref{NN_over} is computed by one star-triangle transformation, in amputated form after \eqref{to_inf}, in formulae:
\begin{align}
\begin{aligned}
\label{amp_star}
\langle \Psi_{1 2}(Y_n)|\Psi_{1 3}(Y_m)\rangle &= \!\! \int d^4 x \frac{[\mathbf{(x-x_2)}]^n [\overline{\mathbf{(x-x_3)}}]^m}{(x-x_2)^{2(1-i \nu)}(x-x_3)^{2(1+i \mu)}}= \frac{[\mathbf{x_{23}}]^n \mathbf{R}_{n,m}(i\nu-i\mu) [\overline{\mathbf{x_{23}}}]^m} {(x_2-x_3)^{2(i\mu- i \nu)}} \times \\  \!\!  \!\!  \!\! &\times\! \frac{\pi^2  \Gamma\left(1+\frac{m}{2}-i\mu\right) \Gamma\left(1+\frac{n}{2}+i\nu\right) \Gamma\left(\frac{m-n}{2}+i\mu-i\nu\right)}{\Gamma\left(1+\frac{m}{2}+i\mu\right) \Gamma\left(1+\frac{n}{2}-i\nu\right) \Gamma\left(1+\frac{m-n}{2}-i\mu+i\nu\right) \! \left(1+\frac{m+n}{2}-i\mu+i\nu\right)}=\\&= \frac{[\mathbf{x_{23}}]^n \mathbf{R}_{n,m}(i\nu-i\mu) [\overline{\mathbf{x_{23}}}]^m}{(x_2-x_3)^{2(i\mu- i \nu)}} \times H(Y_m|Y_n)\,,
\end{aligned}
\end{align}
and in graphical notation (in yellow ``star" region that undergo the transformation):
\begin{center}
\includegraphics[scale=1]{Overlap_N1}
\end{center}
The result for generic $x_1$ is restored by the inverse of transformation \eqref{to_inf}, namely
\begin{equation}
\label{from_inf}
x^{\mu} \, \to \, \frac{x^{\mu}}{x^2} +x_{1}^{\mu}\,,
\end{equation}
and reads
\begin{align}
\begin{aligned}
\langle \Psi_{12}(Y_n)|\Psi_{13}(Y_m)\rangle&= \frac{[\mathbf{x_{12}\overline{x_{23}}x_{31}}]^n \mathbf{R}(Y_n|Y_m) [\mathbf{\overline{x_{13}}x_{32}\overline{x_{21}}}]^m} {(x_{12}^2 x_{13}^2)^{(-i\mu+i \nu)} (x_2-x_3)^{2(i\mu- i \nu)}}  \times H(Y_m|Y_n) \,.
\end{aligned}
\end{align}
\subsection*{$N= M\geq 2$}
For $N=M\geq 2$ the procedure is a bit more cumbersome since the eigenfunctions have a more and more involved expression. In operator form the overlap has a compact expression in terms of the layer operators \eqref{Layer_inhom}, here for $N=2$: \begin{align}
\begin{aligned}
\langle\Psi_{12}(\mathbf{Y}_n)|\Psi_{13}(\mathbf{Y}_m)\rangle&=\frac{E(Y_{m_1})}{E(Y_{n_1})} \bar{\boldsymbol \Lambda}_1(Y_{n_2})  \bar{\boldsymbol \Lambda}_2(Y_{n_1}) {\boldsymbol \Lambda}_2(Y_{m_1})   {\boldsymbol \Lambda}_1(Y_{n_2})\,.
\end{aligned}
\end{align}
Nevertheless, the computation can be carried out quite straightforwardly via iterative applications of the star-triangle identity. Starting from $ \bar{\boldsymbol \Lambda}_2(Y_{n_1}) {\boldsymbol \Lambda}_2(Y_{m_1}) $ we show how it can be rewritten by three of star-triangle transformation (in yellow the region interested by the transformation at each step):
\begin{center}
\includegraphics[scale=0.97]{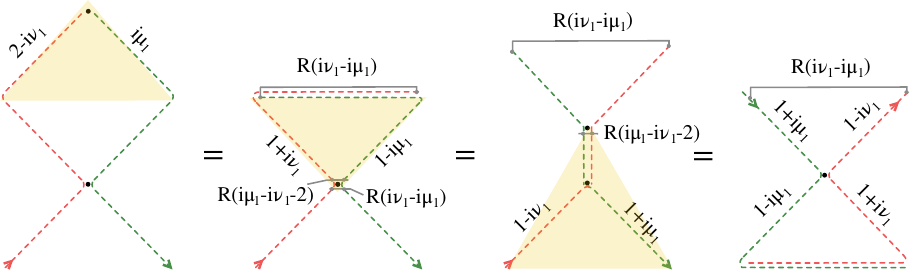}
\end{center}
The last three passages produce a factor:
\begin{equation}
A_{n_1,m_1}(2-i\nu_1,i\mu_1,2-i\mu_1+i\nu_1)= H(Y_{m_1}|Y_{n_1}) \times \frac{E(Y_{m_1})}{E(Y_{n_1})}\,.
\end{equation}
The convolution of the resulting expression with layers $\bar{\boldsymbol \Lambda}_1(Y_{n_2})$ and ${\boldsymbol \Lambda}_1(Y_{m_2})$ is straightforward to compute again via of star-triangle, and reads  
\begin{center}
\includegraphics[scale=1]{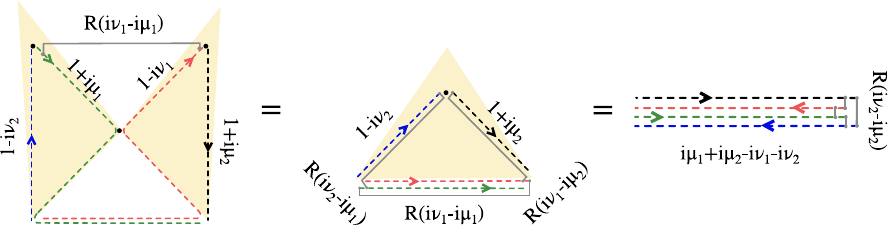}
\end{center}
producing the extra factors
\begin{equation}
H(Y_{m_2}|Y_{n_1}) 
H(Y_{m_1}|Y_{n_2})\,,
\end{equation}
and the last transformation produces one more factor $H(Y_{m_2}|Y_{n_2})$. Overall, the factor that multiplies the picture on the rhs, taking into account also the normalisation of the wave-functions amounts to   
\begin{equation}
H(Y_{m_1}|Y_{n_1})H(Y_{m_1}|Y_{n_2})H(Y_{m_2}|Y_{n_1})H(Y_{m_2}|Y_{n_2})\,,
\end{equation}
whereas the coordinate-dependence in the rhs is
\begin{equation}
\frac{\left[\overline{\mathbf{x_{32}}} \right]^{m_1} \!\left[\overline{\mathbf{x_{32}}} \right]^{m_2}  \! \mathbf{R}_{n_1,m_2}(i\nu_1\! -\! i\mu_2)  \mathbf{R}_{n_1,m_1}(i\nu_1\!-\! i\mu_1)  \mathbf{R}_{n_2,m_2}(i\nu_2\!- \! i\mu_2)  \mathbf{R}_{n_2,m_1}(i\nu_2\!-\! i\mu_1)\! \left[{\mathbf{x_{23}}} \right]^{n_1}\! \left[{\mathbf{x_{23}}} \right]^{n_2}}{(x_{23}^2)^{i\mu_1+i\mu_2 -i\nu_1-i\nu_2}}
\end{equation}
The derivation for $N=2$ can be upgraded to higher $N$ by a similar iteration of star-triangle and star-star transformations; the general formula reads
\begin{equation}
\langle \Psi_{1 2} (\mathbf{Y}_n) |\Psi_{1 3}(\mathbf{Y}_m)\rangle = \prod_{h,k=1}^N   H(Y_{m_h}|Y_{n_k})  \frac{\left[\overline{\mathbf{x_{32}}} \right]^{m_{N-h}}  \mathbf{R}_{n_k,m_{N-h}}(i\nu_k\! -\! i\mu_{N-h})   \left[{\mathbf{x_{23}}} \right]^{n_k}}{(x_{23}^2)^{i\mu_{N-h}-i\nu_k}}\,,
\end{equation}
where the product of two $\mathbf{R}$-matrices shall be understood as matrix product whenever two spaces are the same, i.e. $\mathbf{R}_{n,m}\mathbf{R}_{n',m}$ or $\mathbf{R}_{n,m}\mathbf{R}_{n,m'}$.
In conclusion, using the following \textbf{bold} notations
\begin{equation}
\label{bold_notations}
 H(\mathbf{Y}_m|\mathbf{Y}_n)= \prod_{h,k=1}^N H({Y}_{m_{h}}|{Y}_{n_k})  \,,\,\,\,\,  \mathbf{R}(\mathbf{Y}_n|\mathbf{Y}_m)=\prod_{k=1}^N \prod_{h=1}^M \mathbf{R}({Y}_{n_k}|{Y}_{m_{N-h}})\,,
\end{equation}
and the general $f(\mathbf{Y}_n) = \prod_{k=1}^N f(Y_{n_k})$, the overlap \eqref{NN_over} of a bra/ket wave-functions defined along adjacent cuts reads
\begin{equation}
\langle \Psi_{1 2} (\mathbf{Y}_n) |\Psi_{1 3}(\mathbf{Y}_m)\rangle=  H(\mathbf Y_{m}| \mathbf Y_{n})  \times \frac{ \left[{\mathbf{x_{23}}} \right]^{\mathbf n}  \mathbf{R}(\mathbf{Y}_n|\mathbf{Y}_m)   \left[\overline{\mathbf{x_{32}}} \right]^{\mathbf m }}{(x_{23}^2)^{i\boldsymbol{\mu}-i\boldsymbol{\nu}}}\,,
\end{equation}
or, for a generic point $x_1\neq \infty$,
\begin{equation}
\langle \Psi_{1 2} (\mathbf{Y}_n) |\Psi_{1 3}(\mathbf{Y}_m)\rangle=  H(\mathbf Y_{m}| \mathbf Y_{n})  \times \frac{ \left[\overline{\mathbf{x_{23}}}\mathbf{x_{31}} \right]^{\mathbf n}  \mathbf{R}(\mathbf{Y}_n|\mathbf{Y}_m)   \left[\mathbf{x_{12}}\overline{\mathbf{x_{23}}} \right]^{\mathbf m }}{(x_{12}^2 x_{13}^2/x_{23}^2)^{i\boldsymbol{\mu}-i\boldsymbol{\nu}}}\,.
\end{equation}
\section{Bra/ket overlap of $N=M+1$ magnons}
\label{NMp1}
The first instance of the overlaps of case II in section \ref{sec:overs} is when the width $N$ of the bra wave-function exceeds that of the ket $M$ by one, and thus the last point in the bra wave-function is identified with the endpoint $x_3$ of the ket. In the following we compute explicitly a few cases.\subsection*{$N=2,M=1$}
This overlap in formulae reads
\begin{equation}
 \langle \Psi_{12} ({Y}_{n_1},Y_{n_2}) |\Psi_{13}({Y}_m) \otimes  \delta^{(4)}_{x_3} \rangle = \int d^4 z \, \overline{\Psi_{12}} ({Y}_{n_1},Y_{n_2}|z,x_3) \Psi_{13} ({Y}_{m};z)\,.
\end{equation}
The result of the overlap is a straightforward consequence  of two (amputated) star-triangle identities
\begin{center}
\includegraphics[scale=1]{Overlap_N2M1}
\end{center}
Keeping track of the coefficients and the $\mathbf{R}$-matrices produced by the transformations, and also of the normalisation factors in \eqref{eig_biscalar}, the result reads
\begin{align}
\begin{aligned}
& \langle \Psi_{12} ({Y}_{n_1},Y_{n_2}) |\Psi_{13}({Y}_m) \otimes  \delta^{(4)}_{x_3} \rangle = \\&= \frac{E(Y_{n_1})}{E(Y_m)}  H(Y_m|Y_{n_1})H(Y_m|Y_{n_2}) \! \times\! \frac{ [\mathbf{x_{23}}]^{m}   \mathbf{R}_{n_1,m}(i\nu_1-i\mu) \mathbf{R}_{n_2,m}(i\nu_2-i\mu)   [\overline{\mathbf{x_{23}}}]^{n_1}  [\overline{\mathbf{x_{23}}}]^{n_2} }{(x_{23}^2)^{1+i \mu-i\nu_1-i\nu_2}} E(Y_{n_1})^{-1} =\\ &= \frac{H(Y_m|Y_{n_1})H(Y_m|Y_{n_2})}{E(Y_m)}   \times
 \frac{ [\overline{\mathbf{x_{23}}}]^{n_1}  [\overline{\mathbf{x_{23}}}]^{n_2}   \mathbf{R}(Y_{n_1}|Y_m) \mathbf{R}(Y_{n_2}|Y_m) [\mathbf{x_{23}}]^{m}}{(x_{23}^2)^{1+i \mu-i\nu_1-i\nu_2}}  \,.
\end{aligned}
\end{align}
\subsection*{$N=3,M=2$}
Next, we shall consider the case of a bra of width $N=2$ and with a ket-function of width $M=3$, with the third coordinate of the bra identified with the endpoint $x_3$ of the ket, in formulae
\begin{align}
\begin{aligned}
& \langle \Psi_{12} ({Y}_{n_1},Y_{n_2},Y_{n_3}) |\Psi_{13}({Y}_{m_1},{Y}_{m_2})\otimes \delta^{(4)}_{x_3} \rangle =\\&= \int d^4 z \int d^4 z' \, \overline{\Psi}_{12} ({Y}_{n_1},Y_{n_2},Y_{n_3}| z,z',x_3) \Psi_{13} ({Y}_{m_1},Y_{m_2}|z,z')\,,
\end{aligned}
\end{align}
and in pictures:
\begin{center}
\includegraphics[scale=0.8]{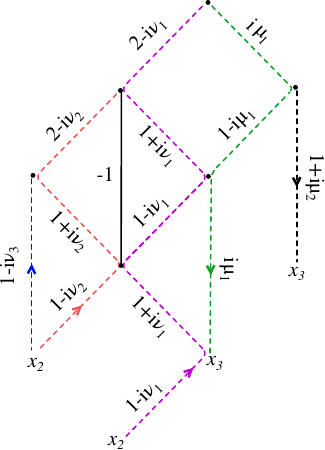}
\end{center}
Let us spell out the star-triangle transformations needed to compute the overlap, enlightening in yellow step-by-step the interested star or triangle region. The first few passages regard only the longest convoluted layers, namely
\begin{center}
\includegraphics[scale=0.8]{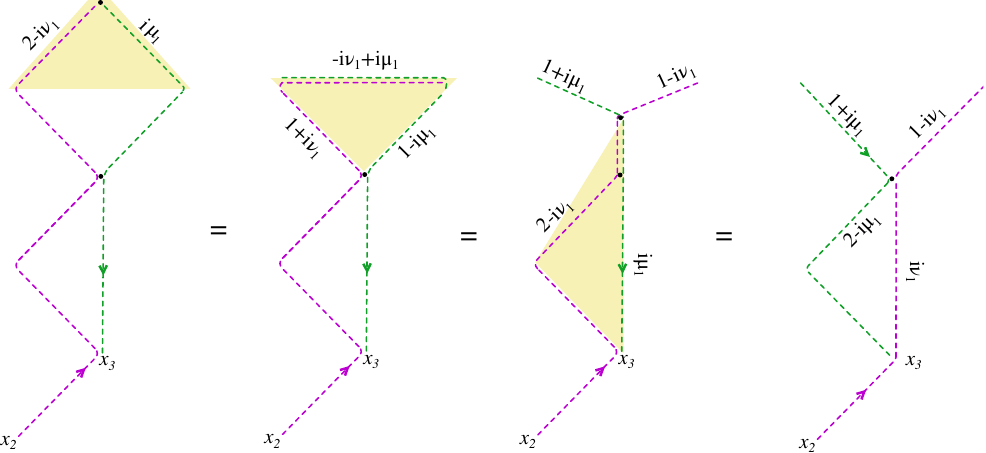}
\end{center}
followed by some additional steps 
\begin{center}
\includegraphics[scale=0.8]{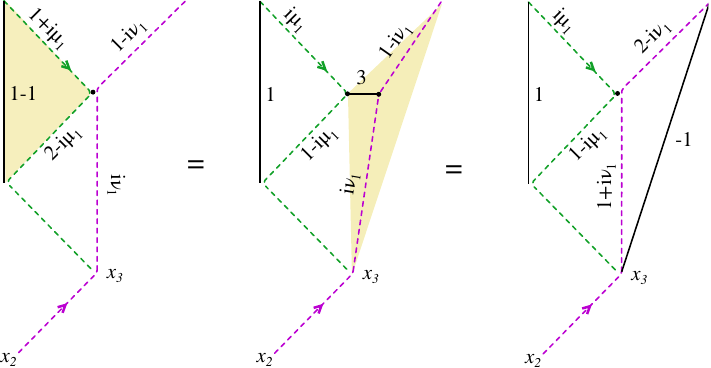}
\end{center}
At this point, gluing back into the picture the shorter layers of the eigenfunctions, one gets
\begin{center}
\includegraphics[scale=1]{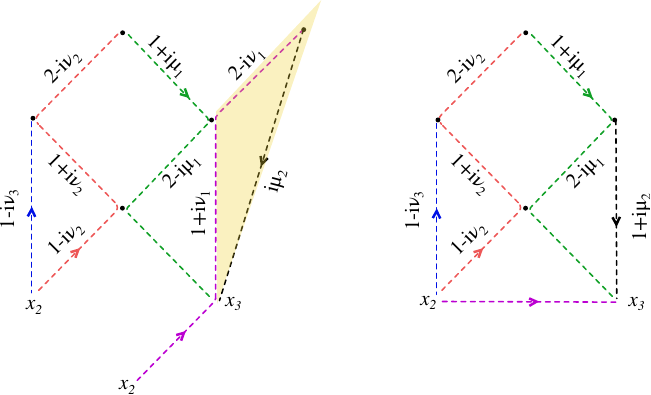}
\end{center}
Now, one can realise that the last picture is nothing but the overlap computed for $N=M=2$ with quantum numbers weights $(Y_{m_1},Y_{m_2})$ and $(Y_{n_2},Y_{n_3})$ respectively. Taking the result of that former computation one gets the analytic form of this overlap, which also accounts for the star-triangle factors and wave-function normalisation:
\begin{align}
\begin{aligned}
&  \langle \Psi_{12} ({Y}_{n_1},Y_{n_2},Y_{n_3}) |\Psi_{13}({Y}_{m_1},{Y}_{m_2})\otimes \delta^{(4)}_{x_3} \rangle= \\ &=
\frac{H_{m_1,n_1}(\mu_1,\nu_1) H_{m_1,n_2}(\mu_1,\nu_2) H_{m_1,n_3}(\mu_1,\nu_3)  H_{m_2,n_1}(\mu_2,\nu_1) H_{m_2,n_2}(\mu_2,\nu_2) H_{m_2,n_3}(\mu_2,\nu_3) }{E(Y_{m_1})E(Y_{m_2})}\times \\& \times \frac{ [\overline{\mathbf{x_{23}}}]^{n_1}[\overline{\mathbf{x_{23}}}]^{n_2}[\overline{\mathbf{x_{23}}}]^{n_3} \mathbf{R}_{n_1,m_1}(i\nu_1-i\mu_1)  \mathbf{R}_{n_1,m_2}(i\nu_1-i\mu_2)  \cdots \mathbf{R}_{n_3,m_2}(i\nu_3-i\mu_2)   [{\mathbf{x_{23}}}]^{m_1}  [{\mathbf{x_{23}}}]^{m_2} }{(x_{23}^2)^{1+i \mu_1+i\mu_2-i\nu_1-i\nu_2-i\nu_3}}  \,.
\end{aligned}
\end{align}
In \textbf{bold} notation, the general results for $N=M+1$ is
\begin{align}
\begin{aligned}
\langle \Psi_{1 2} (\mathbf{Y}_n) |\Psi_{ 1 3}(\mathbf{Y}_m)\otimes \delta^{(4)}_{x_3} \rangle& = 
\frac{{H}(\mathbf{Y}_m|\mathbf{Y}_n)}{E(\mathbf{Y}_{m})} \times \frac{ [\overline{\mathbf{x_{23}}}]^{\mathbf{n}} \mathbf{R}(\mathbf{Y}_n|\mathbf{Y}_m)  [{\mathbf{x_{23}}}]^{\mathbf{m}}}{(x_{23}^2)^{1+i \boldsymbol{\mu}-i\boldsymbol{\nu}}}\,,
\end{aligned}
\end{align}
which one can check to hold for $N>2$ following a similar chain of transformations as for $N=2$.
\section{Bra/ket overlap of $N=M-1$ magnons}
\subsection*{$N=1,M=2$}
We consider the overlap of a bra wave-function of length $N=1$ mirror magnons with a ket of length $M=2$. The second point of the ket is identified with the endpoint $x_2$ of the bra-function, in formulae
\begin{equation}
 \langle \Psi_{12} ({Y}_{n})\otimes \delta^{(4)}_{x_2} |\Psi_{13}({Y}_{m_1},{Y}_{m_2})\rangle = \int d^4 z \, \overline{\Psi}_{12} ({Y}_{n}|z) \Psi_{13} ({Y}_{m_1},Y_{m_2}|z,x_2)\,.
\end{equation}
The overlap is computed via a two star-triangle transformations analogous to the ones depicted for $N=2,M=1$ in section \ref{NMp1}, upon the exchange of bra and ket wave-functions. The result reads
\begin{align}
\begin{aligned}
\label{MN21}
& \langle \Psi_{12} ({Y}_{n})\otimes \delta^{(4)}_{x_2} |\Psi_{13}({Y}_{m_1},{Y}_{m_2})\rangle=\\ &=
E(Y_n)   H(Y_{m_1}|Y_n) H(Y_{m_2}|Y_n) \frac{ [\overline{\mathbf{x_{23}}}]^{n} \mathbf{R}_{n,m_1}(i\nu-i\mu_1) \mathbf{R}_{n,m_2}(i\nu-i\mu_2)   [{\mathbf{x_{23}}}]^{m_1}  [{\mathbf{x_{23}}}]^{m_2} }{(x_{23}^2)^{1+i \mu_1+i\mu_2-i\nu}}  \,.
\end{aligned}
\end{align}

\subsection*{$N=2,M=3$}
In a similar fashion to last section, we repeat the computation for larges sizes $N=2, M=3$
\begin{equation}
\langle \Psi_{12} ({Y}_{n_1},Y_{n_2})\otimes \delta^{(4)}_{x_2} |\Psi_{13}({Y}_{m_1},{Y}_{m_2},{Y}_{m_3}) \rangle
\,.
\end{equation}
The result follows again from an iteration of star-triangle transformations and mimics the case $N=3, M=2$ of section \ref{NMp1}, upon the exchange of bra/ket wave-functions. In formulae, we found
\begin{align}
\begin{aligned}
\label{MN32}
&\,\,\,\,\,{H(Y_{m_1}|Y_{n_1}) H(Y_{m_1}|Y_{n_2}) H(Y_{m_2}|Y_{n_1})  H(Y_{m_2}|Y_{n_2}) H(Y_{m_3}|Y_{n_1}) H(Y_{m_3}|Y_{n_2})}{E(Y_{n_1})E(Y_{n_2})}\times \\& \times \frac{ [\overline{\mathbf{x_{23}}}]^{n_1}[\overline{\mathbf{x_{23}}}]^{n_2} \mathbf{R}_{n_1,m_3}(i\nu_1-i\mu_3)  \mathbf{R}_{n_1,m_2}(i\nu_1-i\mu_2)  \cdots \mathbf{R}_{n_2,m_1}(i\nu_2-i\mu_1)   [{\mathbf{x_{23}}}]^{m_1}  [{\mathbf{x_{23}}}]^{m_2}[{\mathbf{x_{23}}}]^{m_3} }{(x_{23}^2)^{1+i \mu_1+i\mu_2+i\mu_3-i\nu_1-i\nu_2}}  \,.
\end{aligned}
\end{align}
The computations done so far in an explicit fashion by graphical passages (star-triangle relation) can be extended to higher $N$ and $M$ by an easy iteration of passages, despite becoming more cumbersome. The general formula that result is evident from \eqref{MN21} and \eqref{MN32}
\begin{align}
\begin{aligned}
\langle \Psi_{1 2} (\mathbf{Y}_n) )\otimes \delta^{(4)}_{x_2}  |\Psi_{ 1 3}(\mathbf{Y}_m)\rangle& = 
{{H}(\mathbf{Y}_m|\mathbf{Y}_n)}{E(\mathbf{Y}_{n})} \times \frac{ [\overline{\mathbf{x_{23}}}]^{\mathbf{n}} \mathbf{R}(\mathbf{Y}_n|\mathbf{Y}_m)  [{\mathbf{x_{23}}}]^{\mathbf{m}}}{(x_{23}^2)^{1+i \boldsymbol{\mu}-i\boldsymbol{\nu}}}\,.
\end{aligned}
\end{align}

\section{Bra/ket$^2$ overlap of $N=M+R$ magnons}
\label{app:lastover}
We consider the overlap of the one (bra) wave-functions of $N$ magnons relative to a cut, say $(12)$,
\begin{equation}
\mathbf{Y}_n = (Y_{n_1},\dots,Y_{n_N})\,,
\end{equation}
with two wave-functions of respectively $M$ and $R$ magnons such that $M+R=N$
\begin{equation}
\mathbf{Y}_m = (Y_{m_1},\dots,Y_{m_M})\,,\,\, \text{and}\,\, \mathbf{Y}_r = (Y_{r_1},\dots,Y_{r_R})\,,
\end{equation}
and relative to the cuts $x_1x_3$ and $x_2x_3$ respectively:
\begin{equation}
 \langle \Psi_{12}(\mathbf{Y}_n)| \Psi_{13} (\mathbf{Y}_m) \otimes \Psi_{32} (\mathbf{Y}_r) \rangle\,.
\end{equation}
We the standard notation adopted for the magnon's quantum numbers is $Y_a = (\alpha,a) \in \mathbb{R} \times \mathbb{N}$.
\subsection*{$N=2$, $M=1$, $R=1$}
Let us start from the simplest non-trivial case $N=2$ and $M=R=1$, in formulae
\begin{equation}
 \langle \Psi_{12}({Y}_{n_1},{Y}_{n_2})| \Psi_{13} ({Y}_m) \otimes \Psi_{23} ({Y}_r) \rangle= \int d^4 z d^4 z'\, \overline{\Psi}_{12}({Y}_{n_1},Y_{n_2}|z,z') \Psi_{13}({Y}_{m}|z) \Psi_{32}({Y}_{r}|z')\,.
\end{equation}
The computation of the overlap integral is made by five applications of the star-triangle identity,
\begin{center}
\includegraphics[scale=0.9]{Oversplit_21_I}
\end{center}
\begin{center}
\includegraphics[scale=0.9]{Oversplit_21_II}
\end{center}
The expression of the rhs of last picture is a product of spinning propagators of section \ref{sec:eigenf}, whose symmetric spinor indices are contracted with a product of $R$-matrices. This rhs is multiplied by the factors produced at each star-triangle transformation as well as by the normalisation of the eigenfunctions. The product of $R$-matrices in the rhs reads
\begin{equation}
\left(\mathbf{R}_{n_1r}(i\rho-i\nu_1)^{t_r} \mathbf{R}_{n_2 r}(i\rho-i\nu_2)^{t_r} \mathbf{R}_{mr}(i\mu-i\rho-2)^{t_r} \mathbf{R}_{n_2 m}(i\nu_2-i\mu)\mathbf{R}_{n_1 m}(i\nu_1-i\mu)\right)^{t_r}\,,
\end{equation}
where $t_r$ stands for the transposition of the spinor indices of the excitation $Y_r$.
Recalling the crossing property \eqref{unitarity_crossing}, the last expression can be written up to scalar factors as
\begin{equation}
\left((-\boldsymbol{\varepsilon})^{\otimes r } \mathbf{R}_{n_1r}(i\nu_1-i\rho-1) \mathbf{R}_{n_2 r}(i\nu_2-i\rho-1) \mathbf{R}_{n_2 m}(i\nu_2-i\mu)\mathbf{R}_{n_1 m}(i\nu_1-i\mu) \mathbf{R}_{mr}(i\rho-i\mu+1)\boldsymbol{\varepsilon}^{\otimes r }\right)^{t_r}
\end{equation} 
where for the first time in the overlaps one encounters $R$-matrices depending on a difference of rapidities shifter by $\pm i$. Collecting the scalar factors produced by the star-triangle identities, by the application of property \eqref{unitarity_crossing} and by the normalisation of the wave-functions \eqref{eig_biscalar}, one gets:
\begin{align}
\begin{aligned}
&{H(Y_{m_1}|Y_{n_1})  H(Y_{m_1}|Y_{n_2})  H(Y_{m_2}|Y_{n_1})   H(Y_{m_2}|Y_{n_2}) H(Y_{m_3}|Y_{n_1}) H(Y_{m_3}|Y_{n_2}) }{E(Y_{n_1})E(Y_{n_2})}\times \\ \times &\frac{ [\overline{\mathbf{x_{23}}}]^{n_1}[\overline{\mathbf{x_{23}}}]^{n_2} \mathbf{R}_{n_1,m_1}(i\nu_1-i\mu_1)  \mathbf{R}_{n_1,m_2}(i\nu_1-i\mu_2)  \cdots \mathbf{R}_{n_2,m_3}(i\nu_2-i\mu_3)   [{\mathbf{x_{23}}}]^{m_1}  [{\mathbf{x_{23}}}]^{m_2}[{\mathbf{x_{23}}}]^{m_3} }{(x_{23}^2)^{1+i \mu_1+i\mu_2+i\mu_3-i\nu_1-i\nu_2}}  \,.
\end{aligned}
\end{align}

\subsection*{$N=3$, $M=2$, $R=1$}
It is interesting to see the generalization of previous computation in presence of longer wave-functions on the ket site of the overlap, e.g. $M=2$. This computation goes in the direction of a general formula for the case III of section \ref{sec:overs}. In formulae, the overlap is a double convolution of three wave-functions
\begin{equation}
\label{N3M2R1}
\int d^4 z d^4 z'\, \overline{\Psi}_{12}({Y}_{\mathbf n};z,z',z'') \Psi_{31}({Y}_{\mathbf m};z,z') \Psi_{32}({Y}_{r};z'')\,,
\end{equation}
which carry mirror excitations with the following quantum numbers
\begin{equation}
\mathbf{Y}_{ n} =(Y_{n_1},Y_{n_2},Y_{n_3})\,,\,\,\mathbf{Y}_{m} =(Y_{m_1},Y_{m_2})\,,\,\text{and} \,\,Y_{r}\,.
\end{equation}
The computation of the integral \eqref{N3M2R1} is achieved by several star-triangle transformation, that we depict step-by-step starting from the graphical form of \eqref{N3M2R1}. The sign equal in the graphical passages shall be understood modulo scalar factors -- which are reported explicitly under each set of passages.
\begin{center}
\includegraphics[scale=0.77]{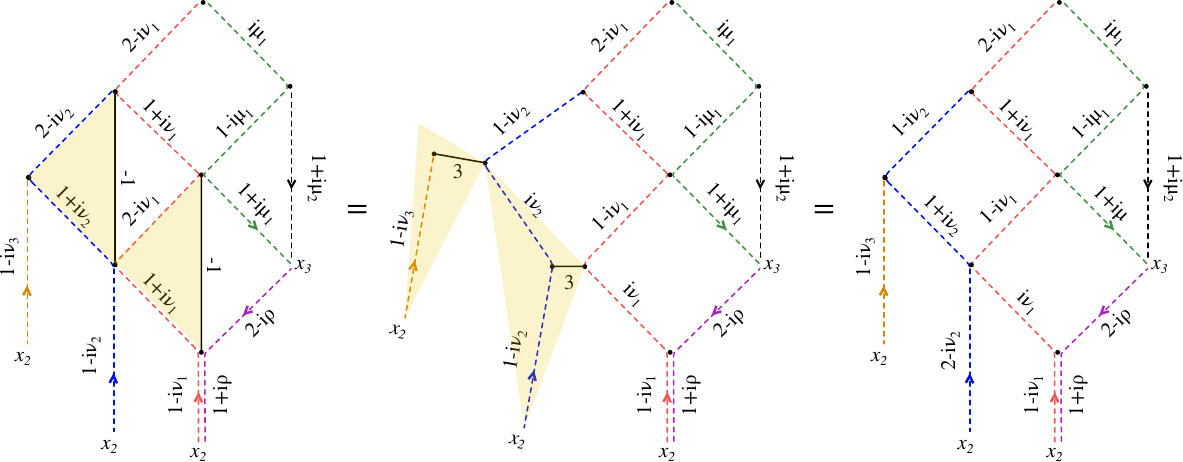}
\end{center}
\begin{equation*}
E(Y_{n_1}) E(Y_{n_3})^{-1}\,,
\end{equation*}
\begin{center}
\includegraphics[scale=0.8]{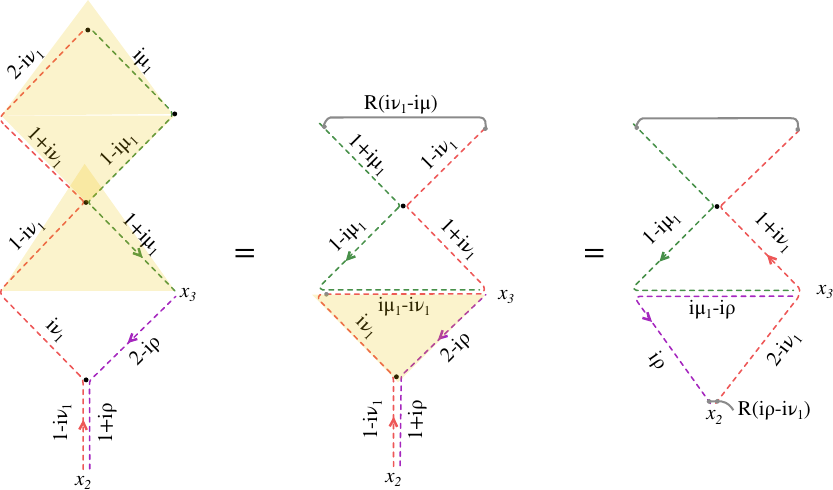}
\end{center}
\begin{equation*}
H(Y_{m_1}|Y_{n_1})E(Y_{n_1}) E(Y_{m_1})^{-1} A_{n_1,r}(i\nu_1,2-i\rho,2-i\nu_1+i\rho) H(Y_{n_1}|Y_{r})E(Y_{n_1})^{-1} E(Y_{r})  \,,
\end{equation*}
\begin{center}
\includegraphics[scale=0.8]{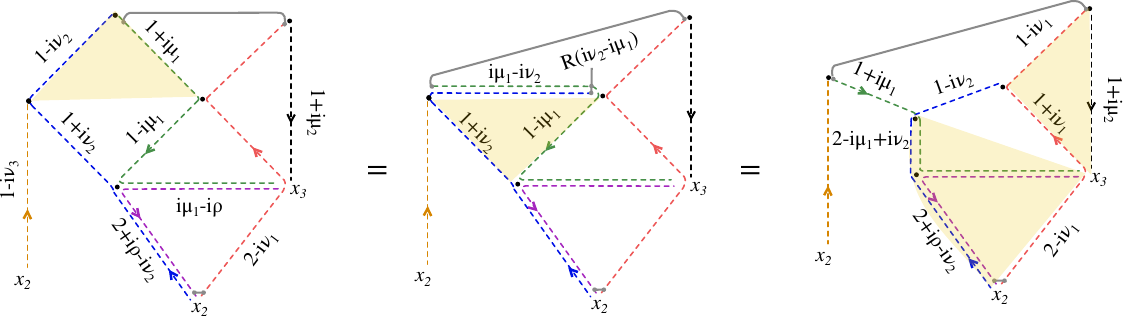}
\end{center}
\begin{equation*}
H(Y_{m_2}|Y_{n_1}) {H(Y_{m_1}|Y_{n_2}) H(Y_{n_2}|Y_{r})} \frac{1}{H(Y_{m_1}|Y_{r})\,r_{m_1r}(i\rho-i\mu_1)}  \,,
\end{equation*}
\begin{center}
\includegraphics[scale=0.82]{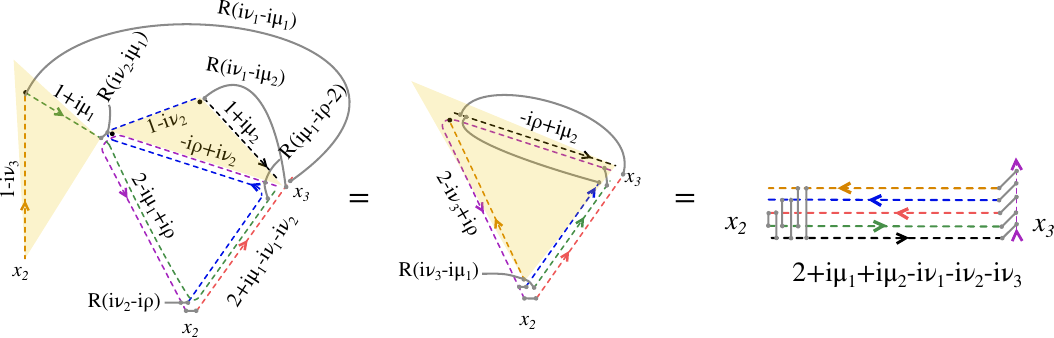}
\end{center}
\begin{align*}
H(Y_{m_2}|Y_{n_3}) H(Y_{m_2}|Y_{n_2}) H(Y_{n_3}|Y_{r}) H(Y_{m_2}|Y_{n_3}) H(Y_{n_3}|Y_{r}) \frac{1}{H(Y_{m_2}|Y_{r})\,r_{m_2 r}(i\rho-i\mu_2)}\,.
\end{align*}
The product of $R$-matrices produced at the end of the transformations can be read out of the rhs of last picture and expressed as follows:
\begin{align}
\begin{aligned}
&\left(\mathbf{R}_{n_1r}(i\rho-i\nu_1)^{t_r} \mathbf{R}_{n_2 r}(i\rho-i\nu_2)^{t_r} \mathbf{R}_{n_3r}(i\rho-i\nu_3)^{t_r} \times\right. \\& \times \mathbf{R}_{n_1 m_2}(i\nu_1-i\mu_2) \mathbf{R}_{n_2 m_2}(i\nu_2-i\mu_2)\mathbf{R}_{n_3 m_2}(i\nu_3-i\mu_2)\times \\&  \times  \mathbf{R}_{n_1 m_1}(i\nu_1-i\mu_1) \mathbf{R}_{n_2 m_1}(i\nu_2-i\mu_1)\mathbf{R}_{n_3 m_1}(i\nu_3-i\mu_1)\times \\& \left.\times \mathbf{R}_{m_2r}(i\mu_2-i\rho-2)^{t_r} \mathbf{R}_{m_1r}(i\mu_1-i\rho-2)^{t_r}\right)^{t_r}\, .
\end{aligned}
\end{align}
Using crossing \eqref{unitarity_crossing} to treat the $R$-matrix transposition, the last expression takes the form
\begin{align}
\begin{aligned}
&\left(\boldsymbol{\varepsilon}^{ r } \mathbf{R}_{n_1r}(-i\rho+i\nu_1-1) \mathbf{R}_{n_2 r}(-i\rho+i\nu_2-1) \mathbf{R}_{n_3r}(-i\rho+i\nu_3-1) \times \right. \\& \times \mathbf{R}_{m_1r}(1-i\mu_1+i\rho) \mathbf{R}_{m_2r}(1-i\mu_2+i\rho) \times  \\& \times \mathbf{R}_{n_3 m_1}(i\nu_3-i\mu_1) \mathbf{R}_{n_2 m_1}(i\nu_2-i\mu_1)\mathbf{R}_{n_1 m_1}(i\nu_1-i\mu_1)\times \\&\left. \times \mathbf{R}_{n_3 m_2}(i\nu_3-i\mu_2) \mathbf{R}_{n_2 m_2}(i\nu_2-i\mu_2)\mathbf{R}_{n_1 m_2}(i\nu_1-i\mu_2) \boldsymbol{\varepsilon}^{ r }\right)^{t_r}\,,
\end{aligned}
\end{align}
and gets decorated by a few crossing factors:
\begin{equation}
c_{n_1r}(i\rho-i\nu_1)c_{n_2r}(i\rho-i\nu_2)c_{n_2r}(i\rho-i\nu_2)c_{m_1r}(-2-i\rho+i\mu_1)c_{m_2r}(-2-i\rho+i\mu_2)\,.
\end{equation}
At this point, collecting all the scalar factors produces along the way, including the normalisation of the wave-functions
\begin{equation}
E(Y_{n_1})^{-2}  E(Y_{n_2})^{-1} E(Y_{m})^2 \,,
\end{equation}
and using the property $r_{nm}(u) = c_{nm}(u)  c_{nm}(-u-2) = 1/ c_{nm}(-u-1)  c_{nm}(u+1)$, leads to the expression
\begin{equation}
E(Y_r) \prod_{i=1}^2\prod_{j=1}^3 \frac{H(Y_{m_j}|Y_{n_i}) H(Y_{n_i} |Y_r)}{H(Y_{m_j}|Y_r)} \frac{c_{rn}(i\rho-i\nu_j)}{c_{rm}(i\rho-i\mu_i)} \frac{1}{E(Y_{n_j})}\,.
\end{equation}
In \textbf{bold} notation, and denoting excitations with shifted rapidity by $Y_r^{\pm} =(\rho\pm i,r)$, the result of the computation of \eqref{N3M2R1} reads
\begin{align}
\begin{aligned}
&\frac{{H}(\mathbf{Y}_{m}|\mathbf{Y}_{n}){H}(\mathbf{Y}_{n}|{Y}_{r})}{{H}(\mathbf{Y}_{m}|{Y}_{r})} \frac{c(Y_{r}|\mathbf{Y}_{n})}{c (Y_{r}|\mathbf{Y}_{m})} \frac{E({Y}_{r})}{E(\mathbf{Y}_{n})}\, \frac{\boldsymbol{\varepsilon}^{r}  [\overline{\mathbf{x_{23}}}]^{\mathbf{n}} \mathbf{R}(\mathbf{Y}_{n}|Y_r^-)  \mathbf{R}(\mathbf{Y}_{n}|\mathbf{Y}_{m})  \mathbf{R}(Y_r^-|\mathbf{Y}_{m})  [{\mathbf{x_{23}}}]^{\mathbf m} \boldsymbol{\varepsilon}^{r} }{(x_{23}^2)^{2-i\boldsymbol{\mu}-i\boldsymbol{\nu}}}\,.
\end{aligned}
\end{align}
This type of computation can be extended to large $N,M,R$ in a systematical way, guided by the procedure of star-triangle transformations explained here in detail. The overlaps of this class are described by a general formula for case III in section \ref{sec:overs} and demonstrates the conjectural formulae of \cite{Basso:2018cvy}, obtained via a guess based on integrability and the decoupling property of \emph{hexagons}.

\section{Hexagonalisation dictionary}
The formulae presented in the text can be compared with the formalism of hexagonalisation \cite{Basso:2018cvy} after establishing a dictionary between the notations. First of all it should be noticed that the quantum numbers (\emph{rapidity} and \emph{bound-state index}) of an excitation $(u,a)$ are translated in the notation of this work as ${Y_{u,a}}=(u-i/2,a-1)$.
The \emph{abelian factor} $H_{a,b}(u,v)$ given in (2.53) of \cite{Basso:2018cvy} in our notation is re-expressed as
\begin{equation}
\tilde{H}(Y_{u,a}|Y_{v,b}) E(Y_{u,a})^{\frac{3}{2}} E(Y_{v,b})^{\frac{3}{2}}= \frac{g^2}{H_{b,a}(v,u)} \,,
\end{equation}
and the energy factors are related by exponentiation modulo $g^2$, namely $E_a(u)=-\tfrac{1}{g^2} \log E(Y_{u,a})$.
As regard the fused SU(2) $R$-matrix, the notations translate as
\begin{equation}
R_{a,b}(u,v)= \mathbf{R}_{a-1,b-1}(-iu+iv) =\mathbf{R}(Y_{v,b}|Y_{u,a})  \,,
\end{equation}
and the crossing-symmetry factor $c_{a,b}(u,v)=c_{a,b}(u-v)$ appearing in (2.36) of \cite{Basso:2018cvy} translates with the function $1/c_{a,b}(iu-iv)=1/c(Y_{u,a}|Y_{v,b})$ in this paper. According to this dictionary, one can check the matrix part \eqref{Mat_part_II} of the overlap translates as
\begin{align}
\begin{aligned}
\mathcal{M}(\mathbf{Y}_{v,b},\mathbf{Y}_{w,c}|\mathbf{Y}_{u,a})&= \frac{c(\mathbf{Y}_{u,a}|\mathbf Y_{w,c})}{{c}(\mathbf{Y}_{v,b}|\mathbf Y_{w,c})} \times\mathbf{R}(\mathbf{Y}_{v,b}|\mathbf Y_{w,c}^{+})  \mathbf{R}(\mathbf{Y}_{v,b}|\mathbf Y_{u,a})  \mathbf{R}(\mathbf{Y}_{w,c}^{+}|\mathbf Y_{u,a})=\\&= \frac{ c(\mathbf v, \mathbf w)}{{c}(\mathbf u ,\mathbf w)} \times {R}(\mathbf w^{++} , \mathbf v)  {R}(\mathbf u, \mathbf v)   {R}(\mathbf u,\mathbf w^{++})\,,
\end{aligned}
\end{align}
that is (2.51) of \cite{Basso:2018cvy}, taking into account that $w^{++}$ and $Y_{w,c}^+$ are the same shift by $+i$ in the rapidity. The conjectured general formula (2.48) in the paper of Basso, Caetano and Fleury is obtained in this work as the product of the dynamical factor $\mathcal{D}(\mathbf{Y}_{v,b},\mathbf{Y}_{w,c}|\mathbf{Y}_{u,a})$ dressed by ``half" of the interacting terms in the measure \eqref{rho_def} over each set of magnons, i.e.
\begin{equation}
 \prod_{k \neq j}^N {H_{n_k n_h}(\nu_k,\nu_h)}=  \prod_{k > j}^N {H_{n_k n_h}(\nu_k,\nu_h)}  \prod_{k < j}^N {H_{n_k n_h}(\nu_k|\nu_h)} \equiv H_{<}(\mathbf{Y}_n|\mathbf{Y}_n) \times H_{>}(\mathbf{Y}_n|\mathbf{Y}_n)\,.
\end{equation}
Thus, (2.48) of \cite{Basso:2018cvy}, taking into account that the authors work in the assumption $M+R=N$, in the formalism of this work translates as
\begin{align}
\begin{aligned}
H(\mathbf u \to \mathbf v | \mathbf w) &= \frac{ \mathcal{D}(\mathbf{Y}_{v,b},\mathbf{Y}_{w,c}|\mathbf{Y}_{u,a}) }{H_{>}(\mathbf{Y}_{u,a}|\mathbf{Y}_{u,a}) H_{>}(\mathbf{Y}_{v,b}|\mathbf{Y}_{v,b})H_{>}(\mathbf{Y}_{w,c}|\mathbf{Y}_{w,c})}  \left(E(\mathbf Y_{v,b}) E(\mathbf Y_{w,c}) E(\mathbf Y_{u,a})\right)^{\frac{3}{2}}\,.
\end{aligned}
\end{align}
\bibliographystyle{JHEP}
\bibliography{biblio_v2}

\end{document}